\documentclass[showpacs,prb,twocolumn]{revtex4-1}

\usepackage{color}\usepackage{enumerate}
\usepackage{amsmath}\usepackage{amssymb}\usepackage{stmaryrd}
\usepackage{multirow}
\usepackage{float}
\usepackage[toc,page]{appendix}
\usepackage{longtable,booktabs}
 \usepackage{epsfig}
\usepackage{graphicx} 
\usepackage{feynmf}
\usepackage{psfrag} 
\usepackage{slashed}
\usepackage{amsmath,amssymb} 
\usepackage{colordvi} 
\usepackage{hyperref}
\usepackage{setspace}
\usepackage[all]{hypcap}
\usepackage{natbib}
\usepackage{subfigure}

\hypersetup{
    bookmarks=true,         
    unicode=false,          
    pdftoolbar=true,        
    pdfmenubar=true,        
    pdffitwindow=false,     
    pdfstartview={FitH},    
    pdftitle={My title},    
    pdfauthor={Author},     
    pdfsubject={Subject},   
    pdfcreator={Creator},   
    pdfproducer={Producer}, 
    pdfkeywords={keyword1} {key2} {key3}, 
    pdfnewwindow=true,      
    colorlinks=true,       
    linkcolor=red,          
    citecolor=green,        
    filecolor=magenta,      
    urlcolor=black           
}
\def\be{\begin{equation}}
\def\bea{\begin{eqnarray}}
\def\eea{\end{eqnarray}}
\def\ee{\end{equation}}
\def\del{\partial}

\begin{document}
\title{Patterns of Electro-magnetic Response in Topological Semi-metals}

\author{Srinidhi T. Ramamurthy and Taylor L. Hughes}

\affiliation{Department of Physics and Institute for Condensed Matter Theory, University of Illinois at Urbana-Champaign, IL 61801, USA}
\begin{abstract}
Topological semimetals are gapless states of matter which have robust and unique electromagnetic responses and surface states. In this paper, we consider semimetals which have point like Fermi surfaces in various spatial dimensions $D=1,2,3$ which naturally occur in the transition between a weak topological insulator and a trivial insulating phase. These semimetals include those of Dirac and Weyl type. We construct these phases by layering strong topological insulator phases in one dimension lower. This perspective helps us understand their effective response field theory that is generally characterized by a 1-form $b$ which represents a source of Lorentz violation and can be read off from the location of the nodes in momentum space and the helicities/chiralities of the nodes. We derive effective response actions for the 2D and 3D Dirac semi-metals, and extensively discuss the response of the Weyl semimetal. We also show how our work can be used to describe semi-metals with Fermi-surfaces with lower co-dimension as well as to describe the topological response of 3D topological crystalline insulators. 
\end{abstract}
\maketitle

\tableofcontents


The discovery of topological band insulators (TIs) and their novel electronic properties has led to a re-examination and search for robust topological features of the electronic structure of many different material types\cite{HasanKane}. Some notable properties of topological insulators include a gapped, insulating bulk interior, protected boundary modes that are robust even in the presence of disorder, and quantized electromagnetic transport. A full (periodic) classification table of non-interacting fermionic states of matter that are protected by time-reversal (T), chiral, and/or particle-hole (C) symmetries has been established\cite{schnyder2008,qi2008,kitaev2009}. Recent work has further augmented the initial periodic table by including the classification of states  protected by spatial symmetries such as translation, reflection, and rotation\cite{FuKaneWeak2007,teo2008,fu2011topological,hughes2011inversion,turner2012, fang2012bulk,teohughes1,slager2012, Benalcazar2013,Morimoto2013,RyuReflection2013,fang2013entanglement,HughesYaoQi13,Sato2013,Kane2013Mirror,jadaun2013}. While these symmetry protected topological phases are theoretically interesting in their own right, this field would not have attracted so much attention if it were not for the prediction and confirmation of candidate materials for many different topological classes. A few examples are the 3D T-invariant strong topological insulator (e.g., BiSb\cite{hsieh08}, Bi$_2$Se$_3$\cite{moore2009topological,xia2009observation,zhang2009experimental}), the 2D quantum spin Hall insulator (e.g., CdTe/HgTe quantum wells\cite{Kane2005A,bernevig2006c,konig2008}), the 2D quantum anomalous Hall (Chern) insulator (e.g., Cr-doped (Bi,Sb)$_2$Te$_3$\cite{haldane1988,chang2013}), and the 3D T-invariant topological superfluid state (e.g., the B-phase of He-3\cite{qi2009,schnyder2008,kitaev2009}). 

All of above work pertains to gapped systems, however, recent theoretical predictions have shown that even materials that are not bulk insulators can harbor robust topological electronic responses and conducting surface/boundary states\cite{nielsen1981,wan2011,turner2013,haldane2014,matsuura2013}. This class of materials falls under the name topological semi-metals, and represents another type of non-interacting electronic structure with a topological imprint. The most well-studied examples of topological semi-metals (TSMs) are the 2D Dirac semi-metal (e.g., graphene\cite{graphenereview}), the 3D Weyl semi-metal (possibly in pyrochlore irridates\cite{wan2011} or inversion-breaking super-lattices\cite{balentshalasz}), and the 3D Dirac semi-metal\cite{youngteo2012,wang2012,liu2013,neupane2013,wang2013,yang2014}. While there are yet to be any confirmed experimental candidates for 3D Weyl semi-metals, their unique phenomenology, including incomplete Fermi-arc surface modes, an anomalous Hall effect, and a chiral magnetic effect has drawn theoretical and experimental attention to these materials. The class of 3D Dirac semimetals have been reported to be found in Refs. \onlinecite{liu2013,neupane2013,wang2013,wang2012} after being proposed and studied in Ref.  \onlinecite{youngteo2012}. In addition to these TSMs there is a large set of symmetry-protected TSMs which rely on additional symmetries for their stability\cite{matsuura2013}. We should also note that there are superconducting relatives of these semi-metal phases called topological nodal superconductors or Weyl superconductor phases that await experimental discovery\cite{meng2012,cho2012,matsuura2013}, though we will not consider them further. 

In this article we explore the quasi-topological response properties of TSMs in the presence of external electromagnetic fields. We present a generic construction of TSMs that can be adapted to model almost any type of TSM. This construction allows us to determine the electromagnetic response properties of the TSMs in question, and to exhibit different patterns that connect semi-metals in different spatial dimensions. In addition, our work nicely complements the extensive recent work studying the topological response properties of Weyl semi-metals\cite{wan2011,turner2013,haldane2014,matsuura2013,hosur2013}. 

 The previous field-theoretic calculations of the response of Weyl semi-metals have predicted a novel electro-magnetic response for the 3D TSMs, but not without some subtlety \cite{tewari2012,zyuzin2012,vazifeh2013,chen2013,haldane2014,chen2013weyl,vanderbilt2014comment}.  Thus, another goal of this article is to address the electro-magnetic (EM) response for various topological semi-metals, and to show the validity and limitations of the field-theory results. To this end, we provide explicit numerical simulations using simple lattice models to complement our transparent analytic discussion. In addition to the discussion of the 3D Weyl semi-metals, we carefully illustrate the pattern of TSM response actions that exist in 1D metallic wires and 2D Dirac semi-metals to establish a unified framework of the EM response of TSMs. We discuss the influence of and, in some cases, the necessity of, anti-unitary and/or spatial symmetries for the stability of the semi-metal phase, and the resultant implications for the EM response. Furthermore, we provide an analytic solution for the boundary modes of the TSMs in our simple lattice models, derive a topological effective response action for the 2D and 3D Dirac semi-metals, calculate the EM response at interfaces between different TSMs, and, where possible, emphasize the important physical quantities of TSMs that can be observed. 

The article is organized as follows: in Section \ref{sec:prelim} we discuss the preliminaries and motivation for the work. This section provides our approach to the characterization of TSMs and further reviews previous work, especially on the response of Weyl semi-metals. After this we begin by discussing one-dimensional semi-metals in Section \ref{sec:oneD} as a warm-up problem for the rest of the article. From this we move on to 2D Dirac semi-metals in Section \ref{sec:twoDDSM}. We discuss the connection between 1D topological insulators and 2D Dirac semi-metals, and discuss the low-energy boundary states of the Dirac semi-metal. We calculate the ``topological" contribution to the electromagnetic response for a TSM with two Dirac points using a field-theoretical calculation, and then go on to generalize the picture to a generic number of Dirac points. We also discuss the microscopic origin and subtleties of the response using lattice model realizations. Next we switch to 3D, and in Section \ref{subsec:threeDWSM} we discuss the response properties of a Weyl semi-metal. Many of the results here are already known, but we present the material from a slightly different perspective, and also include numerical calculations of the response, an analytic description of the boundary modes for a lattice model, and the response behavior at an interface between two different Weyl semi-metals. We also present a discussion of the anomaly cancellation which connects the bulk and surface response. We finish our discussion with Section \ref{subsec:threeDDSM} on 3D Dirac semi-metals. We present a new type of electromagnetic response which appears when the surface of the 3D TSM is in contact with a magnetic layer. Finally in Section \ref{sec:conclusion} we summarize our results, illustrate how to apply our work to calculate the topological response properties of 3D topological crystalline insulators, and how to consider semi-metals with Fermi-surfaces with different co-dimensions. 

\section{Preliminaries and Motivation}\label{sec:prelim}
\subsection{Electromagnetic Response}
One of the primary goals of this work is to produce valuable intuition for understanding the response properties of generic topological semi-metals. In this section we will begin with a simple physical construction that is applicable to different types of topological semi-metals and provides a basis for understanding the EM response of a wide-class of TSMs in a unified manner. In this context we will discuss some of the previous work on the EM response of Weyl semi-metals as an explicit example. Finally, before we move on to more technical calculations, we will illustrate the pattern followed by the electromagnetic response of TSMs in various spatial dimensions.

A simple way to understand a topological semi-metal is as a gapless phase that separates a trivial insulator phase from a weak topological insulator phase. A trivial insulator is essentially a band insulator that is adiabatically connected to the decoupled atomic limit. The electronic structure of trivial insulators does not exhibit any non-vanishing topological properties.  On the other hand, weak topological insulators (WTIs) are anisotropic, gapped topological phases that are protected by translation symmetry and characterized by a vector topological invariant $\vec{\nu}.$ The fact that the topological invariant is a vector, and not a scalar, is an indication that they are essentially anisotropic. This anisotropy can be made more apparent because each WTI phase in $d$ spatial dimensions can be adiabatically connected to a limit of decoupled $d-1$-dimensional systems that are layered perpendicular to $\vec{\nu}.$ The $(d-1)$-dimensional building blocks that make up the $d$-dimensional WTI must each be in a $(d-1)$-dimensional topological insulator phase to generate the higher dimensional WTI phase. Of course one can also construct a $d$-dimensional WTI from $(d-q)$-dimensional ($1<q<d$) topological phases although we will not consider this type in this article. 

The most well-known example of a WTI is a stack of planes of 2D integer quantum Hall states (or 2D Chern insulators) that create the so-called 3D quantum Hall effect\cite{halperin1987}. If the 2D planes are parallel to the $xy$-plane then the vector invariant $\vec{\nu}\propto \hat{z}.$ If the coupling between the planes is weak, then the bulk gap, arising from the initial bulk gaps of the 2D planes, will not be closed by the dispersion in the stacking direction. However, when the inter-layer tunneling becomes strong enough, the system will become gapless and exhibit the so-called Weyl semi-metal phase. Eventually, as the tunneling strength increases, the system will transition to another gapped phase that will either be a different WTI phase or a trivial insulator. 
Thus, in the simplest case, the Weyl semi-metal is an intermediate gapless phase separating a WTI from a trivial insulator.
As we will discuss later, a similar picture can be developed for the 2D Dirac semi-metal which can be adiabatically connected to an array of 1D TI wires that are stacked into 2D. Ultimately, this type of description of TSMs will be very useful since the relevant EM response properties of the lower dimensional TI building blocks are well-known\cite{qi2008}, and the problem of the TSM response is transformed into understanding how the inter-layer coupling affects the EM responses of the TI constituents. 

While it is well-known that TIs and WTIs exhibit topological electromagnetic response properties,  at the transition between trivial and topological phases the relevant topological response coefficients are no longer well-defined. In fact, there is usually a jump from a quantized non-zero value in the topological phase to a vanishing value of the response coefficient in the trivial phase. Therefore, it is a bit surprising that the semi-metal phases intermediate between trivial and topological insulators retain an imprint of the topological response. This is illustrated beautifully in the case of the Weyl semi-metal as we will now discuss. A trivial insulator has no topological component to its EM response, it obeys Maxwell's equations with the conventional insulator constituent relations for polarization and magnetization. On the other hand, the non-trivial WTI represented by the 3D quantum Hall insulator produces a topological response term in the effective action
\begin{equation}
S_{eff}[A_{\mu}]=-\frac{e^2}{2\pi h}\int d^3x dt\; \nu_{\mu}\epsilon^{\mu\sigma\rho\tau}A_{\sigma}\partial_\rho A_{\tau}\label{eq:3dqhe}
\end{equation}\noindent where $\nu_0=0,$ $\nu_i=\tfrac{n}{2}G_i$ are the components of a half-integer multiple $n/2$ of a reciprocal lattice vector $\vec{G},$ and $A_\mu$ are external EM fields. This action implies that spatial planes perpendicular to $\vec{\nu}$ will have a Hall effect, and the 3D Hall conductance is $\sigma_{xy}=-ne^{2}/ha_G$ where $a_G$ is the lattice spacing along $\vec{G}.$ Note that we have chosen the global negative sign to match the convention of Ref. \onlinecite{zyuzin2012}. The trivial insulator phase can be thought of as the case when $\vec{\nu}=\vec{0}.$
 It is clear that the topological response is anisotropic, as the particular $\vec{\nu}=\frac{\vec{G}}{2}$ breaks rotation invariance (and as a consequence Lorentz invariance if we are considering relativistic theories which are a common low-energy description of a TSM). 
 
 \begin{figure}[t]
\label{fig:weylmove}
\centering
   \epsfxsize=3.5in \epsfbox{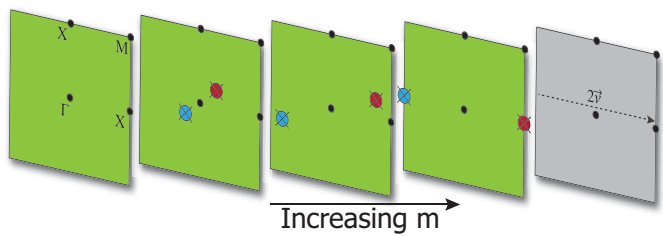}
   \caption{Schematic Illustration of the motion of point-nodes in the $k_z=0$ plane of a cubic, 3D Brillouin zone as a parameter $m$ is adjusted. As $m$ increases two Weyl nodes with opposite chirality (as represented by the color shading) are created in the 2D subspace (i.e., $k_z=0$) of a full 3D Brillouin zone. As $m$ increases further, the nodes move throughout the Brillouin zone, meet at the boundary, and then finally annihilate to create a gapped phase with a weak topological invariant proportional to the reciprocal lattice vector separation $\vec{G}=2\vec{\nu}$ of the Weyl nodes before annihilation. The far left Brillouin zone represents a trivial insulator, the far right represents a weak topological insulator, and the intermediate slices represent the Weyl semi-metal phase.}
\end{figure}
 
 Now that we understand the topological response of the two phases that straddle the Weyl semi-metal phase, we can try to understand the response of the simplest type of Weyl semi-metal, i.e., the kind with only two Weyl nodes (the minimal number). Let us imagine the following process where we begin with a trivial insulator and nucleate two Weyl nodes at the $\Gamma$-point in the 3D Brillouin zone (BZ) by tuning a parameter $m$ (see Fig. \ref{fig:weylmove}). The low-energy $k\cdot P$ Hamiltonian near each Weyl-node is of the form $H_{Weyl}(\mbox{p})=p_{1}\sigma^{1}+p_{2}\sigma^{2}+p_{3}\sigma^{3}$ where $\sigma^a$ are Pauli matrices. As $m$ is further changed, the Weyl nodes will move through the BZ but cannot be gapped (assuming translation invariance) unless they meet each other again, or another node. The reason is that if the Weyl-nodes are separated, then there is no matrix which anti-commutes with $H_{Weyl}(\mbox{p}),$ and thus no perturbation can be added that will open a gap. If the two Weyl nodes (with opposite chirality) meet and become degenerate, then the resulting $4\times 4$ Hamiltonian $H_{Weyl}\oplus \tilde{H}_{Weyl}$ has the Dirac form. In this case one can find an anti-commuting matrix to add that will perturbatively open a gap and  annihilate the nodes. If the Weyl nodes meet at the boundary of the BZ, at points which differ by a reciprocal lattice vector $\vec{G},$ then upon annihilation the system will undergo a change of its weak-invariant, i.e.,  $\Delta\vec{\nu}=\frac{\vec{G}}{2}.$ Thus if the system starts with $\vec{\nu}=0$ then it will have a transition to a non-trivial WTI during this process. 
 
 During the process of tuning $m$ we see that before we nucleate the Weyl nodes there is no topological response, and after they annihilate at the BZ boundary there will be a non-trivial Hall response. We now can ask, what is the response in the gapless semi-metal phase? The answer turns out to be simple, we just have the response of Eq. \ref{eq:3dqhe} with $\vec{\nu}=\vec{b}$ where $2\vec{b}$ is the difference in momentum between the two Weyl nodes\cite{zyuzin2012}. Interestingly, the response coefficient smoothly interpolates between the two insulating end-points. This remarkable result can be extended even further because we also have a notion of a relative energy between the Weyl nodes. Because of this we can generate a coefficient $\nu_0=b_0$ in Eq. \ref{eq:3dqhe} where $2b_0$ is the energy difference between the two Weyl nodes. This enhances the response as now we can have a Lorentz-invariance violating $4$-vector response coefficient $\nu_{\mu}.$ 
 
The addition of a response proportional to $\nu_0$ is a new feature of the semi-metal since one cannot define a notion of $\nu_0$ in the pure WTI because the low-energy theory is gapped. The reason one can have a spatial vector in the gapped WTI is because of the translation symmetry (and continuous rotation symmetry) breaking lattice structure which gives rise to the reciprocal lattice vector(s) $\vec{G}.$ On the other hand, if we had a periodically driven system, i.e., a system evolving according to Floquet dynamics, then, even in the insulating case, we could have a non-zero $\nu_0$ which would be proportional to the driving frequency of the time-dependent field, i.e., the reciprocal lattice vector for time. In the Weyl semi-metal phase, the existence of non-degenerate Weyl nodes immediately gives rise to a Lorentz-breaking 4-vector similar to the kind anticipated by Refs. \onlinecite{carroll1990limits, jackiw2003} for Lorentz-violations in high-energy physics.   The resulting response from Eq. \ref{eq:3dqhe} generates an anomalous Hall effect along with a chiral magnetic effect (CME). The chiral magnetic effect occurs when $b_0\neq 0$ and is anticipated to give rise to a current when a magnetic field is applied to the system, but in the absence of any electric field. Explicitly, the charge density and current are given in terms of $b_\mu$ and the applied EM fields as
\bea
j^{0}&=&\frac{e^{2}}{2\pi h}(2\vec{b})\cdot\vec{B}\\
\vec{j}&=&\frac{e^{2}}{2\pi h}((2\vec{b})\times\vec{E}-(2b_{0})\vec{B}).
\eea
 While the origin and detection of the anomalous Hall current is well understood, there has been some disagreements in the recent literature about the possibility of a non-vanishing CME. To summarize the results so far, the field theoretical results are somewhat ambiguous because of the dependence on a regularization\cite{tewari2012}: a tight-binding lattice calculation has shown a vanishing result\cite{vazifeh2013}, while a more recent calculation has indicated the need for a slowly varying magnetic field that eventually tends toward a uniform/constant field\cite{chen2013}. In Section \ref{subsec:threeDWSM} we comment on these results and note that having an explicit source of  Lorentz violation is a necessity for a non-vanishing CME effect. We also discuss the interpretation of the CME effect from a quasi-1D perspective generated from placing a Weyl semi-metal in a uniform magnetic field.

While we see it is the case for the Weyl semi-metal, it is generically true that the general pattern of EM response for TSMs stems from the existence of the Lorentz-violating vector (or tensor) response coefficient. In systems with translation symmetry, the vector is connected to the momentum and energy difference between non-degenerate point-nodes (e.g., Dirac nodes in 2D and Weyl nodes in 3D).  In general, the vector represents a source of Lorentz-violation in the system. Let us call this vector $b_{\mu}$ using the same notation as above. In odd dimensional space-time ($D+1$ is odd), we have 
\be
S[A]={\cal{A}}_{D}\int \mbox{d}^{D+1}x\,\epsilon^{a_1 a_2\ldots a_{D+1}}b_{a_1}F_{a_2 a_3}\ldots F_{a_{D} a_{D+1}}
\ee
where the ellipses in the above equation represent further factors of the field strength,  and ${\cal{A}}_D$ is a dimension dependent normalization coefficient.
In even space-time dimensions ($D+1$ is even), the bulk electromagnetic response has the following form 
\be
S[A]={\cal{A}}_{D}\int \mbox{d}^{D+1}x\,\epsilon^{a_1 a_2\ldots a_{D+1}}b_{a_1}A_{a_2}F_{a_3 a_4}\ldots F_{a_{D}a_{D+1}}\label{eq:genActionEven}
\ee
where the ellipses in the above equation represent further factors of the field strength. For example in $1+1$-d, we have just $S[A]={\cal{A}}_{1}\int\mbox{d}^{2}x\,\epsilon^{\mu\nu}b_{\mu}A_{\nu}.$ 
In even dimensional space-times the literature differs on the convention for the choice of the action and some sources use
\begin{equation*}
S[A]=\frac{{\cal{A}}_{D}}{2}\int \mbox{d}^{D+1}x\,\epsilon^{a_1 a_2\ldots a_{D+1}}\theta F_{a_1 a_2}F_{a_3 a_4}\ldots F_{a_{D}a_{D+1}}
\end{equation*}\noindent where $\theta\equiv 2b_{\mu}x^{\mu}.$ However this second form, while it looks somewhat nicer as far as gauge invariance is concerned, has an implicit breaking of translation symmetry. This comes from the freedom of the choice of origin in the definition of $\theta$ as we could have alternatively defined $\theta$ to be $\theta \equiv 2b_{\mu}(x^{\mu}+x^{\mu}_{0})$ with some constant $4$-vector $x^{\mu}_{0}.$ Because of this, we will always choose the form Eq. \ref{eq:genActionEven} to avoid the translation symmetry ambiguity. In fact, using the $\theta$-term version of the action leads to spurious effects when the system is not homogeneous, e.g., in the presence of boundaries. 

In general, the pattern of response actions for TSMs with nodal (point-like) Fermi-surfaces is attached to an intrinsic $1$-form $b=b_{\mu}dx^{\mu}$ which emerges from the band structure. This type of $1$-form indicates some inherent anisotropy in the electronic structure, and can appear in any dimension. For example, for a translation invariant 3D material with an even number of Weyl nodes, we can determine
\begin{equation}
\vec{b}=\frac{1}{2}\sum_{a}\chi_{a}\vec{K}_{a},\;\;\; b_0=\frac{1}{2}\sum_{a}\chi_{a}\epsilon_a
\end{equation}\noindent where the sum runs over all of the Weyl nodes, and $\chi_{a},$ $\vec{K}_a,$ and $\epsilon_a$ are the chirality, momentum location, and energy of the $a$-th node respectively. Additionally, from the Nielsen-Ninomiya no-go theorem we know there is also the constraint that $\sum_{a}\chi_{a}=0$\cite{nielsen1981}. 

We also note that because of the lattice periodicity, the vector $2\vec{b}$ is only determined up to a reciprocal lattice vector. Thus, the response of a TSM is only determined up to a quantum determined by the addition of a filled band. For Weyl semi-metals this indeterminacy is due to the possibility of a contribution of an integer Hall conductance (per layer) from filled bands; the low-energy Fermi-surface physics does not contain information about the Hall conductance of the filled bands\cite{haldane2005}. We also note that, more generally, we can have terms in the  effective action which involve an $n$-form in space-time dimensions greater than or equal to $n$ when a $d$-dimensional system has a Fermi-surface with co-dimension less than $d$;  some cases of which will be discussed elsewhere\cite{inpreparation}.

\subsection{Boundary Degrees of Freedom}
The other generic feature of TSM phases is the existence of low-energy boundary modes. It is well-known that topological insulators have robust, gapless boundary modes that exist in the bulk energy gap. A (strong) TI will contain topological boundary states on any surface, while a WTI only harbors topologically protected boundary states on surfaces where $\vec{\nu}$ does not project to zero in the surface Brillouin zone\cite{FKM2007}. This is another clear signature of the anisotropy, and it gets passed on to the TSMs that interpolate between the WTI and trivial insulator phases. TSMs themselves will have low-energy boundary modes, but again, only on surfaces where $b_{i}$ does not project to zero in the surface Brillouin zone. That is, there will be surface states on surfaces where the normal vectors are not parallel to the node separation vector $\vec{b}.$ We note that even in cases where $b_{i}=0$ (or $b_{i}$ projects to zero on a surface) there can still be surface states because $b_{i}$ is only well-defined modulo a reciprocal lattice vector. However, the surface states that exist when $b_{i}=0$ come from fully filled bands and will exist over the entire Brillouin zone. They are not related to the properties of the semi-metal and do not depend on the locations of the nodes as they are continuously deformed. 

The existence of boundary modes in TSMs is most easily illustrated with a simple example. Let us again resort to the picture of a Weyl semi-metal arising out of a stack of identical 2D Chern (quantum anomalous Hall) insulators and, for simplicity, assume that the layers are stacked in the z-direction. Then, for the WTI phase in the completely decoupled limit, each Chern insulator layer contributes one set of chiral edge modes on surfaces with normal vectors in the $\hat{x}$ and/or $\hat{y}$ directions. This is the simple picture of a WTI, and if each layer has a first Chern number $C_1=1,$ then the vector invariant $\vec{\nu}=(0,0,\pi/a)$ where $a$ is the spacing between the Chern insulator layers. If the system has length $L_z=Na$ in the $z$-direction then the total Hall conductance is $\sigma_{ij}=-\epsilon_{ijk}\frac{e^2}{\pi h}\nu^{k}L_z=-N\frac{e^2}{h},$ i.e., an amount $e^2/h$ per stacked layer. When the coupling between layers is turned on, then the bulk and edge states will disperse in the $z$-direction, but as long as the inter-layer coupling does not close the bulk gap, then the system will remain in the WTI phase with the same Hall conductance. 

To be explicit, we can represent this system as a tight-binding model on a cubic lattice where each site contains a single electronic orbital with spin-up and spin-down degrees of freedom. A representative Bloch Hamiltonian is
\begin{eqnarray}
&H(\vec{k})&=A\sin k_x\sigma^x +A\sin k_y\sigma^y\nonumber\\
&+&(2B-m-B\cos k_x -B\cos k_y -C\cos k_z)\sigma^z\label{eq:wsm}
\end{eqnarray}\noindent where $A, B, C, m$ are parameters, $\sigma^a$ represents spin, and we have set the lattice constant $a=1.$ If we choose the parameters $A=B=2m=1$ and $C=0,$ this will represent a WTI phase built from decoupled layers of Chern insulator states as discussed above. We can see this from the fact that when $C=0$ there is no dispersion in the $z$-direction, and thus we have many copies of a two-dimensional system, one for each allowed $k_z$, i.e., one for each layer. The important point is that when $A, B, m$ are tuned as above, then, ignoring the $z$-direction, the resulting two dimensional system is in a Chern insulator phase with $C_1=1$\cite{qi2008}, and thus we have decoupled copies of a non-trivial Chern insulator. When the tunneling between the layers is activated, the parameter $C$ will be non-vanishing. With $A, B, m$ fixed as above then for $-1/2<C<1/2$ the model will remain in the WTI phase.  At $C=1/2$ the bulk energy gap closes at $\vec{k}=(0, 0, \pi).$ If $C$ is further increased then there will be two points where the gap vanishes, i.e., two Weyl-nodes, and they will occur at $\vec{k}=\left(0, 0, \cos^{-1}\left(-\frac{m}{C}\right)\right)$ where we added the dependence for a variable $m$ parameter back in. Accordingly, when $\vert m/C\vert <1$ the system will exhibit a Weyl semi-metal phase if $A=B=1.$

As was shown in Ref. \onlinecite{ran2011} we can use a model like Eq. \ref{eq:wsm} to create a nice description of the Weyl semi-metal phase.  For this picture, it is useful to think about the system as a family of 2D insulators $H_{k_z}(k_x,k_y)\equiv H(k_x,k_y,k_z),$ parameterized by $k_z.$ For parameters representing the fully gapped WTI phase (e.g. $A=B=2m=1$, $C=0$), then for each value of $k_z$ the 2D insulator $H_{k_z}(k_x,k_y)$ is in the Chern insulator phase. 

Now, when we tune the $C$ parameter into the Weyl semi-metal phase then the model will contain gapless Weyl-nodes at $\vec{k}=(0,0,\pm k_{c})$, and a separation vector $\vec{b}=(0,0,k_c).$ To understand the existence of surface states in the semi-metal phase it is again helpful to think of each 2D insulator at fixed $k_z$ being in a trivial $C_1=0$ phase when $\vert k_z\vert >\vert k_c\vert$ and a Chern insulator phase with $C_1=1$ when $\vert k_z\vert <\vert k_c\vert.$ Exactly at $k_z=\pm k_c$ there is a gapless ``transition" as a function of $k_z$ between the trivial 2D insulator with $C_1=0$ and the non-trivial 2D insulator with $C_1=1.$  This illustration shows that in the Weyl semi-metal phase we should only expect boundary states to exist over a finite range of $k_z$, i.e., $\vert k_z\vert <\vert k_c\vert$ for this particular example. For each $k_z$ in the topological range, the 2D insulator $H_{k_z}(k_x,k_y)$  contributes one propagating chiral fermion mode to the boundary degrees of freedom. These chiral boundary states manifest as incomplete surface Fermi-arcs that connect Weyl points in the surface Brillouin zone for surfaces with normal vectors which are not parallel with $\vec{b}.$ The picture of a TSM as a momentum-space transition in a family of lower-dimensional gapped insulators is helpful because similar concepts can be applied to understand the properties of all topological semi-metals.

In summary, we have introduced some important physical intuition and concepts pertaining to 3D Weyl semi-metals, and during this process reviewed some of the previous work describing the EM response and boundary states of these systems. Now we will begin a more in-depth discussion of the response and boundary states of semi-metals in 1D, 2D, and 3D following the outline presented above.

\section{Semimetal in $1+1$-dimensions}
 \label{sec:oneD}
We will begin with a careful study of the properties of a 1D TSM, which in this case is just an ordinary 1D metal, as noted in Ref. \onlinecite{turner2013}. As a representative model we can choose a spinless $1$-band tight-binding model of the form
\begin{equation}
H_{1D}=-\alpha\sum_{n}\left[c^{\dagger}_{n+1}c^{\phantom{\dagger}}_{n}+c^{\dagger}_{n}c^{\phantom{\dagger}}_{n+1}\right]\label{eq:1dtight}
\end{equation}\noindent where the sum over $n$ runs over all of the lattice sites, and we will let the lattice constant be $a.$ This familiar model is easy to diagonalize and the energy spectrum is of the form
\begin{equation}
E(k)=-2\alpha\cos ka
\end{equation}\noindent were $k\in [-\pi/a, \pi/a).$ In the momentum basis the Hamiltonian is just $H_{1D}=\sum_{k}E(k)c^{\dagger}_{k}c^{\phantom{\dagger}}_{k}.$ 

Establishing a chemical potential $\mu$ that lies within the band will fill the system with a finite density of electrons. If we keep translation symmetry we can calculate the number of particles by counting the number of occupied momentum states
\begin{equation}
N=\sum_{k\in occ.}1=\frac{L}{2\pi}\int_{-k_{F}}^{k_{F}}dk=\frac{Lk_F}{\pi}
\end{equation}\noindent which implies a charge density $\rho=e\frac{k_F}{\pi}$ where $k_F$ is the Fermi wavevector and $e$ is the electron charge. In the language of the previous section we note that this density breaks Lorentz invariance because it establishes a preferred frame, i.e., the rest-frame of the fermion density. Thus, we should expect a Lorentz-violating contribution to the effective action. In fact, we can easily write down this contribution as a background charge density just couples to the scalar EM potential $A_0$  to give a potential energy term
\begin{equation}
S[A_0]=-\int dx dt \rho A_0.
\end{equation}\noindent

In addition to the density, there is the possibility of introducing an electric current that will also break Lorentz invariance. 
For a moment, let us consider a generic one-dimensional lattice model with translation invariance, and in the momentum basis. When minimally coupled to an EM field (e.g., through Peierls substitution) we find
\be
H=\sum_{k}c^{\dagger}_{k}H(k-\tfrac{e}{\hbar}A_1)c^{\phantom{\dagger}}_{k}
\ee\noindent where $H(k)$ is a Bloch Hamiltonian. 
The current for this system in the limit $A_1\rightarrow0$ is given by  
\be
j=\lim_{A\rightarrow0}\frac{\partial{H}}{\partial{A_1}}=-\frac{e}{\hbar}\sum_{k}\mbox{Tr}\left[\frac{\partial{H(k)}}{\partial{k}}n_{F}\right]
\ee
where $n_{F}$ is the Fermi-Dirac distribution, which will be a step function at $T=0$. This can be rewritten at zero temperature as 
\be
\label{eq:current}
j=-e\sum_{n\in occ}\int_{BZ} \frac{\mbox{d}k}{2\pi\hbar}\frac{\partial{E_{n}(k)}}{\partial{k}}
\ee\noindent where $n$ runs over the occupied bands.
Specializing to the case of our single-band model, the current is equal to $j=-\frac{e}{2\pi\hbar}(E(k_{F})-E(-k_{F}))$ which is non-zero only if $E(k_{F})\neq E(-k_{F})$. We will discuss two different mechanisms for generating a current in Sections \ref{sec:1dEfield} and \ref{sec:1dNNN}.

\subsection{1D model in an electric field}\label{sec:1dEfield}
One way to generate a non-zero electric current is to apply an external electric field. The electric field is applied by an adiabatic threading of magnetic flux through the hole of the periodic lattice ring via Faraday's law. This is equivalent to  introducing twisted boundary conditions on the wave functions 
\be
\Psi(x+L)=e^{i\Phi(t) L}\Psi(x)
\ee
where 
\be
\Phi(t)=\frac{eEt}{\hbar}
\ee
\noindent for an electric field $E$ at time $t.$ Using Eq. \ref{eq:current} we can easily calculate the electric current to be
\be
\label{eq:four_sawtooth}
j=\frac{2\alpha e}{\pi\hbar}\sin(k_{F}a)\sin(\Phi(t)a).
\ee

\begin{figure}[t!]
\label{fig:elec_field}
\centering
   \epsfxsize=3.5in \epsfbox{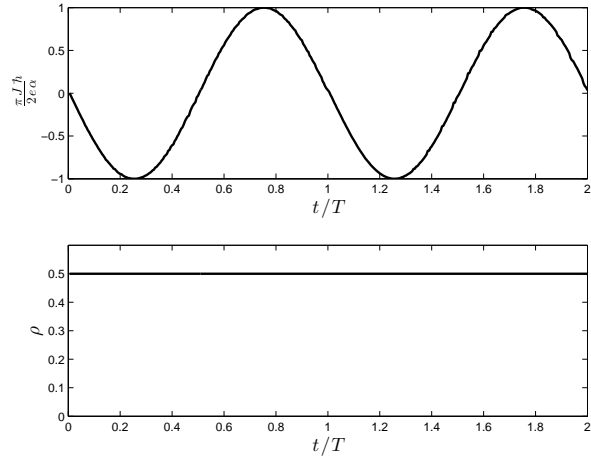}
   \caption{The current and charge density of the 1D (semi-)metal are plotted vs time for half filling and for nearest neighbor hopping $\alpha=1$. The current has a periodic response as expected with a period of 200 time slices for an electric field of strength $E=\tfrac{h}{eT(200a)}$ for some time-scale $T$ that is long.  The charge density is given by $\rho=e\tfrac{k_{F}}{\pi}=e/2a$ as expected and shows no time dependent behavior. }
\end{figure}

For comparison, we numerically calculate the charge density and current for the case when the single band is half-filled. At half-filling $k_F=\pi/2a,$ and thus the density should be uniform, time-independent and equal to $\rho=\frac{e}{2a},$ i.e., half an electron per site.
At half filling, the current reduces to $j=\frac{2\alpha e}{\pi\hbar}\sin(\Phi(t)a).$ The numerical calculations are shown in Fig. \ref{fig:elec_field}, and they agree with the analytic results.

We note in passing that for finite-size lattice models some care must be taken to correctly calculate a smooth electric current response. We have intended to calculate the current of a metallic/gapless system, but there are finite-size gaps in the energy spectrum between each state separated by $\Delta k =2\pi/L.$ Thus, if we want the system to behave as a gapless system should, we must apply a minimum threshold electric field. If too small of an electric field is applied at a given system size, the model will behave like a gapped insulator instead. To avoid this we can simply enforce the canonical momentum $\Pi_{x}=p_{x}-eA_{1}$ to be a multiple of $2\pi\hbar/L$ so that the system remains gapless at each time step. If this is not done, then  the system will behave as gapped insulator and we will see steps in the current response. Ensuring that $\tfrac{e}{\hbar}A_{1}=\tfrac{2\pi m}{L}$ at every time step saves us this trouble, and in our simulations for this section we have always taken $\Phi(t)=eEt/\hbar$ to be a multiple of $2\pi/L$ and never smaller than this value. Physically we understand that, for a system with these finite-size gaps, an infinitesimal adiabatic current-generation will not work. Instead we must turn on a large enough electric field so that there is some non-adiabaticity so that the finite-size gaps can be overcome. 

Although we do not present the results here, we have carried out numerical calculations for various filling factors and electric field strengths, and the analytic results match the numerical simulations. If we change the boundary conditions from periodic to open then the charge density remains the same, but the current vanishes as expected. Hence, we see  that in the presence of an electric field with periodic boundary conditions the response action of the 1D (semi-)metal is 
\begin{equation}
S[A_{\mu}]=\int dx dt\left[-\rho A_0+j A_1\right]=\int dx dt j^{\mu}A_{\mu}\label{eq:1dactionJA}
\end{equation}\noindent where $j^{\mu}=(\rho,j),$ which in our convention already has the electric charge factored in. Other than the presentation, most of what we have done here is elementary, we are just using these results to set the stage for the later sections. 

Now, we can re-write the action in a few suggestive ways. First we can define a new $2$-vector 
\begin{equation}
b_{\mu}=\frac{\pi}{e}\left(j,\rho\right)\end{equation}\noindent such that the action can be re-written
\begin{equation}
S[A_{\mu}]=\frac{e}{\pi}\int dx dt\epsilon^{\mu\nu} b_{\mu}A_{\nu}.
\end{equation}\noindent This is to be compared with Eq. \ref{eq:genActionEven}. Alternatively we can define an axion-like field 
\begin{eqnarray}
\theta(x,t)&\equiv&2b_{\mu}x^{\mu}= \frac{2\pi}{e}(\rho x-jt)\nonumber\\
&=&2k_{F}x-\frac{4\alpha }{\hbar}\sin(k_{F}a)\sin(\Phi(t)a)t
\end{eqnarray} and if the system is homogeneous with no boundaries, we can use $\theta(x,t)$ to rewrite Eq. \ref{eq:1dactionJA} as
\begin{equation}
S[A_{\mu},\theta]=-\frac{e}{4\pi}\int dx dt\;\theta(x,t)\epsilon^{\mu\nu}F_{\mu\nu}.\label{eq:1dactionF}
\end{equation}\noindent As mentioned in Section \ref{sec:prelim}, the choice of $\theta(x,t)$ breaks space-time translation symmetry due to the arbitrary choice of origin, and thus we must be careful to specify that the system is translation invariant when writing down Eq. \ref{eq:1dactionF}, otherwise spurious response terms will be generated at boundaries and interfaces. Physically we can interpret $\tfrac{e\theta}{2\pi}$ as the charge polarization since its space and time derivatives are proportional to the charge density and current respectively. 

While this method of generating an electric current came from an external effect, i.e., an externally applied electric field, we now move on to a discussion of an intrinsic effect that can produce a current in the absence of an external electric field.


\subsection{1D Model with Next-Nearest-Neighbor Hopping}\label{sec:1dNNN}
\begin{figure}[t]
\label{fig:NNNspec}
\centering
   \epsfxsize=3.5in \epsfbox{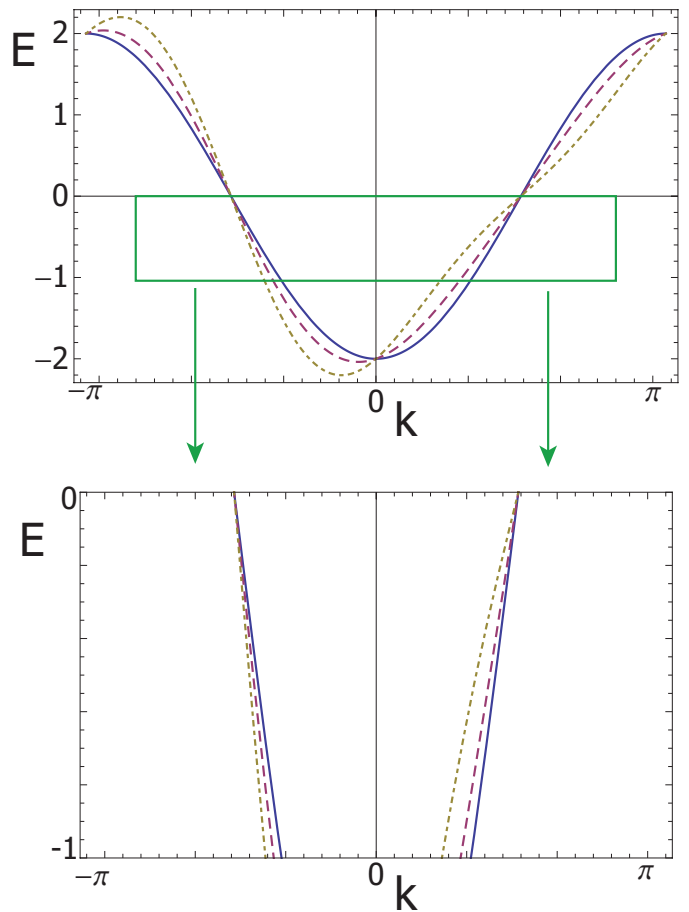}
   \caption{(upper) Energy spectrum of the Hamiltonian $H_{1Dv}$ where each curve represents a different value of $\beta.$ The solid blue line is $\beta=0$, the magenta dashed line is $\beta=0.1$ and the dash-dotted tan line is $\beta=0.25.$ All curves have $\alpha=1.$ (lower) This is a zoomed in region of the upper figure slightly below half-filling, which is the regime for our calculation. Exactly at half-filling $\beta$ has no effect, and the stronger $\beta$ is, the more the Fermi wave vectors and velocities are modified at a fixed $\mu.$}
\end{figure}
In this subsection we illustrate another way to generate a non-vanishing current.  For energies near the Fermi-points, the dispersion of our model is linear, and the modes near each Fermi-point are 1+1-d chiral fermions. In fact, it is well-known that there is a close connection between the physical electric current for a 1D metallic band in an electric field, and the compensating chiral anomalies of the fermion modes near each Fermi-point. The previous subsection explicitly dealt with these issues, albeit using a less elegant  perspective, and in that case an electric current was generated by an external source of Lorentz breaking, i.e., the applied electric field. Here we would like to consider an intrinsic source of Lorentz breaking that will lead to a current as well. By considering this effect, we are trying to make an analogy to the 3D chiral magnetic effect in Weyl semi-metals, where it has been predicted that a current can appear in the presence of an applied magnetic field, but in the absence of an electric field.

The basic idea is that, for the 1D model we have chosen, the chiral fermions near the Fermi-points both have the same velocity, except for the sign, and we want to deform the velocities so that each chiral fermion has a different ``speed of light." This is an obvious way to break Lorentz invariance. If the velocities are different (and the spectra were linear for all energies) then it is clear that we should have $E(-k_F)=v_L k_F\neq v_R k_F=E(k_F)$ which suggests the presence of a current. Physically, this just means that if we have 1+1-d chiral fermions with the same non-zero density, but different velocities, then there will be a non-vanishing current. 
 Since we are in 1D we should find an intrinsic current without the application of a magnetic field or an electric field, and it should be proportional to the response coefficient $b_0.$ In the 3D Weyl semimetal, the number $b_{0}$ represents the energy difference between Weyl nodes and has units of frequency. A simple interpretation of the effect seen here in 1D is that a non-vanishing frequency scale $b_0$ will be generated  by the combination of $\Delta v_F,$ i.e., the velocity difference at the two Fermi points, and a length scale. In our system we have two important length scales: the lattice constant $a$ and the inverse of the Fermi wave vector $k_F.$ To see which one enters the result we will perform an explicit calculation.

To generate the velocity modification effect we deform the tight binding model in Eq. \ref{eq:1dtight} above to include imaginary next-nearest neighbor hopping terms
\begin{eqnarray}
H_{1Dv}=H_{1D}+i\beta\sum_{n}\left[c^{\dagger}_{n+2}c^{\phantom{\dagger}}_{n}-c^{\dagger}_{n}c^{\phantom{\dagger}}_{n+2}\right].
\end{eqnarray}
 The Fourier transform of the Hamiltonian is given by 
\be
H_{1Dv}=-2\sum_{k} (\alpha\cos(ka)-\beta\sin(2ka))c^{\dagger}_{k}c^{\phantom{\dagger}}_{k}.\label{eq:H1dv}
\ee
For $\beta\neq 0,$ inversion symmetry is broken in the model and subsequently we should consider two Fermi-wavevectors $k_{FL}$ and $k_{FR}$ where $k_{FL}\leq k_{FR}$ by definition. Exactly at half-filling $k_{FL}=-k_{FR}=\pi/2a$ (as shown in Fig. \ref{fig:NNNspec}), and thus the electric current is vanishing (since $\beta\sin(2a(\pi/2a))=0$), and the charge density will be $\rho=\frac{e}{2a}$ as was found when no electric field was applied to the model $H_{1D}$ at half-filling. 

Half-filling is just a special point of this model where $\beta$ has no effect because of our choice of next-nearest neighbor hopping.  Instead, let us consider the case where $\mu$ is tuned slightly away from half-filling, i.e., $\mu=0-\delta\mu$ with $\vert\delta \mu\vert\ll\alpha,$ and we will also take $\vert \beta\vert\ll \alpha$ as we want to consider the perturbative effect of turning on this term. We can  define $k_{FL}=-\tfrac{\pi}{2a}+\epsilon_{L}$ and $k_{FR}=\tfrac{\pi}{2a}+\epsilon_{R}.$ By expanding Eq. \ref{eq:H1dv} around the Fermi-points we find that consistency requires
\begin{equation}
\epsilon_{L/R}=\pm\frac{\delta \mu}{2a(\alpha\pm 2\beta)}\approx \pm\frac{1}{2a}\frac{\delta\mu}{\alpha}\left[1\mp \tfrac{2\beta}{\alpha}\right].
\end{equation}\noindent Thus we can determine that 
\begin{equation}
k_{FL/R}\approx\frac{\pi}{2a}\left[\mp 1\pm\frac{\delta\mu}{\pi\alpha}\left(1\mp \tfrac{2\beta}{\alpha}\right)\right]
\end{equation}\noindent and can subsequently define $\kappa_{F}\equiv \tfrac{\pi}{2a}(1-\delta\mu/\pi\alpha),$ which would be the Fermi wavevector if $\beta=0.$ Note that the signs in the previous two equations are correlated. From Fig. \ref{fig:NNNspec} we can see that as $\beta$ is increased the Fermi-wave vector at a fixed $\mu$ (different than half-filling) changes, as well as the velocity of the low-energy fermions. 
From Eq. \ref{eq:current}, the response should be 
\bea
\rho&=& e\frac{k_{FR}-k_{FL}}{2\pi}=\frac{e\kappa_{F}}{\pi}=\frac{e}{2a}\left(1-\tfrac{\delta\mu}{\pi\alpha}\right)\\
j&=& \frac{2e\beta}{\pi\hbar}\sin(2\kappa_{F}a). \label{eq:curr_NNN}
\eea 

\begin{figure}[t]
\label{fig:vel_diff}
\centering
   \epsfxsize=3.5in \epsfbox{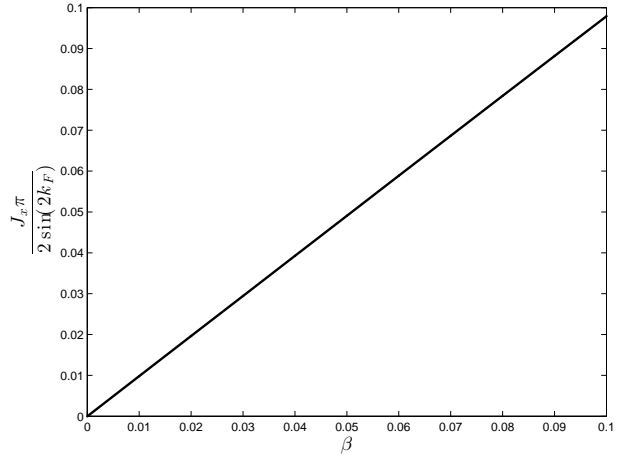}
   \caption{The current of $H_{1Dv}$ is plotted vs  next-nearest neighbor hopping strength $\beta$ near half filling. $\kappa_{F}=\pi/2a-\pi/100a$ was chosen and the nearest-neighbor hopping $\alpha=1$ with periodic boundary conditions. The current increases linearly as a function of $\beta$ as expected from Eq. \ref{eq:curr_NNN}.}
\end{figure}

This result shows that we find a non-zero electric current even in the absence of an applied electric field, and its magnitude is proportional to the inversion breaking parameter $\beta.$ This effect, while simple in origin, is the 1D analog of the 3D chiral magnetic effect. It represents a current proportional to an intrinsic frequency scale, but does not require the application of any external fields. We do note that the definition of the frequency scale does require a non-vanishing Fermi wave-vector, i.e., a non-vanishing background density which cannot arise from a completely empty or filled band. As shown in Fig. \ref{fig:vel_diff}, the numerical calculation of the electric current matches the analytic formula. The response is linear in $\beta$ as expected from Eq. \ref{eq:curr_NNN} and, although we do not show the charge density, it matches as well. The numerical calculations were done for slightly less than half-filling at $\kappa_F=\pi/2a-\pi/100a.$ 

Let us take a closer look at the generation of the electric current. The velocity of the chiral fermions at $\pm \kappa_{F}$ is given by $\hbar v_{\pm}=\pm(2\alpha a\sin(\kappa_{F}a)\mp4\beta a\cos(2\kappa_{F}a)),$ and thus, 
\begin{equation}
\Delta v_{F}=\frac{8\beta a}{\hbar}\cos(2\kappa_{F}a).
\end{equation}  For our choice of the chemical potential, $\kappa_F=\pi/2a+\delta \kappa_F,$ and the current from Eq. \ref{eq:curr_NNN} is approximately 
\begin{equation}
j\approx-\frac{8e\beta}{2\pi\hbar}\delta \kappa_F a=-\frac{e}{2\pi}\frac{8\beta a}{\hbar}\delta \kappa_F=\frac{e}{2\pi}\Delta v_F\delta \kappa_F.
\end{equation}\noindent where we used that near $\kappa_F=\pi/2a$ we have $\Delta v_F\approx -\tfrac{8\beta a}{\hbar}.$ Thus $\Delta v_F\delta\kappa_F$ gives a Lorentz-breaking frequency scale that will give rise to a non-vanishing $b_0$-term in the effective response. In fact, the density and current give us the $2$-vector $b_{\mu}=(\tfrac{1}{2}\Delta v_F\delta \kappa_F,\kappa_F)$ which determines the response action
\begin{equation}
S[A_\mu]=\frac{e}{\pi}\int dx dt\epsilon^{\mu\nu}b_{\mu}A_{\nu}.
\end{equation}
To draw an analogy with the previous literature on the Weyl semi-metal response we could also define a $\theta(x,t)$ by
\begin{eqnarray}
\theta(x,t)&=&\frac{2\pi}{e}(\rho x-jt)=2\kappa_Fx-\frac{4\beta}{\hbar}\sin(2\kappa_{F}a)t\nonumber\\
&\approx &2\kappa_F x-\Delta v_F\delta\kappa_Ft
\end{eqnarray}\noindent which couples into the action
\begin{equation}
S[A_\mu,\theta]=-\frac{e}{4\pi}\int dx dt \theta(x,t)\epsilon^{\mu\nu}F_{\mu\nu}.
\end{equation}


\subsection{Derivation of the effective response}
After our explicit discussion of the different EM responses of the 1D metallic wire, let us elevate our discussion to a field-theoretic calculation. In this section, we use the Fujikawa method to derive the effective response of the low-energy continuum field theory description of the 1D metal in the presence of intrinsic sources of Lorentz invariance violation. The derivation is similar to that for 3D Weyl semi-metals found in Ref. \onlinecite{zyuzin2012}. 

To carry out the calculation let us expand the lattice Bloch Hamiltonian given by $H(k)=-2\alpha\cos k a+2\beta\sin 2ka$ around the chemical potential $\mu=0-\delta\mu$ for $\vert\delta\mu\vert\ll\alpha$ and $\vert\beta\vert\ll\alpha$ as in the previous subsection.  If we expand the right and left-handed chiral branches around $\pm \kappa_F$ respectively we find the approximate continuum Hamiltonian
\begin{equation}
H_{cont}=\left(-\delta\mu+\tfrac{1}{2}\hbar\Delta v_F q\right)\mathbb{I}+\left(\hbar v_F q+\tfrac{1}{2}\hbar \Delta v_F\delta\kappa_F\right)\sigma^z
\end{equation}
\noindent where the upper component represents the fermions near $k_{FR},$ the lower component represents the fermions near $k_{FL},$ $q$ represents a small wavevector deviation, $\hbar v_F\equiv 2a\alpha,$ $\Delta v_F\equiv -\tfrac{8\beta a}{\hbar},$ and $\delta \kappa_F=-\tfrac{\delta\mu}{2a\alpha}.$ The definitions of the parameters are easy to understand by looking at the lattice model in the previous subsection when expanded around $\kappa_{F}.$ Since we know the behavior of the full lattice model, i.e., the high-energy regularization of the continuum model, we can see that our expansion effectively normal orders the current and density with respect to half-filling. The density change away from half filling is given by $\delta\rho=e\frac{-\delta\mu}{2\pi\alpha a}=e\frac{\delta\kappa_F}{\pi}.$ The current change away from half-filling is simply given by $\delta j=\tfrac{e}{2\pi\hbar}\left(E_{R}(q=0)-E_{L}(q=0)\right)=\tfrac{e}{2\pi}\Delta v_F\delta\kappa_F.$ Since the current vanishes exactly at half-filling, the total current is simply $j=\delta j$ which matches the previous subsection. The density does not vanish at half-filling and the full density includes the additional amount $\rho_{0}=\frac{e}{2a}$ that arises from all the occupied states up to half filling. This makes the total density $\rho=\rho_0+\delta\rho=e\frac{k_{FR}-k_{FL}}{\pi}$ as expected.  However, if we are just given the continuum model, without reference to an initial lattice model, it only has information about $\delta\rho$ and $\delta j.$  We note that neither the current, nor the density, depend on the dispersion term $\tfrac{1}{2}\hbar\Delta v_F q\mathbb{I}$ and so we will drop it from further discussion as it is also higher order in the expansion around the Fermi-points.

 From this Hamiltonian it is simple to construct the Lagrangian now using the Dirac matrices $\gamma^{0}=i\sigma^{x}, \gamma^{1}=\sigma^{y}$ and the chirality matrix $\gamma^{3}=\sigma^{z}$. We find
\be
\mathcal{L}=\overline{\psi}\left(i\slashed{\del}-\slashed{b}\gamma^{3}\right)\psi
\ee\noindent where $\slashed{b}=b_{\mu}\gamma^{\mu}$ for $b_{\mu}=\left(\tfrac{1}{2}\Delta v_F\delta\kappa_F,\delta \kappa_F\right).$
If we included the EM gauge field, this Lagrangian would be analogous to the Lagrangian derived for the Weyl semimetal in Ref. \onlinecite{zyuzin2012}, except this is in $1+1$ dimensions.  We can now get rid of the $b_{\mu}$-dependent term by doing a chiral gauge transformation. As is well-known, this transformation can change the measure of the path integral and lead to anomalous terms in the effective action. 

We will use the Fujikawa method to derive the effective response due to this change of measure. Performing a series of infinitesimal chiral transformations parametrized by the infinitesimal $ds$, we can get rid of the $b_\mu$ dependent term:
\bea
\label{eq:chiral_trans}
\psi&\rightarrow& e^{-ids\theta(x)\gamma^{3}/2}\psi\\
\overline{\psi}&\rightarrow&\overline{\psi}e^{-ids\theta(x)\gamma^{3}/2}
\eea
where $\theta(x)\equiv 2b_{\mu}x^{\mu}.$ Note that using this choice of $\theta(x)$ we have made an arbitrary choice of origin which is folded into the calculation. To avoid spurious response terms we need to constrain the system to be homogeneous in space-time so that each choice of space-time origin is equivalent. The Dirac operator $\slashed{D}$ acts as follows
\bea
\slashed{D}&=&i\slashed{\del}-\slashed{A}-\slashed{b}\gamma^{3}(1-s)\\
\slashed{D}\phi_{n}(x)&=&\epsilon_{n}\phi_{n}(x)
\eea
where $A_{\mu}$ is the EM gauge field and $\phi_{n}$ are a complete set of eigenstates of the Dirac operator. Let us write out 
\be
\psi(x)=\sum_{n}c_{n}\phi_{n}(x)\, ,\;\;\; \overline{\psi}(x)=\sum_{n}\overline{c}_{n}\phi_{n}^{*}(x)
\ee
\noindent where $c_{n}$ are Grassman variables, and we can expand $\psi$ in terms of $\phi_{n}$ because they are complete. Considering what the infinitesimal chiral transformation does to the $c_{n}$'s, from Eq.  \ref{eq:chiral_trans}, we see that 
\bea
c_{n}'&=&\sum_{m}U_{nm}c_{m}\, ,\;\;\;\overline{c}_{n}'=\sum_{m}U_{nm}\overline{c}_{m}\\
U_{nm}&=&\delta_{nm}-\frac{ids}{2}\int d^{2}x\phi_{n}^{*}(x)\theta(x)\gamma^{3}\phi_{m}(x).
\eea
The Jacobian of this transformation is $J=\mbox{det}(U^{-2})$. Using the identity that $\mbox{det}(U)=e^{Tr\,\log(U)}$, we see that
\be
J=e^{ids\sum_{n}\int d^{2}x\,\phi_{n}^{*}(x)\theta(x)\gamma^{3}\phi_{n}(x)}.
\ee

The Jacobian due to the chiral rotation thus induces a term in the effective action given by 
\bea
S_{eff}&=&\int_{0}^{1}\mbox{d}s\int\mbox{d}xdt\,\theta(x)I(x)\\
I(x)&=&\sum_{n}\phi_{n}^{*}(x)\gamma^{3}\phi_{n}(x).
\eea
To evaluate $I(x)$, we can use the heat kernel regularization:
\be
I(x)=\lim_{M\rightarrow\infty}\sum_{n}\phi_{n}^{*}(x)\gamma^{3}e^{-\slashed{D}^{2}/M^{2}}\phi_{n}(x)
\ee
to arrive at the well-known result that
\be
I(x)=-\frac{e}{4\pi}\epsilon^{\mu\nu}F_{\mu\nu}.
\ee
So, the effective action is given by 
\be
S_{eff}[A_\mu]=-\frac{e}{4\pi}\int\mbox{d}^{2}x\,\theta(x)\epsilon^{\mu\nu}F_{\mu\nu}.\label{eq:1drespF}
\ee
To remove the dependence on the arbitrary origin we can rewrite the action as  
\be
S_{eff}[A_\mu]=\frac{e}{\pi}\int \epsilon^{\mu\nu}b_{\mu}A_{\nu}.\label{eq:1drespA}
\ee This expression matches the result we determined from simpler calculations of the lattice model in Sections \ref{sec:1dEfield},\ref{sec:1dNNN} if we replace $\rho$ with $\delta \rho$ and $j$ with $\delta j.$

\subsection{Interfaces}
Now that we have derived the EM response via two separate methods, we will put it to use in this section where we calculate the properties of interfaces across which $b_{\mu}$ varies. We will show that the response action in Eq. \ref{eq:1drespA}  predicts  results that match numerical simulations, while the $\theta$-term version in Eq. \ref{eq:1drespF}  gives spurious results due to boundary terms that depend on the arbitrary choice of origin embedded in $\theta(x).$ We want to emphasize that this also happens in the case of the 3D Weyl semimetal and is a generic feature. One might think that one could remove these spurious terms by adding boundary degrees of freedom, however the spurious results to which we refer do not seem to be connected to any anomalies as they can appear on surfaces which do not exhibit gapless boundary modes.  

The form of the action to use when studying inhomogeneous systems (i.e., with relaxed translation invariance) is 
\begin{equation*}
S[A]=\frac{e}{\pi}\int d^{2}x\, \epsilon^{\mu\nu}b_{\mu}A_{\nu}.
\end{equation*} One might complain that this action appears gauge-variant, however, it is not. We note that we can define a current $j_{(b)}^{\mu}=\frac{e}{\pi}\epsilon^{\mu\nu}b_{\nu}$. Therefore, the action itself can be written $S=\int d^{2}x\, j_{(b)}^{\mu}A_{\mu}$. If the current is conserved then the action is gauge invariant due to the continuity equation. For the 1D metal, the current $j_{(b)}^{\mu}$ is exactly the EM charge current and thus is conserved yielding a gauge-invariant response functional.

Now, for the first example of an interface, suppose our 1D metal lies in the spatial region $x>x_{0},$ and there is only vacuum for $x<x_{0}.$ We model this by choosing $b_{\mu}(x)=b_{\mu}\Theta(x-x_{0})$ where $\Theta(x)$ is the step-function, and for simplicity we only turn on a non-vanishing $b_{1}.$ If we look at the charge density the response action would predict, we find
\begin{equation}
	\rho(x)=\frac{e}{\pi}b_{1}\Theta(x-x_{0})
\end{equation}\noindent which is physically correct since the metallic region will have a density equal to this value and the vacuum will have no density. If we had used the axion-action with $\theta(x,t)=2b_{1}(x-x_{1})$ for some arbitrary constant value $x_{1}$ we would have obtained the density
\begin{eqnarray} 
\bar{\rho}(x)&=&\frac{e}{2\pi}\del_{x}\theta(x,t)=\frac{e}{\pi}b_{1}\del_{x}((x-x_1)\Theta(x-x_{0}))\nonumber\\
&=&\frac{eb_1}{\pi}\left[(x_{0}-x_{1})\delta(x-x_{0})+\Theta(x-x_{0})\right].\label{eq:wrongdensity}
\end{eqnarray} This predicts a spurious boundary charge located at the interface $x_0$ and proportional to the distance between the boundary point and our \emph{arbitrary} choice of $x_1.$
This term is clearly unphysical and  simulations show that there is nothing special happening at the interface. Thus, the first action  reproduces the correct response and matches numerics for the 1-band lattice metal.

For a more complicated illustration, consider an interface between two different systems such that $b_{1}$ is non-vanishing in both, and varies in the $x$-direction. This will give an x-dependent charge density. A simple way to implement an $x$-dependent $b_{1}$ is to introduce an on-site energy term which is $x$-dependent. If we had a translationally invariant 1D lattice model with a fixed chemical potential $\mu,$ then shifting the onsite energy up or down will decrease or increase the electron density respectively. Let us consider two 1D segments which have a common boundary. Suppose the onsite energies are constant within each region, but are offset between the two regions by $\epsilon_0.$ To simplify the description we assume that they are glued periodically so, in fact, there are two interfaces. 

\begin{figure}[t]
\label{fig:on_site}
\centering
   \epsfxsize=3.5in \epsfbox{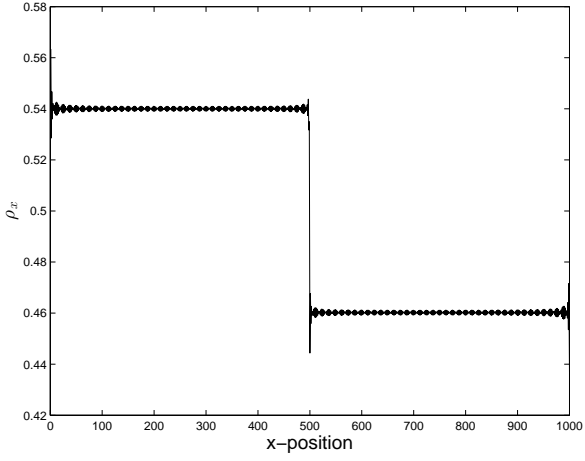}
   \caption{Charge density in units of $e/a$ as a function of position for an inhomogeneous system with $N=1000$ lattice sites where each segment has $L_s=500$ sites. The chemical potential is $\mu=0,$ and if $\epsilon_0$ was tuned to zero the density would be $\rho=e/2a.$    For our choice of $\epsilon_0=0.5t$ we have $b_{1(\ell)}=(\pi/a)0.46,\,b_{1(r)}=(\pi/a)0.54.$
  Away from the interfaces the values match the calculation from the effective response action.  Near the interfaces there are damped oscillations due to finite size effects that are not captured by the analytic calculation. Note that the finite-size boundary effects have nothing to do with the spurious ``interface"-terms in Eq. \ref{eq:wrongdensity}.}
\end{figure}

For an analytically tractable limit, let us study the case when the offset is not too big when compared to the bandwidth of the system, and with the chemical potential fixed at $\mu=0.$ The Hamiltonian is given by
\begin{equation}
H=-t\sum_{n}\left[c^{\dagger}_{n+1}c^{\phantom{\dagger}}_{n}+c^{\dagger}_{n}c^{\phantom{\dagger}}_{n+1}\right]+\sum_{n}\epsilon(n)c^{\dagger}_{n}c^{\phantom{\dagger}}_{n}
\end{equation}\noindent where $\epsilon(n)=\pm \epsilon_0/2$ when $n\leq N/2$ or $n> N/2$ respectively for a system with an even number of sites $N.$ We want to understand what happens to the charge density in the system, and compare it to what is predicted by the EM response action. With this Hamiltonian the system consists of two segments (labelled by $\ell$ and $r$), each of length $L_s=Na/2$ where $a$ is the lattice constant.

We can now compute what $b_{1(\ell)}$ and $b_{1(r)}$ are for each segment since there is a simple relation between charge density and $b_1.$ As the length of the segments approaches the thermodynamic limit, the average charge density will not depend on whether we calculate it with open or periodic boundary conditions, so for simplicity we can calculate the density with periodic boundary conditions for each segment separately.  With $\mu=0$ fixed, the Fermi momentum for the segment $\ell$ with the offset $+\epsilon_0/2$ is given by 
\be
0=\epsilon_{0}/2-2t\cos(k_{F,\ell}a)\implies k_{F,\ell}=\frac{1}{a}\cos^{-1}\left(\frac{\epsilon_{0}}{4t}\right).
\ee
The Fermi momentum for system $r$ is given by 
\be
0=-\epsilon_0/2-2t\cos(k_{F,r}a)\implies k_{F,r}=\frac{1}{a}\cos^{-1}\left(\frac{-\epsilon_0}{4t}\right).
\ee
So, we have $\rho_{r}=e\tfrac{k_{F,r}}{\pi}$ and $\rho_{\ell}=e\tfrac{k_{F,\ell}}{\pi}$ which by definition implies that $b_{1(r/\ell)}=(\pi/e)\rho_{r/\ell}=k_{Fr/\ell}.$ Explicitly we have
\bea
\label{eq:b}
b_{1(\ell)}&=&\frac{1}{a}\cos^{-1}\left(\frac{\epsilon_{0}}{4t}\right)\\
b_{1(r)}&=&\frac{1}{a}\cos^{-1}\left(\frac{-\epsilon_{0}}{4t}\right).
\eea In our geometry we have interfaces at $x=L_s$ and $x=2L_s\equiv0$ and $b_1$ varies across the interfaces. The EM response action predicts
\begin{eqnarray}
\rho&=&\frac{e}{\pi}\left[b_{1(\ell)}(\theta(x)-\theta(x-L_s))\right.\nonumber\\
&+&\left.b_{1(r)}(\theta(x-L_s)-\theta(x-2L_s))\right].
\end{eqnarray} This result matches what is found numerically as shown in Fig. \ref{fig:on_site}.

\subsection{General Comments}
Before we move on to discuss the more interesting higher dimensional semi-metals, we will pause to make a few important comments. 

\emph{(i) Response Action Without Translation Invariance:} Initially we parameterized the EM response of the 1D metal through quantities such as the Fermi-wave vector, and the velocity at the Fermi-points, which can only be clearly defined when there is translation symmetry. That is, when the system is homogeneous we can precisely define momentum space and these two quantities. What we have found is that the response is actually more general because we can define it in terms of the sources of Lorentz violation, i.e., an intrinsic charge density and charge current. These two physical quantities can be defined, and measured, without reference to momentum space and thus we can drop all reference to a Fermi wave vector and a Fermi velocity by using the density and current respectively. The fact that the EM response is accurate even without translation invariance is clearly shown when we have an interface as shown in the previous subsection. 

This physical definition of the response is special to 1D because the semi-metal EM response action is just $\int d^2x j^{\mu}A_{\mu}.$ This type of term will appear in every dimension, but in higher dimensions there are more interesting anisotropic response terms that appear and which we will discuss later. For $d$-dimensional space-time we can introduce a $(d-1)$-form $b_{\mu_1\mu_2\ldots \mu_{d-1}}$ representing a source of Lorentz breaking. We can furthermore take the dual to generate a current $j^{\mu}_{(b)}=\epsilon^{\mu\mu_{1}\ldots\mu_{d-1}}b_{\mu_1\mu_2\ldots \mu_{d-1}}$ which represents an intrinsic charge density or charge current which couples to $A_{\mu}$ minimally. This term yields the higher dimensional analog of the 1D semi-metal EM response. We comment later on the possibility to represent higher dimensional response actions without reference to momentum space. 

\emph{(ii) Response of Filled bands:} As is well-known from elementary solid-state physics, a filled band of electrons in a crystal carries no current. Each filled band also contributes a charge density $\rho_{band}=\frac{e}{a}$ or $e/\Omega$ where $\Omega$ is the size of a unit cell in higher dimensions. The EM response actions of topological semi-metals do not capture density or current contributions from filled bands and thus the response coefficients are ambiguous by a finite quantized amount, i.e., $b_{\mu}$ is ambiguous by the addition of half of a reciprocal lattice vector. 

\emph{(iii) Symmetries of $b_{\mu}$ in 1D:} Let us discuss the transformation properties of $b_{\mu}$ under time-reversal (T), charge-conjugation (C),  and inversion symmetry (P). Since in 1D we know that $b_0$ is proportional to a current and $b_1$ is proportional to a density we can easily determine their symmetry properties: 
\begin{eqnarray}
T&\colon& b_0\to -b_0\nonumber\\
C&\colon& b_0\to b_0\nonumber\\
P&\colon& b_0\to -b_0
\end{eqnarray}\noindent and
\begin{eqnarray}
T&\colon& b_1\to b_1\nonumber\\
C&\colon& b_1\to b_1\nonumber\\
P&\colon& b_1\to b_1.
\end{eqnarray} Note that they are both even under $C$ which is due to the fact that our convention for $b_{\mu}$ defined in terms of the density and current has the electric charge factored out. Subsequently, the response actions will have factors of electric charge in their normalization coefficients. Note that these symmetry properties only hold in 1D because the transformation properties of $b_{\mu}$ under these discrete symmetries are dimension dependent. 

\emph{(iv) Connection between 1D and 3D Semi-metals:} As mentioned in Section \ref{sec:prelim}, the effective response for a 3D Weyl-semi-metal is
\begin{equation*}
S[A_{\mu}]=-\frac{e^2}{2\pi h}\int d^4x \epsilon^{\mu\nu\rho\sigma}b_{\mu}A_{\nu}\partial_\rho A_\sigma.
\end{equation*}\noindent To be explicit, consider a system where $b_{\mu}=(b_0,0,0,b_z)$ in the presence of a uniform magnetic field $F_{xy}=-B_0.$ In this case the action reduces to
\begin{equation}
\frac{e^2\Phi}{\pi h}\int dt dz \epsilon^{ab}b_{a}A_{b}=N_{\Phi}\frac{e}{\pi}\int dt dz\epsilon^{ab}b_{a}A_{b}
\end{equation}\noindent where $\Phi=-B_0L_x L_y$ is the magnetic flux and $a,b=0,z.$ From this we see that the 3D action, for this arrangement of $b_{\mu}$ and $F_{\mu\nu},$ reduces to $N_{\Phi}=\vert\Phi/(h/e)\vert$ copies of the 1D action. This connection hints that it could be possible to define the response of the 3D Weyl semi-metal without reference to momentum space, and instead only using physical quantities, e.g., the charge density and current in a uniform magnetic field. It also shows why the symmetry transformation properties of $b_{\mu}$ in 1D are different than those of $b_{\mu}$ in 3D because of the additional factor of $\Phi$ in 3D which is odd under time-reversal. We will discuss this more in the section on 3D semi-metals.

\section{Dirac Semimetal in $2+1$-dimensions}
\label{sec:twoDDSM}
After our discussion of the simple 1-band metal we will now move on to a discussion of the 2D Dirac semi-metal which has become widely recognized with the experimental discovery of graphene\cite{graphenereview}. Graphene is a honeycomb lattice of carbon atoms with a low-energy electronic structure consisting of four Dirac points. These four Dirac points are located in spin-degenerate pairs at the special points $K$ and $K'$ in the hexagonal Brillouin zone.  For models like graphene with both time-reversal and inversion symmetry, the minimum number of Dirac points that can appear in a 2D lattice model is two, and graphene has twice this amount because of the spin-1/2 degeneracy of the electrons due to the time-reversal symmetry with $T^2=-1.$ For our purposes, we will focus on a reduced case of spinless (or spin-polarized) electrons in which $T^2=+1.$ To recover results for graphene one could trivially add in the degenerate spin degree of freedom. 
\subsection{Dirac Semi-metal From Layered Topological Insulators}

\subsubsection{Topological Insulator in 1D Protected by C or P Symmetry}
As discussed in Section \ref{sec:prelim}, each TSM can be illustrated as a collection of lower dimensional TIs which are stacked and then coupled; the Dirac semi-metal (DSM) is no different. To generate a DSM this way we must begin with 1D TI wires. From the classification of 1D TIs we know that to have a robust, non-trivial state we must require the presence of a symmetry to protect the state\cite{qi2008,schnyder2008,kitaev2009periodic}. This is inherently different than the 3D Weyl semi-metal, which is constructed from stacks of 2D Chern insulators that require no symmetry to have protected topological phases. There are two possibilities for an appropriate 1D TI symmetry: (i) charge-conjugation symmetry ($C$) or (ii) inversion/reflection symmetry ($P$). For $C$-symmetry the 1D topological wire lies in class D of the Altland-Zirnbauer classification\cite{AZ,schnyder2008,qi2008}, and there is a $Z_2$ topological invariant that controls the EM response. For $P$-symmetry the wire belongs to the set of inversion-symmetric insulators and also has a $Z_2$ topological invariant\cite{zak1989berry,pollmann2012symmetry,hughes2011inversion}. In both cases we will call the invariant ${\cal{Z}}_1.$ If ${\cal{Z}}_1$ takes its trivial (non-trivial) value  ${\cal{Z}}_1=+1 ( {\cal{Z}}_1=-1)$ then the insulator will have a bulk charge polarization of $P_{1}= ne\;{\rm{mod}\; \mathbb{Z}e} (P_1=(n+1/2)e\;{\rm{mod} \;\mathbb{Z}e}),$ and will exhibit an even (odd) number of low-energy fermion bound-states on each boundary point. Let us note that we will use $P$ to label refection symmetries (inverting a single coordinate) and ${\cal{I}}$ to represent inversion symmetry (reflection in all coordinates). Of course in 1D they are the same so we will simply use $P$ for 1D systems. 

Since it will become important, let us review the EM response of the 1D TI. The response is captured by the effective action 
\begin{equation}
S_{eff}[A_\mu]=\frac{1}{2}\int d^2 x P_1\epsilon^{\mu\nu}F_{\mu\nu}
\end{equation}\noindent where $P_1$ depends on the insulating phase as given above. The requirement of either $C$ or $P$ symmetry enforces a quantization of the polarization in units of half an electron charge. Naively these symmetries should forbid a non-zero $P_1$ since $P_1\to -P_1$ under $C$ and under $P.$ However, since the polarization in 1D crystalline insulators is only well-defined modulo integer charge, the allowed values of $P_1$ are $0$ and $e/2$ which both satisfy $P_1=-P_1$ modulo integer electron charges\cite{vanderbilt1993,qi2008}. Another way to say this is that 1D insulators with polarizations that differ by an integer electron charge are topologically equivalent (stably equivalent).   

A simple model which exhibits a 1D TI phase is given by the 1D lattice Dirac model. For translationally invariant systems this model has a Bloch Hamiltonian
\begin{equation}
H_{1DTI}(k)=(A\sin ka)\sigma^y+(B-m-B\cos ka)\sigma^z\label{eq:Ham1DTI}
\end{equation}\noindent where $A,B,m$ are model parameters (we set $A=B=1$ from now on), $a$ is the lattice constant, and $\sigma^a$ are the Pauli matrices representing some degrees of freedom within the unit cell. The phases of this model are controlled by the parameter $m$ and for $m<0$ or $m>2$ the system is a trivial insulator with ${\cal{Z}}_1=+1.$ For $0<m<2$ the system is in a TI phase with ${\cal{Z}}_1=-1.$ A benefit of this model is that we can judiciously choose a $C$ operator and a $P$ operator such that the Hamiltonian has that symmetry. For example, if we pick $C=\sigma^yK$ where $K$ is complex conjugation, then $CH_{1DTI}(k)C^{-1}=-H_{1DTI}(-k),$ and if we pick $P=\sigma^z$ then $PH_{1DTI}(k)P^{-1}=H_{1DTI}(-k).$ So, as written, this model is simple enough to have both $C$ and $P$ symmetry, and thus can exhibit a protected topological phase.  If we add perturbations to the model that break one of the symmetries but preserve the other, then the topological phase will remain stable. It is only if we break both symmetries that we can destabilize the 1D TI phase. Usually, for insulators, a $C$-symmetry only exists when the model is fine-tuned, but inversion/reflection symmetry can be approximately preserved in real materials. In what follows we will emphasize the inversion or reflection symmetric cases as it is more relevant when considering semi-metal phases. We note that this model also has time-reversal symmetry with $T=K$ ($T^2=+1$), although this symmetry is not important for the 1D classification.

\subsubsection{Weak Topological Insulator in 2D Protected by C, $P,$ or  {\cal{I}} Symmetry}

Before we approach the DSM let us consider the 2D WTI phase generated by stacking a weakly coupled set of 1D TI wires. To be explicit, suppose that the wires are oriented parallel to the $x$-axis and stacked perpendicularly to spread into the $y$-direction. In the limit of decoupled wires, we can determine that the system will have a charge polarization in the $x$-direction, and will have low-energy boundary states on boundaries with a normal vector in the $x$-direction. In this limit, a 2D Hamiltonian representing this phase is just multiple copies of $H_{1DTI}$ with a fixed value of $0<m<2$ for each wire. These characteristics remain as long as the coupling between the wires does not close the bulk gap, and preserves the relevant 1D symmetry. We can model this using a square-lattice Bloch Hamiltonian
\begin{equation}
H_{2DWTI}(\vec{k})=\sin (k_x a) \sigma^y+(1-m-\cos (k_x a) -t_y\cos (k_y a) )\sigma^z\label{eq:Ham2DWTI}
\end{equation}\noindent for a lattice constant $a$ and a new tunneling parameter $t_y.$ Again, this model has both $C$ and $P_x$ symmetry (reflection with $x\to -x$) with the same operators as above since the inter-wire tunneling term $-t_y\cos (k_ya)\sigma^z$ preserves both. It also has time-reversal symmetry $T=K$, reflection symmetry in the $y$-direction with $P_y=\mathbb{I},$ and inversion symmetry with $\mathcal{I}=\sigma^z.$ If we pick $0<m<2$ then the model remains in the WTI phase as long as no solutions for one of
\begin{eqnarray}
\cos (k_y a)&=&-\frac{m}{t_y}\nonumber\\
\cos (k_y a)&=&\frac{2-m}{t_y}\label{eq:DSMcon}
\end{eqnarray}\noindent can be found. We immediately see that as long as $\vert t_y\vert<\vert m\vert$ and $\vert t_y\vert<\vert(2-m)\vert,$ then the bulk gap will remain open, and for $0<m<2$ the model will be in the WTI phase.

 This WTI is characterized by a 2D topological vector invariant $\vec{\nu}=\left(0,\tfrac{\pi}{a}\right)$ which is a half-reciprocal lattice vector. The EM response of the 2D WTI is just
\begin{equation}
S_{eff}[A_\mu]=\frac{e}{4\pi}\int d^3 x\,\nu_{\mu}\epsilon^{\mu\nu\rho}F_{\nu\rho}\label{eq:2DWTIResponse}
\end{equation}\noindent where $\nu_0=0.$ This response represents the contribution of a charge polarization $\vec{P}_1$ to the action where $P^{i}_{1}=\frac{e}{2\pi}\epsilon^{ij}\nu_{j}=(\frac{e}{2a},0).$ The magnitude of the polarization is due to a contribution of $e/2$ charge per wire, as expected, and the total charge on a boundary with normal vector $\hat{x}$ will be $N_y\tfrac{e}{2}$ where $N_y$ is the number of wire layers. As discussed in Section \ref{sec:prelim}, the WTI phase does not give rise to $\nu_{0}$ because there is no Lorentz-breaking in the time-direction for a filled band. One could generate a $\nu_0$ in an insulator by applying a time-dependent periodic field to generate Floquet dynamics, or by coupling the system to a varying adiabatic parameter that will drive cyclic adiabatic charge pumping\cite{thouless1983}. This will drive a constant, quantized current along the wires which will result in a non-vanishing $\nu_0$ proportional to the charge pumping frequency. As we will discuss further below, just like $\nu_{i}$ is connected to the intrinsic charge polarization, $\nu_0$ is related to the intrinsic magnetization, which is why producing currents will generate such a term. 

\subsubsection{From 2D Weak Topological Insulator To Dirac Semi-metal}
\begin{figure}[t]
\label{fig:graphene_2band}
\centering
   \epsfxsize=3.5in \epsfbox{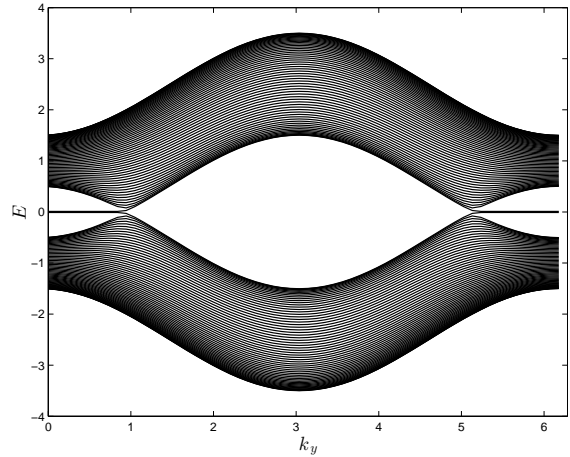}
   \caption{The energy spectrum for the Hamiltonian in Eq. \ref{eq:Ham2DWTI} tuned into the 2D Dirac semi-metal. The figure shows exact diagonalization of this model in a strip geometry (x-direction with open boundaries, and y-direction with periodic boundaries) with $\pm k_{yc}=\pm \pi/3$ and $b_{y}=\pi/3.$ The flat band of states stretched between the Dirac nodes are edge modes.}
\end{figure}

We will now show an explicit example of a Dirac semi-metal and discuss its physical response properties and characteristics before deriving a result for a generic Dirac semi-metal in the following sections. It is easy to construct an explicit example of a DSM phase from the WTI model by choosing $m$ and $t_y$ such that at least one of Eq. \ref{eq:DSMcon} has a solution. To be concrete let $m=1/2, t_y=-1,$ and $a=1$  for which $\cos k_y=-m/t_y$ has two solutions:  $\pm k_{yc}=\pm \pi/3,$ which implies there are Dirac points at $\vec{k}=(0,\pm k_{yc}).$ If we expand the Hamiltonian in Eq. \ref{eq:Ham2DWTI} around these points, we find the continuum Hamiltonians
 \begin{equation}
 H_{2Dcon}=\delta k_x\sigma^{x} \pm \frac{\sqrt{3}}{2}\delta k_y\sigma^z\label{eq:Ham2Dcont}
 \end{equation}\noindent which are anisotropic Dirac points with $\delta k_x$ the deviation from $k_x=0,$ and $\delta k_y$ the deviation from $k_{y}=\pm k_{yc}.$ If we tuned the velocity parameter $A$ in Eq. \ref{eq:Ham1DTI} to be $\sqrt{3}/2$ we would find isotropic Dirac points. In Fig. \ref{fig:graphene_2band} we show the energy spectrum of this model at the parameter values given above in a strip/cylinder geometry with open boundary conditions in the $x$-direction and periodic boundary conditions in the $y$-direction. We see the Dirac points at the predicted values, and also a flat-band of mid-gap states which are exponentially localized on the edges of the strip. 
 
 Despite some superficial differences, the square-lattice model for the DSM captures the same physics as the honeycomb graphene model. In fact, in Appendix \ref{app:graphene} we show that the square lattice model for the DSM can be deformed to the honeycomb graphene model, and thus graphene can be constructed from layers of 1D TIs if we trivially add degenerate spin copies. This matches the well-known result that graphene has anisotropic boundary states that appear only on zig-zag edges and not arm-chair edges, which is a consequence of this layered structure and the close connection to the WTI model\cite{graphenereview}.

Following the pattern discussed in Section \ref{sec:prelim}, when the DSM is formed, we expect there to be a source of Lorentz violation proportional to the momentum and energy differences between the Dirac nodes. For our explicit choice of parameters we should have $b_{\mu}=(0,0,\pi/3).$ As we will prove in Section \ref{sec:2dresponse}, one contribution to the EM response is the analog of Eq. \ref{eq:2DWTIResponse} for the WTI phase, that is:
\begin{equation}
S_{eff}[A_{\mu}]=\frac{e}{4\pi}\int d^3 x\,b_{\mu}\epsilon^{\mu\nu\rho}F_{\nu\rho}.\label{eq:2DDSMResponse}
\end{equation}\noindent From the interpretation of the 2D WTI response above, this implies a non-zero charge polarization 
\begin{equation}
P^{i}_{1}=-\frac{e}{2\pi}\epsilon^{ij}b_{j}.
\end{equation}
To see the origin of the polarization we should heuristically think that the DSM model Hamiltonian represents a family of 1D insulators, one for each value of $k_y.$ That is, each value of $k_y$ (except $k_{y}=\pm k_{yc}$) represents a 1D insulating wire oriented in the x-direction. The 1D wires with $k_y$ values on opposite sides of a Dirac point have opposite values of ${\cal{Z}}_1,$ and thus their contributions to the overall charge polarization differ by a quantized amount. For the gapped WTI phase each wire contributes $e/2$ charge (modulo $ne$) to an edge normal to the $x$-axis, but for the DSM only the fraction of the wires between the Dirac nodes contribute $e/2$ while the remainder contribute charge $0\mod e.$

\begin{figure}[t]
\label{fig:rho_x}
\centering
   \epsfxsize=3.5in \epsfbox{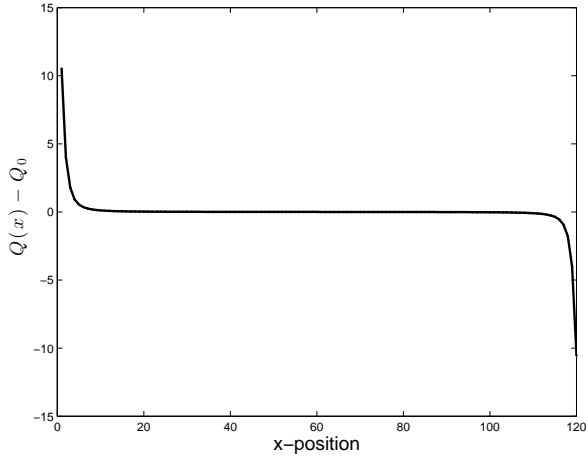}
   \caption{We have plotted the deviation of the charge density from the average for $L_{x}=L_{y}=120$ at half-filling in a 2D Dirac semi-metal with $b_{y}=\pi/3$ (i.e., same parameters as in the previous figure). The average background charge per site is  $Q_{0}=120e.$  We notice peaks at the boundaries of the system due to the charge carried by localized mid-gap modes. The charge density exponentially decays to the value of $Q_{0}=120e$ within a few lattice sites. The total charge at the boundary calculated from summing the boundary charge near the right edge is $Q_{b}=-19.6e$ which matches the expected result $Q_{b}=P_{1}^{x}L_y=-\frac{e}{6a}120a=-20e.$ The deviation from $-20$ is a finite-size effect and the result will converge to the analytic value as the system size increases.}
\end{figure}

Physically, the bulk polarization manifests as a bound charge on the sample edges.  In Fig. \ref{fig:rho_x} we show the charge density as a function of position along the open boundary direction for the cylinder geometry mentioned above. We have subtracted off the average background charge, and two peaks in the charge density can be seen; one on each end of the sample. The amount of charge localized on each end matches the charge density calculated from Eq. \ref{eq:2DDSMResponse} at an interface where the polarization changes from $P^{x}_{1}=-\tfrac{e}{2\pi}\frac{\pi}{3a}=-\tfrac{e}{6a}$ to zero (we have temporarily restored the lattice constant).

There are two important subtleties to consider when calculating the polarization. The first subtlety has to do with which direction the polarization should point, for example, what determines which boundary has the positive charge in Fig. \ref{fig:rho_x}, and which end has a negative charge? The answer to this question is well-known: to uniquely specify the polarization we must apply an inversion-breaking (or C-breaking) field that picks the direction of the polarization, and then take the limit as the system size goes to infinity before setting the symmetry breaking perturbation to zero. This is the conventional paradigm for spontaneous symmetry breaking. Thus, in order to uniquely specify the sign of the polarization, and thus the sign of the effective $b_{i},$ we must turn on a small symmetry breaking perturbation before we calculate, and take the limit in which this perturbation vanishes. This issue will arise in the next section when we try to calculate Eq. \ref{eq:2DDSMResponse} using field-theoretical methods. To be consistent with the notation in the next section, we will call the inversion symmetry breaking parameter $m_A.$

The second subtlety is similar in nature: in a bulk sample without boundary, since the Brillouin zone is periodic and we have no edge states to reference, we cannot determine a unique value for the polarization. For example, how do we determine the magnitude of the polarization if we do not have a preferred way to take the momentum difference between the Dirac nodes? This is a problem because there are multiple ways to subtract the momenta in a periodic BZ. For our concrete example our nodes lie at $\vec{k}=(0,\pm \pi/3),$ and so we could let $\vec{b}=\tfrac{1}{2}(0,2\pi/3)$ or, e.g., we could subtract the nodes across the Brillouin zone boundary to find $\vec{b}'=\tfrac{1}{2}(0,-4\pi/3).$ The resolution of this problem is also clear because the two results differ by the contribution of an entire filled band, i.e., the vectors differ by a half-reciprocal lattice vector $\vec{b}-\vec{b}'=(0,\pi)=\tfrac{1}{2}\vec{G}_y.$ Thus we see another indication that $b_{\mu}$ is only uniquely determined up to the contributions of filled bands.

 Before we move on to discuss the topological response due to the time-component $b_0$, some comments about symmetry protection are in order. For the 1D TI, and the 2D WTI constructed from stacks of these 1D TI wires, we have only required inversion symmetry to have a well-defined electromagnetic response. This symmetry quantizes the 1D polarization to be $0$ or $e/2$ on each wire, and as shown in Refs. \onlinecite{hughes2011inversion,pollmann2012symmetry}, this symmetry is also enough to quantize the polarization (per wire) for the 2D WTI. However, it is well-known\cite{bhbook} that for local stability of the Dirac nodes in a DSM one needs the composite $T\mathcal{I}$-symmetry (for $T^2=+1$). We would like to understand the importance of this difference. First, for 1D wires $T\mathcal{I}$ also quantizes the polarization since $P_1$ is odd under this symmetry. Thus, we could have already constructed a 1D TI and a 2D WTI using this symmetry instead. However, in dimensions greater than one, $T\mathcal{I}$ does something else important: it constrains the Berry curvature to satisfy $F(k_x,k_y)=-F(k_x,k_y).$ Since the Berry curvature flux is only defined modulo $2\pi$ on a lattice, this requires that for gapped systems either (i) $F(k_x,k_y)=0$ or (ii) $F(k_x,k_y)=\pi$\footnote{The authors do not know of any models which realize the latter case. One must also worry about the fact that the total flux must be a multiple of $2\pi$ and thus, to be well-defined we must have an even number of discretized momentum points. This constraint seems a bit artificial so we will not consider this case further.} and is constant throughout the Brillouin zone. However, if $F(k_x,k_y)$ is not required to be smooth, we can have singular points in momentum space where $F(k_{xc},k_{yc})=\pi$; these are exactly the set of Dirac node locations. Since the Berry flux that passes through a closed manifold must be a multiple of $2\pi$ this implies that there are an even number of singular points, i.e., fermion doubling. 
 
 This constraint, and thus the $T\mathcal{I}$ symmetry itself, is also essential for the 2D charge polarization response of the DSM. Let us illustrate the idea. Suppose we wish to calculate the charge polarization of a crystalline DSM. The physical consequence of a non-vanishing polarization is a boundary charge, so let us specify a particular boundary with a normal vector ${\bf{G}}_N$ in the reciprocal lattice. Let  ${\bf{G}}_F$ be the dual vector to ${\bf{G}}_N$, i.e., $G_{F}^{i}=\epsilon^{ij}G_{N}^{j}.$ Then ${\bf{G}}_F$ is the normal vector to a set of lattice lines whose ends terminate on the surface normal to ${\bf{G}}_N.$ For example, pick ${\bf{G}}_N=2\pi\hat{x}$ and ${\bf{G}}_F=2\pi\hat{y}.$ In this case our choice picks out a family of 1D wires parallel to the $x$-direction and stacked in the $y$-direction. Consequently this gives rise to a family of 1D Bloch Hamiltonians parameterized by the momentum along ${\bf{G}}_F.$ In this example we have the family  $H_{k_y}(k_x)$ which is parameterized by $k_y.$ 
 
 To calculate the charge polarization of the DSM with our choice of ${\bf{G}}_N$ (i.e., the polarization parallel to ${\bf{G}}_N$), we can start by asking an important question: how much does the charge polarization of the family of 1D systems $H_{k_y}(k_x)$ vary as $k_y$ is varied? We find
  \begin{eqnarray}
&& P_{1}^x(k_{y2})- P_{1}^x(k_{y1})\nonumber\\
 &=&\frac{e}{2\pi}\int_{-\pi}^{\pi} dk_x a_{x}(k_x,k_{y2})-\frac{e}{2\pi}\int_{-\pi}^{\pi} dk_x a_{x}(k_x,k_{y1})\nonumber\\
&=&\frac{e}{2\pi}\int_{-\pi}^{\pi} dk_{x}\int_{k_{y1}}^{k_{y2}} dk_{y} {\cal{F}}(k_x,k_y)\nonumber\\
&=&\frac{e}{2}\sum_{a=1}^{N_{enc}}\chi_a
 \end{eqnarray}\noindent where we have used Stokes theorem to replace the line integrals over the Berry connection ${\bf{a}}({\bf{k}})$ by an area integral over ${\cal{F}}(k_x,k_y)=\partial_{k_x}a_y-\partial_{k_y}a_x$, i.e., the Berry curvature, and we have assumed only one occupied band for simplicity. In the last equality we used the fact that for systems with  $T\mathcal{I}$-symmetry the Berry curvature is  equal to the contribution from the singular Dirac points. The quantity $\chi_a=\pm 1$ indicates whether the flux carried by the node is $\pm \pi.$ Thus, two 1D Hamiltonians that are members of the parameterized Hamiltonian family specify cycles in the Brillouin zone, and from this result  we see that the polarization can only change if the area of the Brillouin zone enclosed between those two 1D cycles contains Dirac nodes. This is the importance of the $T\mathcal{I}$-symmetry. Since the BZ is a closed manifold It is also important to note that it does not matter which region we choose, out of the two possible choices, to be the region ``enclosed" by the cycles. The two choices differ in a change of the polarization by an integer per unit cell, which is the same ambiguity we have seen for $b_{i}$ changing by half a reciprocal lattice vector. 
 
 This result is generically true given a general family of Bloch Hamiltonians (with $T\mathcal{I}$-symmetry) with some orientation specified by ${\bf{G}}_N,$ and parameterized by momentum along ${\bf{G}}_F.$ In fact, given two 1D cycles that are members of a parameterized Hamiltonian family in the Brillouin zone, then any deformation/rotation of the orientation of the lines, i.e., variation of the choice of the direction vector  ${\bf{G}}_N$ will not change the difference in polarization between the two parallel lines unless the lines cross Dirac points during the deformation process. This implies that the \emph{changes} in polarization are always quantized.  
 
 Even though the changes in polarization are quantized we might now ask what about the absolute magnitude of the polarization? Since each 1D subspace is mapped onto itself by $T\mathcal{I},$ but the polarization of that 1D system is odd under $T\mathcal{I}$, we see that the polarization of each of the wires/cycles is quantized to be $0$ or $e/2.$ The other wires in the family of Hamiltonians either have exactly the same polarization, or they differ by a quantized amount. This argument shows that the (fractional part of the) boundary charge, up to an integer per unit cell, is completely determined by the length of $b_{i}$ that projects onto the edge BZ which is what is predicted by Eq. \ref{eq:2DDSMResponse}.
 It is important to note that the distinction between whether the polarization of some set of wires is $0$ or $e/2$ will dictate where the edge modes occur in the edge Brillouin zone in a clean system. However, as far as the total response is concerned we could switch all the wires with $e/2$ polarizations to $0$ and all the ones with $0$ to $e/2$ and the total topological response will only change by the addition of a half-reciprocal lattice vector to $b_{\mu}$, i.e., by the topological response of a fully occupied band. This will switch the way that the edge states connect the Dirac nodes in the edge Brillouin zone, but as far as the bulk response is concerned it  is equivalent to the ambiguity in how we assign a difference in momentum to the Dirac nodes since there is not a unique way to subtract numbers on a circle.

\begin{figure}[t!]
\label{fig:DSMmAmB}
\centering
\includegraphics[width=3.1in]{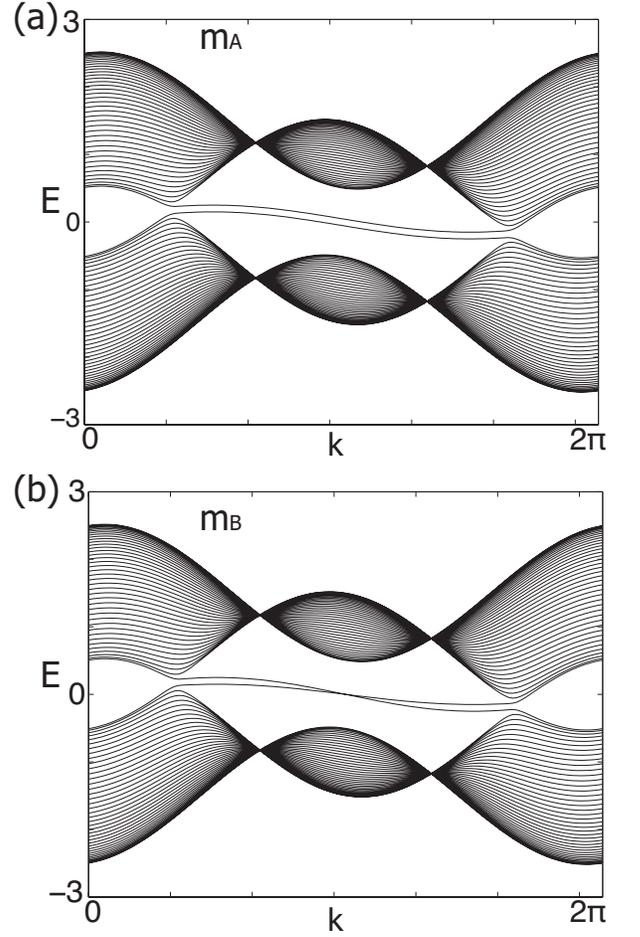}
   \caption{The Energy spectrum is shown for the DSM with $b_{0}\neq0$ and different masses turned on. (a) With $m_{A}\neq 0$, we see that the edge modes split and don't cross as they move between the Dirac nodes. (b) With $m_{B}\neq0$, it looks like the edge mode dispersion of a Chern insulator and they cross at $k=0$.  }
\end{figure}

   \begin{figure}[t!]
\centering
   \epsfxsize=10cm \epsfbox{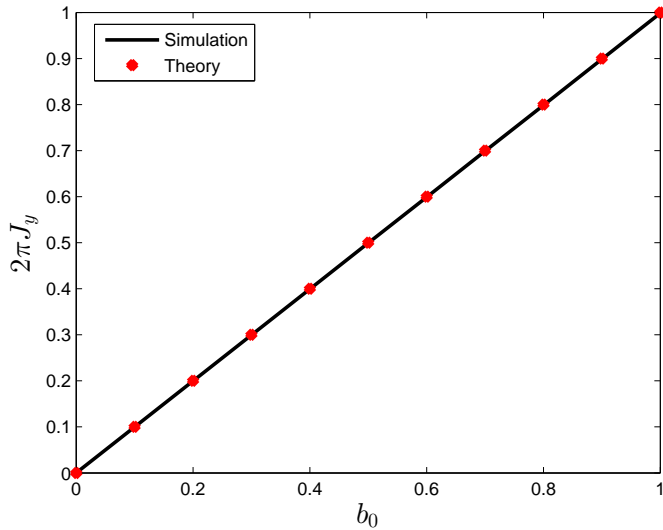}
   \caption{ The bound current $J_{y}$ localized near a single edge vs. $b_{0}$ is plotted for the model in Eq. \ref{eq:Ham2DWTI} with $b_{y}=\tfrac{\pi}{3}, m_{A}=10^{-3}, L_{x,y}=120,$ and periodic boundary conditions in the $y$-direction. The current matches the field theory prediction.
   }
\label{fig:b0_smooth}
   \end{figure}

We have seen that the spatial part of $b_{\mu}$ can be interpreted as a charge polarization, and, as will be shown below, the component $b_{0}$ represents an orbital magnetization. A non-vanishing magnetization implies a circulating current bound at the edges of the sample. In Eq. \ref{eq:Ham2DWTI} we can generate a $b_{0}$ by adding the term $\gamma\sin k_{y}\mathbb{I}.$ The value of $b_{0}$ generated would be $b_{0}=(\gamma/\hbar)\sin k_{yc}$ where $k_{y}=k_{yc}$ is the location of the Dirac node (and consequently $-k_{yc}$ is the location of the other node). The dispersion of the edge modes attached to the Dirac points is what generates the current on the edges which harbor topological bound states.  Again, to properly calculate the response numerically, there is a subtlety about how to  fill the edge sates. To do this properly we again need to choose a small non-zero inversion-breaking mass to fill the edge modes. In the language of Ref. \onlinecite{vanderbiltchern}, to properly fill the edge modes in the presence of a non-vanishing $m_A$ we need to use the adiabatic filling, not the thermal filling, if we want to calculate the magnetization. One can see the energy spectrum for $b_0\neq0$ in Fig. \ref{fig:DSMmAmB}a with a finite $m_A$ parameter. Adiabatic filling implies filling all of the states, including the edge modes, in the lower half of the spectrum below the energy gap induced by $m_A.$ In Fig. \ref{fig:b0_smooth} we plot the boundary current localized near a single edge vs. $b_0.$ The bound edge current is exactly $\tfrac{eb_0}{2\pi}.$

\begin{figure}[t]
        \centering
                \includegraphics[width=9cm]{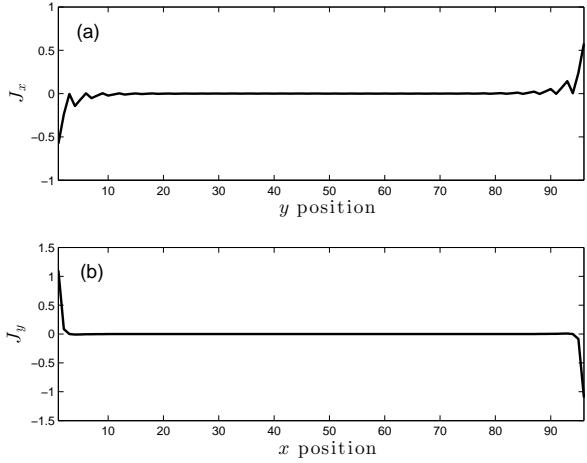}
        \caption{Plots of (a) $J_{x}$  vs. $y$ which is the current on the non-topological edge and (b) $J_{y}$ vs $x$ which is the current on the topological edge. This is for the Dirac semi-metal considered in previous figures but with a non-zero $b_0$. For this system $b_{y}=\tfrac{\pi}{3}, \gamma=0.1, m_{A}=0.1m,$ and $L_{x,y}=96.$ There are open boundary conditions in both directions. We note that $j_{y}$ is exponentially localized whereas $j_{x}$ is less-sharply localized and oscillates as it decays into the bulk. The oscillation wavelength  coincides with the wave-vector location of the Dirac nodes in $k$-space. With open boundary conditions, we must be careful to properly fill the edge states by using a non-zero inversion breaking mass term $m_A.$ }\label{fig:cur_xy}
\end{figure}

\begin{figure}[t]
\centering
   \epsfxsize=9cm \epsfbox{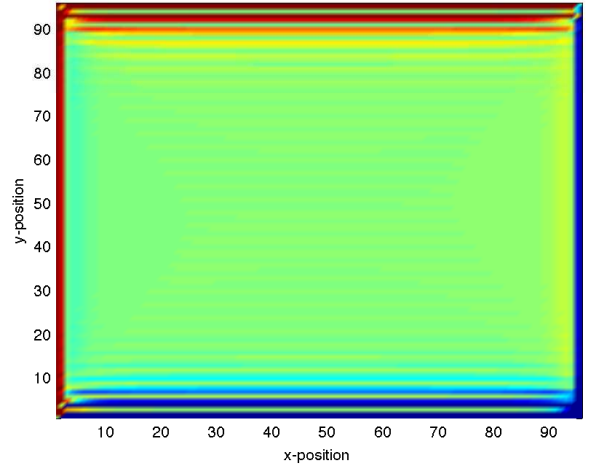}
   \caption{With a similar set up to the previous figure,  we use a density plot for the current vs $x,y$ position for the 2D Dirac semi-metal with $b_{y}=\tfrac{\pi}{3}, \gamma=m_{A}=0.1, L_{x,y}=96,$ and we have open boundary conditions in both directions. We calculated the current-density in the x-direction and summed it with the current density in the y-direction to produce this pseudo-color plot. We see that the currents are spatially localized at the edges, strongly for the one moving along the edges parallel to the $y$-axis  and less-strongly and oscillatory for the one moving along the edges parallel to the $x$-direction. The total magnitude of the current in the neighborhood of each edge is the same, and the current circulates around the boundaries of the sample. }
\label{fig:pcolor}
   \end{figure}

It is interesting to note that in the model in Eq. \ref{eq:Ham2DWTI} the $x$ and $y$ directions are very different since we have topological wires oriented along $x$ that are stacked in $y.$ This should be contrasted with the fact that an orbital magnetization in 2D implies the existence of a bound current on \emph{any} edge (i.e., the interfaces where the magnetization jumps from a finite value to zero). For the ``topological" edges with normal vectors parallel to the $x$-direction, a non-zero $b_0$ gives the edge modes a non-zero dispersion as shown in Fig. \ref{fig:DSMmAmB}a. The dispersing edge states produce an exponentially localized current $j^{y}_{bound}$ that corresponds to the change in magnetization at the edge.  However, in the $y$-direction there are no topological edge modes, and it is interesting to consider what happens to $j^{x}$ on these edges. We show the result of a numerical calculation in Figs. \ref{fig:cur_xy} and \ref{fig:pcolor}. In the former, we compare the current profiles of the different edges in a completely open geometry. In Fig. \ref{fig:cur_xy}a we show the current on a non-topological edge ($J_x$ on an edge normal to $\hat{y}$) which we see is localized on the boundary but now has an oscillatory decay. The wavelength of the oscillation in fact matches the wavelength of the Dirac node wave-vectors in momentum space. In Fig. \ref{fig:cur_xy}b we show the current localized on topological edges ($J_y$ on an edge normal to $\hat{x}$) and we can see that each edge carries exponentially localized current with opposite currents on opposite edges. In Fig. \ref{fig:pcolor} we show the current density on an open sample, where we see that all of the current is localized near the edges. The colors are associated to the magnitude of the current parallel to a given edge.  Essentially this is just a different presentation of the data that shows that on both sets of edges there is a bound current, as expected from the orbital magnetization, but on the non-topological edge the current oscillates as it decays. The magnitude of the current localized near edges of either type is identical, so indeed, even though the model is highly anisotropic, the bulk orbital magnetization generates bound currents on all edges, not just topological ones. 

Now that we have motivated the electromagnetic response of the DSM using some analytic and numeric results on an example model, we will now prove these claims using a Dirac semi-metal model with two nodes and then go on to generalize to a generic even number of nodes. 

%
%
%

%

\subsection{Derivation of Response for Continuum Dirac Semi-metal in 2D}\label{sec:2dresponse}
In the previous subsection we posited a form for the EM response action of the DSM and gave some concrete examples in which the numerical simulations in lattice models matched the response derived from the effective action in Eq. \ref{eq:2DDSMResponse}. In this subsection we will derive the EM response from a continuum model of the DSM using standard linear response techniques. We derived an example of a continuum Hamiltonian for the DSM in Eq. \ref{eq:Ham2Dcont}. After tuning the velocity coefficients to be isotropic, we can write the Hamiltonian for two Dirac cones that exist at the same point in the Brillouin zone as
\begin{equation}
H=k_x\mathbb{I}\otimes \sigma^x+k_y\tau^z\otimes\sigma^z.
\end{equation}\noindent To this Hamiltonian we will add two perturbations, the first of which is a splitting vector $b_\mu=(b_0,b_x,b_y)$ that shifts the two cones apart in momentum (by $2\vec{b}$) and energy (by $2b_0$). With the inclusion of this vector, which we will soon allow to be slowly varying in space-time, the Hamiltonian becomes
\begin{equation}
H=k_x\mathbb{I}\otimes \sigma^x-b_x\tau^z\otimes \sigma^x+k_y\tau^z\otimes\sigma^z-b_y\mathbb{I}\otimes\sigma^z+b_0\tau^{z}\otimes\mathbb{I}.
\end{equation}\noindent The second perturbations we will allow for are the coupling to external EM fields which enter the Hamiltonian via minimal coupling ${\bf{k}}\to{\bf{k}}-(e/\hbar){\bf{A}}.$

To calculate the linear response we need the current operators that will enter the Kubo-formula calculation. For the EM field the current operators are 
\begin{eqnarray}
J_{A}^{x}&=&\frac{\delta H}{\delta A_x}=\frac{e}{\hbar}\mathbb{I}\otimes \sigma^x\equiv \frac{e}{\hbar}\Gamma^x\\
J_{A}^{y}&=&\frac{\delta H}{\delta A_y}=\frac{e}{\hbar}\tau^z\otimes \sigma^z\equiv \frac{e}{\hbar}\Gamma^y\\
J_{A}^{0}&=&\frac{\delta H}{\delta A_0}=\frac{e}{\hbar}\mathbb{I}\otimes \mathbb{I}.
\end{eqnarray}\noindent For the splitting vector $b_{\mu}$ the associated currents are
\begin{eqnarray}
J_{B}^{x}&=&\frac{\delta H}{\delta b_x}=-\tau^z\otimes\sigma^x\equiv \Lambda^x\\
J_{B}^{y}&=&\frac{\delta H}{\delta b_y}=-\mathbb{I}\otimes \sigma^z\equiv \Lambda^y\\
J_{B}^{0}&=&\frac{\delta H}{\delta b_0}=\tau^z\otimes \mathbb{I}\equiv \Lambda^0.
\end{eqnarray}

We want to calculate the ``topological" response terms for the DSM and, in 2+1-d, such response terms will break either time-reversal or inversion symmetry. It is well known that Dirac fermions in 2+1-d exhibit a parity anomaly that gives rise to a Chern-Simons contribution to the effective action that encodes a non-vanishing Hall conductivity\cite{redlich}. There is a subtlety: to calculate the non-vanishing coefficient one must introduce a finite, symmetry breaking mass parameter that is taken to vanish at the end of the calculation. Since the resulting response coefficient ends up being proportional only to the sign of the symmetry breaking parameter, it remains non-zero even in the limit where the symmetry breaking is removed. This effect is a manifestation of a quantum breaking of symmetry, i.e., an anomaly. To calculate the responses due to $A_\mu$ or $b_\nu$ perturbations, we will need to introduce two different types of mass terms
\begin{eqnarray}
\Sigma_{A}&=&\mathbb{I}\otimes \sigma^y\\
\Sigma_{B}&=&\tau^z\otimes \sigma^y.
\end{eqnarray}\noindent These two different mass matrices commute and thus they are \emph{competing} mass terms. They both separately anti-commute with the kinetic part of the Dirac Hamiltonian (including a constant momentum shift $\vec{b}$), and thus the spectrum will be gapped as long as the coefficients $(m_A,m_B)$ of $(\Sigma_A,\Sigma_B)$ are not equal in magnitude. Explicitly, if both mass terms are activated, the energy spectrum is $\pm E_{\pm}=\pm\sqrt{(k_x-b_x)^2+(k_y-b_y)^2+(m_A\pm m_B)^2}$ which is gapped unless $\vert m_A\vert=\vert m_B\vert.$ These mass terms are very familiar in the literature: $\Sigma_A$ is essentially the inversion-breaking Semenoff mass term\cite{semenoff1984}, and $\Sigma_B$ is the continuum version of the time-reversal breaking Haldane mass term. 

Generically we will find contributions to the effective action of the form 
\begin{equation}
S_{eff}[A_\mu,b_{\nu}]=\int \frac{d^3 p_1}{(2\pi)^3}{\cal{A}}^{a}_{\mu}(p_1)\Pi^{\mu\nu}_{ab}(p_1){\cal{A}}^{b}_{\nu}(-p_1)
\end{equation}\noindent which has been written in the Fourier-transformed basis, and where $a,b=A,B$ and ${\cal{A}}^{A}_{\mu}=A_{\mu}$ and ${\cal{A}}^{B}_{\mu}=b_{\mu}.$
The linear response calculation (or equivalently the calculation of the quadratic term in the effective action) amounts to the calculation of the long-wavelength, DC limit of the generalized polarization tensor
\begin{equation}
\Pi^{\mu\nu}_{ab}(\nu,{\bf{q}})=\frac{\hbar}{2}\int \frac{d\omega d^2 p}{(2\pi)^3}{\rm{tr}}\left[J^{\mu}_{a}G(\omega+\nu,{\bf{p}}+{\bf{q}})J^{\nu}_{b}G(\omega,{\bf{p}})\right]
\end{equation}\noindent where $\mu,\nu=0, x, y;$ $a,b=A,B,$ and $G(\omega,{\bf{p}})$ is the space and time Fourier transform of the single-particle Green function of the unperturbed ($b_{\mu}=0, A_{\mu}=0$) Dirac model. The calculation of $\Pi^{\mu\nu}_{ab}$ is sensitive to the choice of symmetry breaking masses $m_{a}.$ Since we are only interested in extracting the topological terms, we can consider two cases (i) $\vert m_A\vert >\vert m_{B} \vert$ and (ii) $\vert m_B\vert>\vert m_{A}\vert.$ Since the system is gapped as long as $\vert m_{A}\vert \neq \vert m_{B}\vert,$ then the coefficients of the topological terms in the effective action will not depend on the magnitudes of $m_A$ and $m_B$ in cases (i) and (ii). Thus to simplify the calculation, for case (i) we can choose $m_B=0$ and for case (ii) we can choose $m_{A}=0.$

The Fourier transform of the unperturbed Green function in either of these limits will be
\begin{eqnarray}
G(\omega,p)&=&\frac{1}{\omega-p_x\Gamma^x-p_y\Gamma^y-m_c\Sigma^c}\nonumber\\&=&\frac{\omega+p_x\Gamma^x+p_y\Gamma^y+m_c\Sigma^c}{\omega^2-\vert {\bf{p}}\vert^2-m_{c}^2}
\end{eqnarray}\noindent where the label $c=A,B$ and is not summed over. The topological terms in the polarization tensor can be calculated by extracting the terms proportional to odd powers of the symmetry breaking mass:
\begin{eqnarray}
\Pi^{\mu\nu}_{ab}(\nu,{\bf{q}})&=&\frac{\hbar}{2} \int \frac{d\omega d^2 p}{(2\pi)^3}f(\omega+\nu,{\bf{p}}+{\bf{q}})f(\omega,{\bf{p}})\nonumber\\
&\times&{\rm{tr}}\left[J^{\mu}_{a}m_c\Sigma_c J^{\nu}_{b}(\omega+p_x\Gamma^x+p_y\Gamma^y)\right.\nonumber\\
&+&\left.J^{\mu}_{a}(\omega+\nu+(p_x+q_x)\Gamma^x\right.\nonumber\\
&+&\left.(p_y+q_y)\Gamma^y)J^{\nu}_{b}m_c\Sigma^c\right]\\
f(\omega,{\bf{p}})&=&\frac{1}{\omega^2-\vert {\bf{p}}\vert^2-m_{c}^2}.
\end{eqnarray} 

Now, to be explicit, let us consider case (i) where $m_{A}$ is the non-vanishing mass term. We can extract the leading term in the external frequency/momentum which  we find to be
\begin{eqnarray}
\Pi^{\mu\nu}_{ab}(\nu,{\bf{q}})&=&4 \frac{e}{2}  m_A\epsilon^{\mu\rho\nu}(iq_{\rho})\sigma_{ab}\int \frac{d\omega d^2 p}{(2\pi)^3}\left[f(\omega,{\bf{p}})\right]^2\nonumber\\
&=&\frac{4\pi^2}{(2\pi)^3} \frac{e}{2} \frac{m_A}{\vert m_A\vert}\epsilon^{\mu\rho\nu}(iq_{\rho})\sigma_{ab}\nonumber\\
&=&\frac{e}{4\pi}({\rm{sgn}}\; m_A) \epsilon^{\mu\rho\nu}(i q_{\rho})\sigma_{ab}\end{eqnarray}\noindent where $q_{\rho}=(\nu,{\bf{q}})$ is the external 3-momentum,  $\sigma_{AB}=\sigma_{BA}=1,$ and $\sigma_{AA}=\sigma_{BB}=0.$ This leads to a term in the effective action
\begin{equation}
S^{(A)}_{eff}[A_\mu,b_{\nu}]=\frac{e}{2\pi}({\rm{sgn}}\; m_A)\int dt d^2x \epsilon^{\mu\nu\rho}A_{\mu}\partial_\nu b_{\rho}.
\end{equation}\noindent This result exactly matches Eq. \ref{eq:2DDSMResponse} except for the factor of ${\rm{sgn}}\; m_A$ which we already motivated as being necessary to pick the sign of the charge polarization. 

If we repeat this calculation for case (ii), where $m_{B}$ is non-vanishing, the result is almost identical except the replacement of the matrix $\sigma_{ab}$ by the Kronecker $\delta_{ab},$ i.e., the polarization tensor is
\begin{equation}
\Pi^{\mu\nu}_{ab}(\nu,{\bf{q}})=\frac{\hbar e_{a}^2}{4\pi}({\rm{sgn}}\; m_B) \epsilon^{\mu\rho\nu}(i q_{\rho})\delta_{ab}
\end{equation}\noindent where the charge $e_{A}=e/\hbar$ and $e_{B}=1.$ Now this gives rise to two terms in the effective action
\begin{eqnarray}
S^{(B)}_{eff}[A_\mu,b_{\nu}]&=&\frac{e^2}{2h}({\rm{sgn}}\; m_B)\int dt d^2x \epsilon^{\mu\nu\rho}A_{\mu}\partial_\nu A_{\rho}\nonumber\\
&+&\frac{\hbar}{4\pi}({\rm{sgn}}\; m_B)\int dt d^2x \epsilon^{\mu\nu\rho}b_{\mu}\partial_\nu b_{\rho}.
\end{eqnarray}\noindent The first term is the conventional Chern-Simons term which yields a Hall conductivity of $\sigma_{xy}=\tfrac{e^2}{h}({\rm{sgn}}\; m_B)$ which consists of $\tfrac{e^2}{2h}({\rm{sgn}}\; m_B)$ from each of the two Dirac cones.  The second term, which does not yield an electromagnetic response since it is independent of $A_\mu,$ will be discussed later in Appendix \ref{app:Kmatrix}

\subsection{Physical Interpretation of the Dirac Semi-metal Response}
The topological EM response of the DSM is more complicated than the 1D band metal because the response density and current depend on derivatives of $b_{\mu},$ not just the vector itself. When the time-reversal mass term $m_B$ dominates then we just generate the well-known Chern insulator phase\cite{haldane1988} when we have only two nodes. We will thus leave further discussion of the time-reversal breaking mass to the more complicated cases with four or more nodes. In this section we will detail the less well-understood case of when $m_A$ dominates. 

\subsubsection{Response When $m_A$ Dominates}
Let us consider the limit in which the inversion breaking mass $m_A$ dominates over the time-reversal mass $m_B$ and then send them both to zero. In that limit the response that we derived is given by 
\begin{equation*}
S^{(A)}_{eff}[A_\mu,b_\mu]=\frac{e}{2\pi}({\rm{sgn}}\; m_A)\int dt d^2x \epsilon^{\mu\nu\rho}A_{\mu}\partial_\nu b_{\rho}.
\end{equation*}
The current in this model is given by 
\begin{eqnarray}
j^{\alpha}&=&\frac{e}{2\pi}({\rm{sgn}}\; m_A)\epsilon^{\alpha\mu\nu}\del_{\mu}b_{\nu}\\
\implies\rho&=&\frac{e}{2\pi}({\rm{sgn}}\; m_A)(\partial_x b_y-\partial_y b_x)\nonumber\\
j^{i}&=&\frac{e}{2\pi}({\rm{sgn}}\; m_A)\epsilon^{ij}(-\partial_0 b_{j}+\partial_{j}b_0).\nonumber
\end{eqnarray}\noindent To simplify let us assume that $m_A\to 0^{+}$ so that we can replace ${\rm{sgn}}\, m_A=+1.$ 

These equations can be more easily interpreted if we replace $b_{i}$ via the polarization $P^{i}_{1}=-\tfrac{e}{2\pi}\epsilon^{ij}b_{j}$ to generate
\begin{eqnarray}
\rho&=&-\partial_{i}P^{i}_{1}\nonumber\\
j^{i}&=&\partial_0 P^{i}_{1}+\frac{e}{2\pi}\epsilon^{ij}\partial_{j}b_0.\nonumber
\end{eqnarray}\noindent We immediately recognize these equations as the contributions to the charge density and current from gradients and time-derivatives of the polarization. It is also easy to interpret the term involving $b_0$ as it just represents the contribution to the current from gradients in the magnetization. We can let $M=\frac{e}{2\pi}b_0$ be the out-of-plane magnetization from which we finally arrive at
\begin{eqnarray}
\rho&=&-\partial_{i}P^{i}_{1}\nonumber\\
j^{i}&=&\partial_0 P^{i}_{1}+\epsilon^{ij}\partial_{j}M\label{eq:boundcurrent}
\end{eqnarray}\noindent which are the familiar constituent relations for bound charge density and bound charge current in 2D. Thus we see that in the limit where $m_A$ dominates over $m_B$ and then tends to zero, the DSM will exhibit an effective polarization and magnetization if $b_i$ and $b_0$ are non-zero respectively. Bound charge and current manifest at interfaces or boundaries where the bulk values of $b_{\mu}$ are changing and are the consequence of the topological response. 

The relation between $b_{\mu}$ and the bulk magnetization and polarization makes an important physical connection between generic electromagnetic quantities ($P^{i}_{1}, M$) and quantities that are defined as momentum and energy differences between momentum resolved Dirac points in the electronic spectrum $(b_{i}, b_0).$ Accordingly, we can rewrite the effective action as
\begin{equation}
S^{(A)}_{eff}[A_\mu,b_{\mu}]=({\rm{sgn}}\; m_A)\int dt d^2x \left[MB+P^{i}_{1}E_{i}\right].
\end{equation}\noindent This result is interesting because it shows that the DSM can have a well-defined polarization which is something that is usually reserved for insulators. In fact, one possible signature of a clean DSM would be a semi-metal with $T{\cal{I}}$-symmetry and a non-vanishing charge polarization/magnetization. 

Using the model for the DSM introduced above, we can explicitly understand the origin of the bound charge and bound current from a more microscopic picture. 
To get a non-zero charge density, we need $b_{y}$ to change with $x$ or vice versa. To get a non-zero current, we need $b_{0}$ to vary with $x$ or $y.$ The easiest way to do this is to have an interface or boundary. First, suppose we have a boundary where $b_{y}$ changes with $x$ as $b_{y}=b_{y}\Theta(x-x_{0})$ where $\Theta(x)$ is a step-function. From the response action we should have a bound charge density
\be
\rho=({\rm{sgn}}\; m_A)\frac{eb_{y}}{2\pi}\delta(x-x_{0}).
\ee
The magnitude of the charge density determined by the bulk response action exactly matches the boundary charge we find in the DSM model from the edge modes  stretched between the two nodes. The choice of the $({\rm{sgn}}\; m_A)$ fixes which edge has the occupied states. Due to the inversion breaking mass, each  boundary state on one edge will be occupied and contribute $e/2$ charge on that boundary for each edge mode. On the other edge all of the boundary modes will be unoccupied and each contributes a deficit charge of $-e/2.$  The total number of occupied of states on the edge is given by the distance between the two nodes multiplied by $\tfrac{L_{edge}}{2\pi}$, which is $L_{edge}\tfrac{2b_{y}}{2\pi}$. So, the total charge at the positive edge is given by $L_{edge}\tfrac{e}{2}\times\tfrac{b_{y}}{\pi}=L_{edge}\frac{eb_y}{2\pi}$. This implies a polarization of $\frac{eb_y}{2\pi}$ as expected. Thus, we see that while the charge response in the 1D semi-metal is controlled by the bulk states, here it is contributed by the boundary modes. 

The bound current that exists on interfaces when $b_0$ is non-vanishing, i.e., when there is a bulk magnetization is more delicate.  For example, the magnetization, as far as the 2D system is concerned, is isotropic and thus should give rise to bound currents on \emph{any} interface, not just an edge with low-energy modes. We already showed in Fig. \ref{fig:cur_xy} that, even though the DSM model we have chosen is inherently anisotropic,  there
are bound currents on all of the edges. Let us now prove that this indeed is connected to the bulk orbital magnetization. First, to generate a non-vanishing $b_0$ in the DSM model we can add a kinetic energy term $\epsilon(k)=\gamma\sin k_y\mathbb{I}$ to the Hamiltonian $H_{2DWTI}(k)$ in Eq. \ref{eq:Ham2DWTI}. If the Dirac nodes are separated in the $k_y$ direction and located at $\vec{k}=(0,\pm k_{yc})$, as for our earlier parameter choice, then this simple kinetic term will generate an energy difference of $2\gamma\sin k_{yc}\equiv 2\hbar b_0$ between the Dirac nodes. Note that this term breaks both $T$ and ${\cal{I}}$ but preserves the composite symmetry $T{\cal{I}}$ which is required for the local stability of the Dirac nodes. 

 We can calculate the magnetization for this model according to the results of Refs. \onlinecite{niuvalley,vanderbiltmag} using
\begin{eqnarray}
M&=&\frac{e}{2\hbar}\int \frac{d^2k}{(2\pi)^2}{\rm{Im}}\left[\langle\partial_x u_{-}\vert( H(k)+E_{-}(k))\vert\partial_y u_{-}\rangle\right.\nonumber\\
&-&\left.\langle\partial_y u_{-}\vert (H(k)+E_{-}(k))\vert\partial_x u_{-}\rangle\right]
\end{eqnarray}\noindent where $E_{-}(k), \vert u_{-}\rangle$ are the energy and Bloch functions of the lower occupied band, $H(k)=\epsilon(k)+H_{2DWTI}(k),$ and the derivatives are with respect to momentum. To properly calculate this quantity we need to turn on a small but finite $m_A$ and then set it to zero at the end of the calculation. From symmetry, and from the fact that the extra kinetic term is proportional to the identity matrix,  the only terms that contribute to the non-vanishing magnetization are those proportional to $\epsilon(k)$, and we find the simplification
\begin{eqnarray}
M=\frac{e}{2\hbar}\int \frac{d^2 k}{(2\pi)^2}2\epsilon(k)F_{xy}(k)
\end{eqnarray}\noindent where $F_{xy}(k)$ is the Berry curvature. For small $m_A$ we know that $F_{xy}$ is sharply peaked at each of the two Dirac nodes, since when $m_A=0$ then $T{\cal{I}}$ is preserved and the Berry curvature is a $\delta$-function source at each node. When $m_A\neq 0$ the contributions of the two Dirac points have \emph{opposite} signs. Thus,  we can see that if $\epsilon(k)$ had the same value for both Dirac nodes then $M$ would vanish. In the semi-metallic limit $m_A\to 0$, which is the limit of physical interest, the magnetization becomes
\begin{equation}
M=({\rm{sgn}}\,m_A)\frac{e\Phi_{Dirac}}{4\pi^2\hbar}\sum_{a=1}^{N_{Dirac}}\epsilon(\vec{K}_{a})\chi_{a}
\end{equation}\noindent where $\vec{K}_a$ is the location of the $a$-th Dirac point, $\epsilon(\vec{K}_a)$ is the energy of the $a$-th Dirac point, $\chi_{a}$ is the sign of the Berry phase around the Fermi-surface of each Dirac point for an infinitesimally positive chemical potential, and $\Phi_{Dirac}$ is the constant Berry curvature flux carried by each Dirac point in the gapless limit, i.e., $\Phi_{Dirac}=\pi.$ In terms of $b_0$ for our single pair of Dirac points we find $M=({\rm{sgn}}\,m_A)\frac{e}{2\pi}b_0$ as expected. 

 Now that we have explicitly determined the relationship between bulk magnetization and $b_0,$ let us try to connect the response to the edge state properties. Consider the DSM model with $m_A>0$ on a cylinder with periodic boundary conditions in the trivial direction ($y$-direction) and open boundary conditions in the topological direction ($x$-direction) so that the system will exhibit gapless boundary modes. Let us add in the term $\epsilon(k)=\gamma\sin k_y\mathbb{I}$ to generate a non-vanishing $b_0.$ The sample thus has $b_{0}=b_{0}(\Theta(x)-\Theta(x-L_x))$ where we have chosen the cylinder to lie between $x=0$ and $x=L_x.$ The current density near the left-edge ($x=0$) is given from the response action by 
\be
j^{y}_{L}=-\frac{e}{2\pi}b_{0}\delta(x).
\ee The total current traveling within a region near $x=0$ is simply $J^y_{L}=\int_{-\delta}^{\delta}dx j^y_{L}=-\tfrac{eb_0}{2\pi}.$ Of course, the \emph{total} current in the $y$-direction will vanish once we take both edges into consideration. 

Now we can use this result to compare to the current carried by the edge states. In Fig. \ref{fig:DSMmAmB}a we show the energy spectrum  for the DSM in a cylinder geometry for a non-zero $\gamma$ and a non-zero $m_A>0.$ We see that the edge states are attached to the  Dirac nodes (slightly gapped by $m_A$) and their dispersion is essentially $\epsilon_{edge}(k_y)=-\gamma\sin k_y$ (for a derivation see Appendix \ref{app:edgeDSM}). When $m_A$ is identically zero, then at half-filling each edge branch will be occupied up to $E=0$ (which happens at $k_y=\pi$ for our model), and the boundary currents vanish. When $m_A\neq 0$ then the remaining states on the left edge become occupied which generates a current; the other edge will now have an excess of unoccupied (hole) states which produce a current in the opposite direction. 

Explicitly, the current on the left edge when all of the boundary modes are occupied is 
\begin{eqnarray}
J^y_{L}&=&\frac{e}{2\pi\hbar}\int_{k_{y0}}^{k_{yc}} dk_y \frac{\partial \epsilon_{edge}(k_y)}{\partial k_y}\nonumber\\
&=&-\frac{e\gamma}{2\pi\hbar}[\sin k_{yc}-\sin k_{y0}]\nonumber\\&=&-\frac{e}{2\pi}[\tfrac{\gamma}{\hbar}(\sin k_{yc}-\sin k_{y0})]=-\frac{eb_0}{2\pi}
\end{eqnarray}\noindent where $k_{y0}$ is the energy up to which the edge state is occupied when $m_A=0,$ and $k_{yc}$ is the point up to which the additional occupied states are filled when the entire edge branch is occupied. Thus we see, that on sides of the system that have edge states, the current is completely accounted for by the boundary modes. 
As discussed above, the non-vanishing bulk magnetization also implies there should be bound currents on edges that do not have low-energy boundary modes. Current conservation also indicates that on finite-sized systems, where all boundaries are open, the edge currents from a gapless edge must flow somewhere after hitting a corner. Indeed this is confirmed in Figs. \ref{fig:cur_xy},\ref{fig:pcolor}. Though we do not have a simple argument to derive the magnitude of the edge current on non-topological edges, we found numerically that the magnitudes of the currents localized on each edge are the same.

\subsubsection{Four node case}

\begin{figure}[t!]
\label{fig:dsm_4node}
\centering
   \epsfxsize=9cm \epsfbox{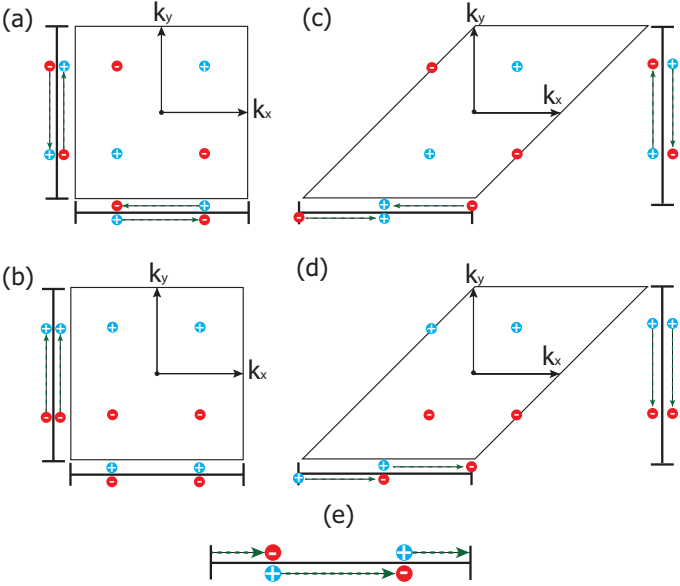}
   \caption{Subfigures (a) and (c) are the Dirac nodes in the model Eq. \ref{eq:4node1}. (a) shows the BZ and projected nodes when ${\bf{G}}_N=2\pi\hat{x}$ or $2\pi\hat{y}$ and (c) shows the BZ and projected nodes when ${\bf{G}}_N=2\pi(\hat{x}+\hat{y})$ and ${\bf{G}}_F=2\pi\hat{x}$ and vice-versa. Subfigures (c) and (d) show the same for the model in Eq. \ref{eq:4node2}. The dotted arrows indicate one possible set of edge state branches for the given nodes. We have drawn the arrows connecting positive $\chi$ nodes to negative $\chi$ nodes although, since the edge states are only $Z_2$ stable, there is not a real distinction between edge states with opposite arrow orientations.  Since all of the edge states overlap in pairs they effectively cancel to give a trivial response. (e) An alternative way to connect the edge states in subfigure (a) so that they connect across the BZ boundary. For the Dirac semi-metal these two alternatives are equivalent though they differ by a topological contribution to the polarization coming from an occupied band. }
\end{figure}

We will now generalize this discussion to the case with four Dirac nodes before giving the fully general results. 
Let us consider the following model
\begin{equation}
H^{(4)}_{1}=\cos k_{x}\sigma^{x}+\cos k_{y}\sigma^{z}.\label{eq:4node1}
\end{equation}
It has Dirac nodes whenever we have $\cos k_{x}=\cos k_{y}=0$, which happens at $K_1=(\tfrac{\pi}{2},\tfrac{\pi}{2}),K_2=(-\tfrac{\pi}{2},\tfrac{\pi}{2}),K_3=(\tfrac{\pi}{2},-\tfrac{\pi}{2}),$ and $K_4=(-\tfrac{\pi}{2},-\tfrac{\pi}{2})$. The nodes $K_1$ and $K_4$ have $\chi_a=+1$ and the other two have $\chi_a=-1.$ We want to understand the polarization response, and thus we want to take the limit where the local mass term at each Dirac node approaches zero with the same sign. Let $g_a\equiv {\textrm{sgn}}\; m_a$ be the sign of the mass for the $a$-th Dirac node. Without loss of generality let $g_a=+1$ for $a= 1, 2, 3, 4.$
For this model we see that  for this choice of mass terms $b_{i}=\tfrac{1}{2}\sum_{a=1}^{4}b_{(a)i}=0,$  where we define ${\bf{b}}_{(a)}=g_a\chi_a{\bf{K}}_a,$ ($a$ is not summed over). 

Since the total $\bf{b}$ vanishes, we expect a vanishing polarization (modulo an integer charge per cell). Indeed, if one diagonalizes this model with open boundary conditions in either the $x$ or $y$ directions, then the observed polarizations are zero as there is no bound edge charge. However,  since we have Dirac nodes separated in momentum space, there is still a possibility for edge states. But we have to remember that the edge states of a 2D DSM are of the $\mathbb{Z}_{2}$ type, and two of them which overlap in the edge Brillouin zone at the same edge momentum can generically gap each other out.  In Fig. \ref{fig:dsm_4node}a,c we show two different examples of 2D Brillouin zones for this model and examples of  possible edge state projections into edge Brillouin zones. We have chosen to draw the edge state projections as oriented lines connecting nodes with $\chi=+1$ to nodes with $\chi=-1$ although, since the edge states are only $Z_2$ stable, this orientation does not hold a physical meaning. In simple lattice models, however, it is possible that even if pairs of edge states overlap in the edge BZ they might still appear gapless in the spectrum, but they should not be stable to generic perturbations. 

We will also consider a model with the following Hamiltonian
\be
\label{eq:4node2}
H^{(4)}_{2}=(\cos k_{x}-\cos k_{y})\sigma^{x}+(\cos k_{x}+\cos k_{y})\sigma^{y}.
\ee
This model also has nodes at $K_1=(\tfrac{\pi}{2},\tfrac{\pi}{2}),K_2=(-\tfrac{\pi}{2},\tfrac{\pi}{2}),K_3=(\tfrac{\pi}{2},-\tfrac{\pi}{2}),$ and $K_4=(-\tfrac{\pi}{2},-\tfrac{\pi}{2})$, however for this model $\chi_{1}=\chi_{2}=+1$ and $\chi_{3}=\chi_{4}=-1$. Here we also choose to study the response  when  $g_a=+1$ for $a= 1, 2, 3, 4.$ This model has $b_{y}=\tfrac{1}{2}\sum_{a=1}^{4}b_{(a)y}=\pi=0\,\mbox{mod}\,\pi,$ and  $b_{x}=0.$ We see that $b_y$ is equal to a half-reciprocal lattice vector. This implies that even though the sum is non-zero, the contribution to the polarization is still trivial, i.e., equivalent to a set of filled bands. In Fig. \ref{fig:dsm_4node}b,d we show two different examples of 2D Brillouin zones for this model and different possibilities for the respective edge state projections into edge Brillouin zones.

Let us try to understand  why the polarization in these two cases vanishes in more detail. Because of the $T{\cal{I}}$ symmetry it makes sense to talk about a well-defined polarization in any direction that is commensurate with a set of lattice lines, thus it is helpful to  consider these systems from a layering perspective. Since the physical consequence of a charge polarization is to generate a boundary charge we can begin by specifying a boundary via a boundary normal vector ${\bf{G}}_N=h{\bf{G}}_1+k{\bf{G}}_2$ as mentioned earlier. This is a reciprocal lattice vector which is not a multiple of a shorter reciprocal vector ($h, k \in \mathbb{Z}$  are relatively prime), where ${\bf{G}}_{1,2}$ are a set of basis vectors of the reciprocal lattice. If we want to know the polarization along  ${\bf{G}}_N$ (which is equivalent to the boundary charge on the edge normal to ${\bf{G}}_N$) we can study the family of Hamiltonians oriented parallel to ${\bf{G}}_N$ and parameterized by a momentum coordinate along the dual vector ${\bf{G}}_F.$ The reciprocal lattice vector ${\bf{G}}_F$ defines the corresponding set (foliation) of lattice lines in space which have ${\bf{G}}_F$ as their ``normal" vector.  Generically the choice of such a ${\bf{G}}_F$ also determines a family of Bloch Hamiltonians parameterized by the momentum in the ${\bf{G}}_F$ direction. We note that this discussion does not depend on whether or not we actually constructed an anisotropic model from coupled wires.   In the case where we have constructed a model from couples wires  there will be a special choice for ${\bf{G}}_F$ which corresponds to a real stacking direction along which the description might simplify.

Let us consider an example: take ${\bf{G}}_N=2\pi\hat{x}$ and  $\mathbf{G}_{F}=2\pi\hat{y}$ so that the set of spatial wires are aligned parallel to the $\hat{x}$ direction. This is represented in Fig. \ref{fig:dsm_4node}a,b for the two different models respectively. The Brillouin zone is a square and we have the corresponding Hamiltonian family $H_{k_y}(k_x).$ Given this choice of 1D sub-manifolds of the BZ, and our models above, we can now consider the charge polarization $P_{1}^{x}(k_y)$, i.e., the polarization in the direction dual to ${\bf{G}}_F$ and parameterized by $k$ parallel to ${\bf{G}}_F.$ We find that, in both models, $P_{1}^x$ is the same for each value of $k_y$ because for a locally stable DSM the polarization can only change when passing through a Dirac point, and for this choice of ${\bf{G}}_F$ it always passes through two Dirac points simultaneously in $k_y.$ Since the polarization is effectively a $Z_2$ quantity in this system, changing it twice (either oppositely or the same way) is equivalent to leaving it unmodified. Thus, the fractional piece of $P_{1}^x$ is trivial. The family of Hamiltonians for the model $H_{1}^{(4)}$ all have vanishing polarization, while that for $H_{2}^{(4)}$ all have a polarization of $e/2$ and thus the total polarization is that of a WTI. This difference is characterized by the vanishing and non-vanishing $b_y$ components respectively. If we switch the vectors so that ${\bf{G}}_N=2\pi\hat{y}$ and ${\bf{G}}_F=2\pi\hat{x}$ then we can still use the same BZ and the the polarizations in both cases vanish identically since $b_x=0$ for both.

From the boundary perspective we can project the energy spectrum onto the edge BZ to look for edge modes that will represent bound charges on the ends of the wires stacked in the ${\bf{G}}_F$ direction. Thus, we project onto the edge Brillouin zone spanned by ${\bf{G}}_F$ by turning on open boundaries in the direction parallel to the wires, i.e. parallel to ${\bf{G}}_N.$  Let us consider the two cases in the preceding paragraph. In Fig. \ref{fig:dsm_4node}a,b we see that this projection causes pairs of Dirac nodes to overlap at the points $k=\pm\pi/2$ in the edge Brillouin zone. Since any boundary states must stretch from negative to positive Dirac nodes then the boundary modes will either completely overlap in the edge BZ and will generically be gapped/absent, or span the entire Brillouin zone in which case they will contribute the polarization due to a filled band, not the fractional piece determined by a DSM. Since the polarization does not change as a function of $k_y$, the former result is if the polarizations are all trivial, and the latter case occurs when the polarizations are all non-trivial and the behavior is like that of a WTI. The same results will hold true if we pick ${\bf{G}}_F=2\pi\hat{x}$ which will show that $P_{1}^{y}(k_x)$ is constant as a function of $k_x$ and thus yields a trivial result. The exact details of the boundary modes depend on how the edge is terminated as well as the topological properties of the occupied bands. For example, by changing the WTI invariant of the occupied bands the edge states can change from lying fully within the edge BZ as shown in Fig. \ref{fig:dsm_4node}a to spreading out over the entire edge BZ as shown in Fig. \ref{fig:dsm_4node}b. These differ by the addition of a half reciprocal lattice vector to ${\bf{b}}.$

Let us take one more explicit example with ${\bf{G}}_{N}=2\pi(\hat{x}+\hat{y}).$ To construct the necessary BZ we need to transform from the basis $\mathbf{G}_{1}=2\pi\hat{x},\mathbf{G}_{2}=2\pi\hat{y},$ which defines a square BZ, to the basis $\mathbf{G}_{N}=h\mathbf{G}_{1}+k\mathbf{G}_{2}, \mathbf{G}_{F}=l\mathbf{G}_{1}+m\mathbf{G}_{2}$ where where $h,k,l,m\in \mathbb{Z}$ and we need to find $\mathbf{G}_{F}$. In general, if we are transforming between a basis $\mathbf{G}_{1},\mathbf{G}_{2}$ to a new one given by $\mathbf{G}_{N}$ and $\mathbf{G}_{F},$  we must satisfy the constraint that the area of both BZs defined by the two different bases are the same i.e. $|\mathbf{G}_{1}\times\mathbf{G}_{2}|=|\mathbf{G}_{F}\times\mathbf{G}_{N}|$. This property tells us that $hm-kl=1,$ which gives us the result that the group structure behind these transformations is $SL(2,\mathbb{Z}).$ In this example, we have $h=k=1$ and $l,m$ need to be determined. The only constraint we have is that $hm-kl=1\implies m-l=1$. We can choose $m=q+1,l=q$ for any $q\in\mathbb{Z}$, but we will examine the case of $q=0$ where $\mathbf{G}_{F}=2\pi\hat{x}.$ The discussion for general $q$ can be found in Appendix \ref{app:BZ} though none of the conclusions change. We could also consider the opposite case when  $\mathbf{G}_{N}=2\pi\hat{x}$ and $\mathbf{G}_{F}=2\pi(\hat{x}+\hat{y})$ and we show possible edge state projections for both of these cases in Fig. \ref{fig:dsm_4node}c,d.  In this new basis, the Dirac nodes at $\pm(\tfrac{\pi}{2},\tfrac{\pi}{2})$ lie on the BZ boundary and the nodes at $\pm(\tfrac{\pi}{2},-\tfrac{\pi}{2})$ lie in the interior of the BZ. We can see that the projection of the Dirac nodes at $(\pm\tfrac{\pi}{2},\tfrac{\pi}{2}),(\pm \tfrac{\pi}{2},-\tfrac{\pi}{2})$ onto the boundary BZ along $\mathbf{G}_{F}$ coincide, and depending on the Hamiltonian $H_{1}$ or $H_{2}$, we will get a Polarization which is identically zero, or we get a case which looks similar to a weak TI respectively. In all of these cases we see that the polarization is simply related to the projection of ${\bf{b}}$ onto the different boundary directions as expected.

The discussion in this subsection has focused on the polarization, the magnetization, on the other hand, is an isotropic quantity and would be non-zero regardless of which edge the system has as long as $b_{0}\neq0$. It is also something which is not dependent on the choice of ${\bf{G}}_F, {\bf{G}}_N$ that connects to a 1D description. Now that we understand this more complicated DSM we can proceed to the general structure. 

\subsection{General formulation of response for 2D DSM}
Let us consider a generic $T{\cal{I}}$-invariant DSM which harbors an even number of Dirac cones. Each Dirac cone $D_a$ ($a=1,2\ldots 2N$) in the semi-metal is specified by the data $(\chi_a,\hbar\bar{{\bf{K}}}_a,\epsilon_a,g_a)$ which are the helicity,  momentum-space location of the Dirac node, energy of the node, and the sign of an infinitesimal local mass term at the Dirac point respectively. The helicity indicates whether the winding of the (psuedo)-spin around a Fermi-surface at a Fermi-energy above the node gives rise to a Berry phase of $\pm \pi$ (i.e. $\chi_a=\pm 1$). All of the response coefficients in which we are interested arise from anomalous terms which, even for gapless Dirac nodes, depend on how the gapless point was approached from a gapped phase; this is why we must include the $g_i.$  As an example, the first Chern number is determined by contributions from each $D_a$ and can be written
\begin{equation}
C_1=\frac{1}{2}\sum_{a=1}^{2N}\chi_{a}g_a.
\end{equation}\noindent There is also the generic constraint $\sum_{a}\chi_a =0$ coming from the $T{\cal{I}}$-symmetry.


Following Ref. \onlinecite{haldane2014}, in the ultra-clean limit we can associate a conserved current $j^{\mu}_{(a)}$ to each Dirac cone and a matching gauge field $A_{(a)\mu}.$ Each Dirac cone contributes a term to the effective response action of the form
\begin{equation}
\label{eq:CI}
S^{(a)}_{eff}[A_{(a)}]=\chi_a g_a\frac{e^2}{4h}\int d^3 x \epsilon^{\mu\nu\rho}A_{(a)\mu}\partial_\nu A_{(a)\rho}.
\end{equation} This gauge field contains two pieces (i) the contribution from the electromagnetic gauge potential and (ii) the energy-momentum shift of each Dirac node. Thus, we have $A_{(a)\mu}=A_{\mu}+\tfrac{\hbar}{e}\bar{K}_{(a)\mu}$ where $\bar{K}_{(a)\mu}$ tells us the energy-momentum location of the node such that  $\bar{K}_{(a)0}=\epsilon_a/\hbar.$ With this specified, we can rewrite the action in a more transparent manner:  
\be
S[A]=\frac{e^{2}}{4h}\sum_{a=1}^{2N}\chi_{a}g_{a}\int d^{3}x\,\epsilon^{\mu\nu\rho}(A_{\mu}+\tfrac{\hbar}{e}\bar{K}_{(a)\mu})\partial_{\nu}(A_{\rho}+\tfrac{\hbar}{e}\bar{K}_{(a)\rho}).
\ee
With the definition that $b_{\mu}=\tfrac{1}{2}\sum_{a=1}^{2N}\chi_{a}g_{a}\bar{K}_{(a)\mu}$, we can write the action as a sum of two terms $S_{1}[A,b]$ and $S_{2}[A,b]$ where (while ignoring a boundary term)
\bea
S_{1}[A,b]&=&\frac{Ce^{2}}{2h}\int d^{3}x\,\epsilon^{\mu\nu\rho}A_{\mu}\partial_{\nu}A_{\rho}\label{eq:genAction1}\\\nonumber
&&+\frac{h}{16\pi^{2}}\int d^{3}x\,\epsilon^{\mu\nu\rho}\sum_{a}\chi_{a}g_{a}\bar{K}_{(a)\mu}\partial_{\nu}\bar{K}_{(a)\rho}\nonumber\\
S_{2}[A,b]&=&\frac{e}{2\pi}\int d^{3}x\,\epsilon^{\mu\nu\rho}b_{\mu}\partial_{\nu}A_{\rho}.\label{eq:genAction2}
\eea
This is the more general formulation of the two node formulae we had derived previously, the magnetization and polarization organized as  $eb_{\mu}=2\pi(M,\epsilon_{ij}P_{1}^{j})$ are now given in the general case as
\be
\label{eq:pol_mag}
b_{\mu}=\tfrac{1}{2}\sum_{a=1}^{2N}\chi_{a}g_{a}\bar{K}_{(a)\mu}.
\ee\noindent Eqs. \ref{eq:genAction1}, \ref{eq:genAction2}, and \ref{eq:pol_mag} are the general results.

To check that the quantity $b_{\mu}$ is physically meaningful we need to make sure it is invariant under a shift of the origin of the Brillouin zone, and a shift of the reference energy, because as it is written this is not apparent. To illustrate the point, let us take $\bar{K}_{(a)\mu}\rightarrow \bar{K}_{(a)\mu}+\Delta k_{\mu}$. The results that follow have been discussed extensively in Refs. \onlinecite{vanderbiltmag,vanderbiltchern}, and we go through their arguments here for completeness. Let us consider the spatial components of $b_{\mu}$ first, which are related to the polarization $\vec{P}_1$. We can write down the polarization in terms of Bloch wave functions as
\be
\vec{P}_{1[k_{0}]}=\frac{e}{(2\pi)^{2}}\mbox{Im}\int_{[k_{0}]} d^{2}k\,\langle u_{k}|\nabla_{k}|u_{k}\rangle
\ee
where we have included the dependence of the origin of the BZ by $\vec{k}_{0}$. Under a change of the origin from $\vec{k}_{0}\rightarrow \vec{k}_{0}+\Delta\vec{k}$, it can be shown generally that\cite{vanderbiltchern} the polarization changes by 
\be
\label{eq:deltaP}
\vec{P}_{1[k_{0}+\Delta k]}=\vec{P}_{1[k_{0}]}-\frac{eC_1}{2\pi}\hat{z}\times\Delta\vec{k}
\ee\noindent where $C_1$ is the first Chern number. This was discussed in Ref. \onlinecite{vanderbiltchern} which discusses how to define the charge polarization in a Chern insulator. To make sense of this, those authors showed that we need to recall that what is physically meaningful is the \emph{change} in polarization under an adiabatic change of an internal parameter of the system. They show that as long as the same origin in the BZ is used for measuring the initial and final polarization of the system, the results remain consistent. In our case we find that shifting $k_0$ in Eq. \ref{eq:pol_mag} is exactly the same as what was shown in Ref. \onlinecite{vanderbiltchern}. That is, under $\bar{\bf{K}}_{(a)}\rightarrow \bar{\bf{K}}_{(a)}+\Delta {\bf{k}}$, we see that 
\be
\Delta P^{i}_{1}=\frac{e\epsilon^{ij}}{4\pi}\sum_{a=1}^{2N}\chi_{a}g_{a}\Delta k_{j}=\frac{eC_1\epsilon^{ij}\Delta k_{j}}{2\pi}
\ee
which is the same as Eq. \ref{eq:deltaP}. Thus, $\vec{b}$ can change when the origin of the BZ is re-defined, but only if the Chern number is non-vanishing. In this case it is shifted according to the formula derived in Ref. \onlinecite{vanderbiltchern} for the charge polarization in a Chern insulator. 

 Now, we look into what happens with the time component of $b_{\mu}$. Increasing $b_{0}$ at a Dirac node is equivalent to reducing the chemical potential at the node or shifting the reference of zero-energy. So, we can interpret shifting $b_{0}$  as a global change to the chemical potential for the overall system. The magnetization for a Bloch system is defined to be 
\begin{eqnarray}
&&M=\frac{e\epsilon^{ij}}{2\hbar}\int\frac{d^{2}k}{(2\pi)^{2}}\nonumber\\&&\times\mbox{Im}\sum_{n}\int_{\epsilon_{nk}\leq\mu}\langle\del_{k_i}u_{nk}|H_{k}+\epsilon_{nk}-2\mu|\del_{k_j}u_{nk}\rangle.
\end{eqnarray}
Following Ref. \onlinecite{vanderbiltmag}, we see from this relation that 
\bea
\frac{dM}{d\mu}&=&-\frac{eC_1}{h}\nonumber\\
\implies \Delta M&=&-\frac{eC_1\Delta\mu}{h}.
\eea
This is exactly what we get from our definition of $b_{\mu}$. Under $\bar{K}_{(a)0}\rightarrow \bar{K}_{(a)0}-\tfrac{\Delta\mu}{\hbar}$, we see that 
\bea
\Delta b_{0}&=&-\frac{1}{2\hbar}\sum_{a=1}^{2N}\chi_{a}g_{a}\Delta\mu\nonumber\\
\implies \Delta M&=&-\frac{eC_1\Delta\mu}{h}.
\eea\noindent Thus, we again see that $b_0$ changes under a redefinition of the origin of energy, but only when the Chern number is non-zero. In this case it changes in the exact same way as a non-trivial Chern insulator. 

We have not talked about the extra term in the $S_{1}[A,b]$ which is independent of $A_\mu.$ To understand this better we can reformulate the response theory using a K-matrix formalism familiar from the Abelian FQH states\cite{wen1992classification}. This discussion lies outside the main scope of the text and we defer it to Appendix \ref{app:Kmatrix}.

\subsection{General Comments about the 2D Dirac Semi-Metal Response}
\emph{(i) Symmetries of $b_{\mu}$ in 2D:} Let us discuss the transformation properties of $b_{\mu}$ under time-reversal (T), charge-conjugation (C),  and inversion symmetry (P). Since in 2D we know that $b_0$ is proportional to a magnetization and $b_i$ is proportional to a polarization can easily determine their symmetry properties: 
\begin{eqnarray}
T&\colon& b_0\to -b_0\nonumber\\
C&\colon& b_0\to b_0\nonumber\\
P&\colon& b_0\to -b_0
\end{eqnarray}\noindent and
\begin{eqnarray}
T&\colon& b_i\to b_i\nonumber\\
C&\colon& b_i\to b_i\nonumber\\
P&\colon& b_i\to b_i.
\end{eqnarray} Note that they are both even under $C$ which is due to the fact that our convention for $b_{\mu}$ in 2D still has the charge factored out. The other thing to note is that $M\sim {\rm{sgn}}(m_A) b_0$ and $P_{1}^{i}\sim {\rm{sgn}}(m_A)\epsilon^{ij}b_j$ and $ {\rm{sgn}}(m_A)$ is odd under parity (or inversion). When this is taken into account we find that $M$ and $P_{1}^{i}$ transform appropriately. In fact, the symmetry properties of $b_{\mu}$ in 2D match those in 1D.

\emph{(ii) Comments on the electromagnetic response:} The response actions in this section all essentially depend on derivatives of $b_{\mu}.$ Thus, for a homogeneous system there is no charge or current response. This pattern alternates between spatial dimensions. In 1D, 3D, 5D,\ldots the electromagnetic response will be a bulk phenomena that does not depend on derivatives of $b_{\mu}$ whereas in 2D, 4D, 6D,\ldots the response depends on derivatives of $b_{\mu}$ which are most commonly generated at interfaces and boundaries.

\section{3D Topological Semi-Metals}

There has been a series of recent works that lay out the theory of electromagnetic response in Weyl semi-metals(WSM)\cite{zyuzin2012,tewari2012,vazifeh2013,chen2013,haldane2014,chen2013weyl,vanderbilt2014comment} and build on the seminal ideas of Nielsen and Ninomiya from three decades ago\cite{nielsen1981}. We will compliment these results in two ways. First, we discuss a way to interpret the 3D WSM response in terms of the 1D semi-metal response covered in Section \ref{sec:oneD} which leads to a separate possible mechanism for generating the Chiral Magnetic Effect (CME). Next we discuss the response properties of interfaces between two different WSMs using explicit numerical calculations, and through anomaly cancellation.  

Following this we move on to consider the response of 3D Dirac semi-metals\cite{youngteo2012} which are reported to have been realized in $\mbox{Cd}_{3}\mbox{As}_{2}$ and $\mbox{Na}_{3}\mbox{Bi}$ \cite{liu2013,neupane2013,wang2013,wang2012}. The DSM in three dimensions is closely related to the WSM and is essentially a time reversal and inversion symmetric version of the WSM where we have two copies of  Weyl nodes of opposite chirality at the same point in momentum space, i.e., 3D Dirac nodes. To guarantee local stability of the Dirac nodes one must  require several preserved spatial symmetries and only certain crystalline space groups support stable nodes\cite{youngteo2012,yang2014} though we will not discuss much about this in our work. We will predict a topological electromagnetic response for these materials which is related to the quantum spin Hall insulator.  In particular we discuss the response of the 3D DSM when there is a magnetic film in contact with the sample surface. Magnetization domain walls on the surface can generate a line of zero modes along the domain wall and hence give rise to some transport phenomena in these materials including bound charge and currents.

\subsection{Response for 3D Weyl Semi-Metal}
\label{subsec:threeDWSM}
\label{subsec:two_band}

A simple model for the WSM phase can be formulated with two bands 
\begin{align}
H_{WSM}=&\gamma\sin k_{z}\mathbb{I}+\sin k_{x}\sigma^{x}+\sin k_{y}\sigma^{y}\nonumber \\
&+(2-m-\cos k_{x}-\cos k_{y}-\cos k_{z})\sigma^{z}.
\end{align}
This model has two Weyl nodes at $(k_{x},k_{y},k_{z})=(0,0,\pm \cos^{-1}(-m))$. The identity matrix term generates a difference in energy between the nodes. Around the two nodes, we have linear dispersion $\epsilon_{\pm}\approx \pm v_{F}|\mathbf{k}|,$  and each of the nodes acts as a monopole of Berry curvature. The Berry curvature flux contained in a Fermi-surface surrounding each node can be $\pm 2\pi$ depending on whether the node enclosed is of positive or negative chirality. This property also leads to surface states whose Fermi-surfaces consist of open line-segments traveling between the projection of the nodes onto the surface\cite{wan2011}. As mentioned before, we follow the convention used in Ref. \onlinecite{zyuzin2012} and define $\vec{b}$ as half the momentum separation in the Weyl nodes, and $b_{0}$ to be half the energy difference between them. So, in the two band model we have here, $b_{z}=\cos^{-1}(-m)$ and $b_{0}=(\gamma/\hbar)\sin b_{z}$. 


To calculate the response we can use a continuum description of two Weyl nodes. Following the calculation in Ref. \onlinecite{zyuzin2012}, in the continuum approximation we have the following low-energy four-band Hamiltonian
\be
H=\tau^{z}\vec{\sigma}\cdot\vec{k}+\tau^{z}b_{0}+\vec{\sigma}\cdot\vec{b}.
\ee
When written as a Lagrangian density coupled to an electromagnetic gauge field with an appropriate choice of Dirac matrices, the four vector $b_{\mu}=(b_{0},\vec{b})$ appears as an axial gauge field in the action just as it does in the one dimensional case
\be
S[b,A]=-\int d^{4}x\,\overline{\psi}(i\slashed{\partial}-e\slashed{A}-\slashed{b}\gamma^{5})\psi.
\ee
The action has a chiral symmetry and we can use the Fujikawa method to evaluate the chiral anomaly which appears due to the non-invariance of the measure under a finite chiral transformation. This is very similar to the derivation we had for the one dimensional model. This calculation gives us a hint that  breaking Lorentz invariance as we have done in the 1D model is an essential part of the mechanism which ends up producing a non-zero response. The response action was calculated to be\cite{zyuzin2012} 
\be
S_{eff}[A]=-\frac{e^{2}}{2\pi h}\int d^{4}x\, \epsilon^{\mu\nu\rho\sigma}b_{\mu}A_{\nu}\del_{\rho}A_{\sigma}.
\ee
This looks like an interpolation between the WTI phase generated from a stack of 2D Chern insulators and the normal insulator phase, as was discussed in Section \ref{sec:prelim}. The current and charge density, assuming $b_{\mu}$ is homogeneous in space-time, are given by 
\bea
\label{eq:WSM_response}
\rho&=&\frac{e^{2}}{\pi h}\vec{b}\cdot\vec{B}\\
\vec{j}&=&\frac{e^{2}}{\pi h}\left(\vec{b}\times\vec{E}-b_{0}\vec{B}\right).
\eea
The term in the current involving the electric field is the anomalous QHE of the WSM. The other terms depend on the magnetic field $\vec{B}$ and can be easily interpreted using an analogy to the 1D semi-metal. 

\subsubsection{Understanding the Weyl Semi-metal Response Using a Quasi-1D Description}
To make the mapping to the 1D system we need to apply a uniform magnetic field to the 3D WSM.  Consider the two band model with $b_{z}\neq 0$. Let us assume that we have a magnetic field turned on in in the $z$-direction so that we have $F_{xy}=-B_{z}$. It is well-known, and we reproduce the calculation below, that a Weyl node in a uniform magnetic field has a low-energy zeroth Landau level with dispersion $E_{0}=\chi k_{z}-b_{z}$ near the Weyl node with chirality $\chi.$  It is this level that is responsible for the low-energy electromagnetic response in Eq. \ref{eq:WSM_response}. We see that the zeroth Landau level only disperses along the magnetic field direction and passes through the Weyl node with the direction of the Fermi velocity given by the chirality of the node. Thus, the application of the uniform magnetic field generates a quasi-1D mode at low-energy. For a pair of Weyl nodes, as would be found in the simplest WSM, there are two low-energy branches, which, together,   effectively form a 1D semi-metal. This is almost identical to the previous 1D semi-metal description except that each state has a degeneracy which is set by the total flux of the magnetic field through the x-y plane. We denote this degeneracy by $N_{\Phi}=\frac{B_{z}L_{x}L_{y}}{\Phi_{0}}$ where $\Phi_{0}=\frac{h}{e}$ is the fundamental flux quantum. Thus, in a uniform magnetic field, the low-energy physics of the WSM is equivalent multiple copies of the 1D semi-metal. As will be seen below, the description is even more apt because, in a lattice regularized model, the zeroth Landau level modes arising from each Weyl node connect at high energy and form multiple copies of the usual 1D tight-binding bandstructure. 

Let us try to reproduce the charge density we get in Eq. \ref{eq:WSM_response} by using the 1D model. There is a subtlety as to how the states are filled. Of course, if the zeroth Landau level is completely filled or completely empty, then there will be no interesting response. In this case there will be a background charge density of some integer charge per unit cell, but no current will flow in the filled band, and thus there will be no static chiral magnetic effect. This was discussed in detail in Ref. \onlinecite{vazifeh2013}. While a filled band can give rise to Lorentz violation because of the inherent lattice structure, the field theory calculations for the semi-metal are not sensitive to this. In fact, they can only predict the response from a partially filled band which provides an explicit fractional amount of Lorentz violation. This is similar to the idea of Ref. \onlinecite{haldane2005} in which the low-energy structure only determines the fractional part of the response. To match the field-theory calculation we need to assume that the zeroth Landau level is only filled to a chemical potential $\mu=0$ which implies the band is partially filled. For example, to calculate the density response we need to count the number of states filled in the zeroth Landau level which is simply
\begin{align}
Q&=N_{\Phi}eL_{z}\int^{b_{z}}_{-b_{z}} \frac{dk_{z}}{2\pi}\\
\implies \rho&=\frac{e^{2}b_{z}B_{z}}{\pi h}
\end{align}\noindent which matches Eq. \ref{eq:WSM_response}.
Before we attempt to understand the properties which lead to a nonzero current, let us look at the zeroth Landau level structure of the WSM in more detail to see how $b_{0}$ fits into the discussion.

\subsubsection{Zeroth Landau Level Structure in a Weyl Semi-metal}
In this Section, we proceed to show that $b_{0}$ can be thought of in a similar way as what we discussed in Sec. \ref{sec:oneD} for the 1D model. In the usual case a $b_{0}$  is produced by shifting the Weyl nodes with respect to each other in energy. We will show that when this is the case the zeroth Landau level is shifted in momentum parallel to the magnetic field. So, shifting the nodes in energy acts like an electric field ($k$ is shifted) on the zeroth Landau level. As in 1D we can also generate a $b_{0}$ by adding an intrinsic term which generates a velocity difference in the dispersion at the two Weyl nodes; we will discuss this case as well. We show some continuum calculations to justify these statements and then reproduce the same by a simple numerical lattice calculation.

Consider a four band continuum model for the Weyl semimetal (a single pair of nodes) with just $b_{z}\neq 0$\cite{zyuzin2012}. We find the eigenvalues for this model, and then add in a $b_{0}$ which is trivial because the term that generates $b_{0}$ commutes with the remaining Hamiltonian. The Hamiltonian is given by
\be
\label{eq:ham1}
H=\tau^{z}\otimes\sigma^{x}k_{x}+\tau^{z}\otimes\sigma^{y}k_{y}+\tau^{z}\otimes\sigma^{z}k_{z}+b_{z}\mathbb{I}\otimes\sigma^{z}.
\ee
We need to include a magnetic field with $k_{i}\rightarrow k_{i}-eA_{i}$ and $A_{y}=B_z x$, where $B_z$ is the uniform magnetic field in the $z$-direction. We have to note that we have broken translation invariance in the $x$ direction and the eigenvalue equation will be a differential equation in $x$ where we have to replace $k_{x}\rightarrow -i\partial_{x}$. From now on, this is implicitly assumed. The time independent Schrodinger equation reads
\be
\label{eq:ho1}
H\psi=E\psi.
\ee
Following the usual strategy, we can apply $H$ to $\psi$ again to produce $H^{2}\psi=E^{2}\psi.$ We can evaluate the left hand side to find 
\begin{align}
\label{eq:ho}
H^{2}\psi=&\lbrack k_{x}^{2}+eB_z\mathbb{I}\otimes\sigma^{z}+(eB_z)^{2}(x+k_{y}/eB_z)^{2}\nonumber\\
&+k_{z}^{2}+2b_{z}k_{z} \tau^{z}\otimes\mathbb{I}+b_{z}^{2}\rbrack\psi.
\end{align}
The wave function $\psi$ can be taken to be an eigenstate of $\sigma^{z}$ for the spin sector, and $\tau^{z}$ for the orbital sector. Let us denote the eigenvalue of $\sigma^{z}$ as $\zeta=\pm 1$ and the eigenvalue of $\tau^{z}$ as $\chi=\pm 1$. Eq. \ref{eq:ho} is just the harmonic oscillator eigen-equation and has the following energies:
\be
E_{n}(\zeta,\chi,k_{z})=\pm\left\lbrack 2eB_z(n+\frac{1}{2})+(k_{z}+\chi b_{z})^{2}+eB_z\zeta\right\rbrack^{1/2}
\ee
with the corresponding wave functions given by 
\be
\Phi_{n}(\zeta,\chi,\vec{x})=N_{n}\zeta\mbox{e}^{-ik_{y}y-i(k_{z}+\chi b_{z})z}F_{n}(x+k_{y}/eB_z)\times\eta
\ee
where $N_{n}$ is a normalization constant, $F_{n}(x)$ are the Hermite polynomial wave functions, and $\eta=\Lambda(\sigma^{z})\otimes\Lambda(\tau^{z})$ is a four-component spinor where $\Lambda(\pm1)$ mean the eigenvectors of $\sigma^{z},\tau^{z}$ given by $\left(\begin{smallmatrix}1\\0\end{smallmatrix}\right),\left(\begin{smallmatrix}0\\1\end{smallmatrix}\right)$ . 

\begin{figure}[h!]
\label{fig:ll_b0}
\centering
   \epsfxsize=8cm \epsfbox{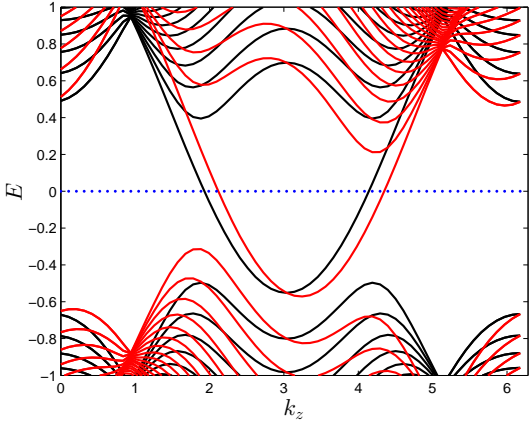}
   \caption{The zeroth Landau level of the Weyl semi-metal in a uniform magnetic field is plotted vs $k_{z}$ before(in black) and after(in red) switching on a $\gamma$ which gives us $b_{0}=(\gamma/\hbar)\sin2\pi/3=0.17$. The blue line is shown to indicate $E=0$. The model parameters have $b_{z}=2\pi/3, m=1/2,$ and $L_{x}=L_{y}=L_{z}=60$ with the magnetic flux per unit cell given by $\phi=2\pi/60$. $b_{0}$ was then switched on to plot the curve in red. We see that the Landau level is simply shifted in momentum space and is akin to turning on an external electric field in the 1D model.}
\end{figure}

To be precise we need to verify that all of these solutions satisfy Eq. \ref{eq:ho1}. This consistency check eliminates half of the zero mode solutions and we end up with the result that the zeroth Landau levels have energy
\be
E_{0}=\chi k_{z}-b_{z}
\ee\noindent which depends on the chirality $\chi$ of the Weyl node.
This dispersion hits zero energy at $k_{z}=\pm b_{z}$, i.e., the location of the Weyl nodes, as expected. These modes also have a degeneracy of $N_{\Phi}$ for each value of $k_z$ as noted above. In a lattice regularization the zeroth Landau levels of the two Weyl nodes will be connected to each other at high-energy (c.f. the energy spectrum in Fig. \ref{fig:ll_b0}). 

Now to turn on a $b_{0}$ we add the extra term $\delta H=b_{0}\tau^{z}\otimes\mathbb{I}$ which commutes with the initial Hamiltonian. We note that acting on the zeroth Landau level wavefunctions the energy is shifted by  $b_{0}\chi$, thus leading to the dispersions
\be
E_{0}=\chi(k_{z}+b_{0})-b_{z}.
\ee
This is just a shifted version of the original zeroth Landau level dispersions,  and they cross zero energy when $k_{z}=-b_{0}\pm b_{z}$. So, the conclusion is that $b_{0}$ shifts the low-energy spectrum of the zeroth Landau level to the right in momentum space, which is the same effect that an external electric field $E_z$ would have. Thus, if the band is partially filled, i.e., when we have explicit Lorentz violation due to the charge density, this will lead to a non-vanishing current in the absence of an applied electric field, which is essentially the chiral magnetic effect.

Further pushing the 1D description,  let us also show that modifying the relative velocities of the two Weyl points will lead to a similar effect. Consider the Hamiltonian given by 
\be
H=\tau^{z}\otimes\sigma^{x}k_{x}+\tau^{z}\otimes\sigma^{y}k_{y}+\tau^{z}\otimes\sigma^{z}k_{z}+\mathbb{I}\otimes\sigma^{z}\alpha k_{z}+\mathbb{I}\otimes\sigma^{z}b_{z}
\ee
where $\alpha\ll 1$. This $\alpha$-dependent term modifies the velocities of propagation in the $z$-direction of the two Weyl nodes. It effectively changes $b_{z}\rightarrow b_{z}+\alpha k_{z}$ from our previous analysis. The entire argument for the energies of the zeroth Landau levels from before carries through here too and we find a modified zeroth Landau level dispersion of 
\be
E_{0}=\chi k_{z}-b_{z}-\alpha k_{z}.
\ee
This dispersion crosses zero at $k_{z}=b_{z}/(\chi-\alpha)\approx \chi b_{z}-\alpha b_{z}+\mathcal{O}(\alpha^{2})$. So, near zero energy this  term behaves like a momentum shift in the Landau level, and this should give us a non-zero current as we have shown in the 1D model in Section \ref{sec:oneD}.

\begin{figure}[h!]
\label{fig:ll_b0_vc}
\centering
   \epsfxsize=8cm \epsfbox{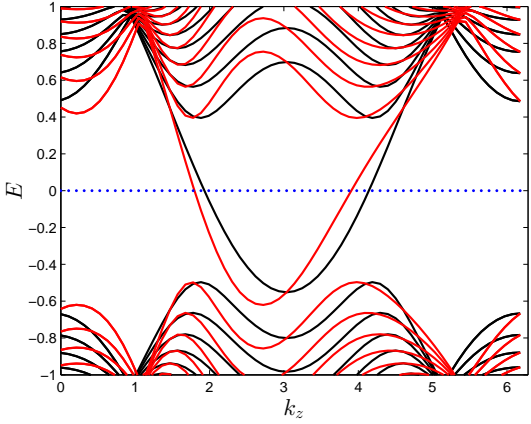}
   \caption{The zeroth Landau level is plotted vs $k_{z}$ before(in black) and after(in red) switching on a $b_{0}$ using the NNN velocity term. The blue line is shown to indicate $E=0$. The model had $b_{z}=2\pi/3,m=1/2,$ and $L_{x}=L_{y}=L_{z}=60$ with $\phi=2\pi/60$. We then switch on a term to change the velocity of the two Weyl nodes with $t_{NNN}=0.2$. The shift we expect is then given by $2t_{NNN}m\approx0.2$ as seen in the figure. In effect, near $E=0$ the zeroth Landau level is shifted.}
\end{figure}

To verify these results, we can perform calculations using a simple lattice regularization of the above continuum model.  The Hamiltonian is given by 
\begin{align}
\label{eq:ll_sim}
H&=\gamma\sin k_{z}\mathbb{I}+\sin k_{x}\sigma^{x}+\sin k_{y}\sigma^{y}\nonumber\\
&+(2-m-\cos k_{x}-\cos k_{y}-\cos k_{z}-t_{NNN}\sin 2k_{z})\sigma^{z}
\end{align}
where the term proportional to $\gamma$ will cause a shift in energy of the Weyl nodes, and the next nearest neighbor term proportional to $t_{NNN}$ causes a change in the velocity of the zeroth Landau level near the two Weyl nodes. 

For $\gamma=0$ and $t_{NNN}\neq 0$ the Weyl nodes are given by solving
\be
\cos k_{z}+t_{NNN}\sin 2k_{z}=m,
\ee
which gives us two solutions for $k_{z}$. Let us try to extract the low-energy Hamiltonians near the nodes in the limit that $t_{NNN}\ll 1$ by writing the two solutions as $k_{z}=\pm \kappa_{z}+\delta k$. 
\be
\cos(\pm \kappa_{z}+\delta k)+t_{NNN}\sin(\pm2\kappa_{z}+2\delta k)=m.
\ee
We can subtract the two equations to find 
\be
2\sin \kappa_{z}\sin\delta k-2t_{NNN}\cos 2\delta k\sin 2\kappa_{z}=0.
\ee
Using the small angle approximations $\sin\delta k\approx\delta k, \cos 2\delta k\approx 1$, we are left with 
\be
\delta k=2t_{NNN}\cos\kappa_{z}\approx 2t_{NNN}m
\ee
Thus we see that a non-zero velocity change will lead to a momentum shift of $2t_{NNN}m$ at the nodal energies. Comparing with the  continuum calculation we see that $\alpha b_z=-2t_{NNN}m.$

We show the numerical results of $\gamma=0.2,$ $t_{NNN}=0$ in Fig. \ref{fig:ll_b0}, and $\gamma=0,$ $t_{NNN}=0.2$ in Fig. \ref{fig:ll_b0_vc}. In both cases we see that near $E=0$ the zeroth Landau levels are shifted.

\subsubsection{Response and Anomaly Cancellation in Weyl Semimetals with Inhomogeneous $b_{\mu}$}
\label{sec:weyl_anomaly}
So far, all of the response properties that we have considered for the WSM have assumed $b_{\mu}$ was constant in space-time. This will not be the case in systems which have boundaries or interfaces across which $b_{\mu}$ will naturally change. 
In this section, we closely examine what the bulk action implies for the surface/interface action, and how the whole system remains gauge invariant. We recall that the response action is
\be
\label{eq:action_mod}
S=-\frac{e^{2}}{2\pi h}\int \mbox{d}^{4}x\,\epsilon^{\mu\nu\rho\sigma}b_{\mu}A_{\nu}\del_{\rho}A_{\sigma}.
\ee
Now when we take the functional derivative of $S$ with respect to $A_{\alpha}$ to extract the current we have to be careful about the behavior of $b_{\mu}$
\bea
\label{eq:current_wsm}
 j^{\alpha}=\frac{e^{2}}{\pi h}\epsilon^{\alpha\mu\rho\sigma}b_{\mu}\del_{\rho}A_{\sigma}+\frac{e^{2}}{2\pi h}\epsilon^{\alpha\mu\rho\sigma}A_{\sigma}\del_{\rho}b_{\mu}.
\eea
This gives us the usual current we expect for the AQHE and CME along with a term which depends on derivatives of $b_{\mu}$ but is not manifestly gauge invariant since it depends directly on $A_{\mu}.$ This signals the presence of an anomaly that will arise whenever $b_{\mu}$ changes. 

The Callan-Harvey mechanism provides a straightforward way of understanding this result\cite{callanharvey}. To be explicit, let us assume we have an interface in the x-direction located at $x=x_0$ where $b_{z}$ jumps from a finite value to zero. This is the case in the lattice models we studied in the previous section.  Under a gauge transformation ($A_{\mu}\to A_{\mu}-\partial_\mu\lambda$) the action transforms as 
\begin{eqnarray}
\delta_\lambda S&=&-\frac{e^{2}}{2\pi h}\int \mbox{d}^{4}x\,\epsilon^{\mu\nu\rho\sigma}b_{\mu}(-\partial_\nu\lambda)\del_{\rho}A_{\sigma}\nonumber\\&=&-\frac{e^{2}}{2\pi h}\int \mbox{d}^{4}x\,\epsilon^{\mu\nu\rho\sigma}\partial_{\nu}b_{\mu}\del_{\rho}A_{\sigma}\lambda\nonumber\\&=& \frac{e^{2}}{2\pi h}\int \mbox{d}^{4}x\,\epsilon^{zx\rho\sigma}b_z\delta(x-x_0)\del_{\rho}A_{\sigma}\lambda\nonumber\\&=&\frac{e^{2}L_z b_z}{2\pi h}\int \mbox{d}y dt\,\epsilon^{\rho\sigma}\del_{\rho}A_{\sigma}\lambda\neq 0.
\end{eqnarray} 

Thus, in order for the system to be gauge invariant there must be localized fermion modes where $b_z$ jumps. In the case of the simple WSM models we have considered, we know that there are such surface/interface states and they are just straight-line Fermi-arcs that stretch between the Weyl nodes projected onto the surface/interface BZ. For a non-zero $b_z$ on a surface with normal vector $\hat{x}$ (just like the interface considered in the previous paragraph) the surface states have a chiral dispersion given by $E(k_y,k_z)= k_{y}$ at low-energy. These chiral modes give rise to the usual chiral anomaly. 
There is an independent chiral fermion for each value of $k_z,$ but the surface states only exist in between the Weyl nodes, i.e., only for $-b_{z}\leq k_{z}\leq b_{z}.$ Each 1D chiral mode generates an anomalous contribution to the variation of the boundary/interface action under a gauge transformation\cite{hughesonkarleigh,callanharvey}
\begin{equation}
\delta_{\lambda}S_{bdry}=-\frac{e^2}{2h}\int dy dt \epsilon^{\rho\sigma}\del_{\rho}A_{\sigma}\lambda
\end{equation}\noindent where $\rho,\sigma = 0, y.$ To calculate the total variation due to all of the modes we can convert the sum over the independent $k_z$ modes to an integral which generates a factor of $\tfrac{L_z}{2\pi}2b_z.$ We thus find
\begin{equation}
\delta_{\lambda}S^{(Tot)}_{bdry}=-\frac{e^2L_z b_z}{2\pi h}\int dy dt \epsilon^{\rho\sigma}\del_{\rho}A_{\sigma}\lambda\label{eq:consistent}
\end{equation} \noindent which exactly cancels the variation coming from the bulk action. Eq. \ref{eq:consistent} is called the consistent anomaly. The consistent anomaly leads to an anomalous Ward identity for current conservation on the edge
\begin{equation}
\partial_\mu j^{\mu}_{bdry}=-\frac{e^2 L_z b_z}{2\pi h} \epsilon^{\rho\sigma}\del_{\rho}A_{\sigma}=-\frac{e^2N_c}{2h} \epsilon^{\rho\sigma}\del_{\rho}A_{\sigma}
\end{equation}\noindent where $N_c$ is the total number of modes in the interface/boundary Fermi-arc.

Going back to the bulk current response in Eq. \ref{eq:current_wsm} we see that the current naturally splits into two terms $j_{bulk}^{\alpha}= \frac{e^{2}}{\pi h}\epsilon^{\alpha\mu\rho\sigma}b_{\mu}\del_{\rho}A_{\sigma}$ and $\tilde{j}^{\alpha}_{bdry}=\frac{e^{2}}{2\pi h}\epsilon^{\alpha\mu\rho\sigma}A_{\sigma}\del_{\rho}b_{\mu}.$ For our interface configuration we find
\be
\tilde{j}^{\alpha}_{bdry}=-\frac{e^{2}}{2\pi h}\epsilon^{\alpha zx\sigma}A_{\sigma}b_z\delta(x-x_0).
\ee If we integrate this current density over $x$ and $z$ we can combining this current with the current from the consistent anomaly to arrive at the Ward identity for the covariant anomaly (the anomaly that contains all contributions to the boundary current)
\begin{equation}
\partial_{\alpha}(j^{\alpha}_{bdry}+\tilde{j}^{\alpha}_{bdry})=-\frac{e^{2}L_zb_z}{\pi h}\epsilon^{\alpha\sigma}\partial_\alpha A_{\sigma}.
\end{equation}
This covariant anomaly precisely matches the bulk-current inflow from $j_{bulk}^{x}$ into the boundary/interface. Note that although we have assumed a model which has simple Fermi-arcs, the chiral anomaly result is very robust and does not depend on the exact form of the surface state dispersion, or any other details,  only that the states are chiral. Thus we expect it to hold in any generic model, even in the cases when the Fermi-arcs are not straight line segments, but are curved.  This result clearly shows that while the bulk action would predict a gauge-variant  response, it is compensated by the surface Fermi-arcs states. The same is true when we do not have a physical boundary, but a region in which $b_{\mu}$ varies in space-time. When $b_{\mu}$ varies there are two contributions to the boundary current, one arising from the bulk action itself, and the other from the consistent anomalous current required of the boundary states in order to preserve gauge invariance of the bulk and boundary. 


\subsubsection{Numerical Results}

\begin{figure}[t]
\label{fig:db_b_0}
\centering
   \epsfxsize=3.5in \epsfbox{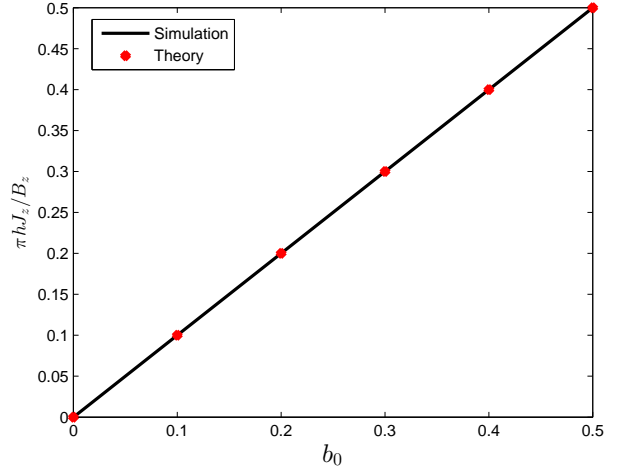}
   \caption{The current is plotted vs $b_{0}$ for the two band model of the Weyl semimetal. The current is linear and the slopes match almost exactly. This plot is generated for $L_{x}=30$ and the flux per plaquette as $\phi=-2\pi/L_{x}$. We use $L_{y}=30$, $L_{z}=30$, and $b_{z}=\tfrac{\pi}{2}$ to generate this plot.}
\end{figure}

We are now prepared to numerically probe two effects (i) the CME which we have tried analyzing using a mapping to the 1D model and (ii) the charge density response in a system with an inhomogeneous $\vec{b}.$ We do this in the context of the two band WSM lattice model
\begin{align}
H&=\gamma\sin k_{z}\mathbb{I}+\sin k_{x}\sigma^{x}+\sin k_{y}\sigma^{y}\nonumber\\
&+(2-m-\cos k_{x}-\cos k_{y}-\cos k_{z})\sigma^{z}
\end{align}
where $\gamma$ generates a nonzero $b_{0}$. It is important to note that to perform our numerical calculations we fill the states up to $E=0$, i.e.  all states with $E\leq 0$ are filled. To illustrate an example of the CME, in Fig. \ref{fig:db_b_0}, we have plotted the current along the $z$ direction as a function of $b_{0}$ in the presence of a uniform magnetic field, but no electric field. The predicted current density from the model, assuming a magnetic field in the $z$ direction,  is given by
\be
j_{z}=-\frac{eb_{0}B_{z}}{\pi h}.
\ee
The lattice calculation is shown in Fig.  \ref{fig:db_b_0} and we find exactly this result. For this calculation the magnetic field is implemented using Peierls substitution.  We use a Landau gauge to retain translation invariance in one of the directions in the $xy$ plane, and the $z$ direction is also translation invariant. The magnetic field is restricted to have rational flux per unit cell for the spectrum to remain periodic in momentum.

\begin{figure}[t]
\label{fig:srho}
\centering
   \epsfxsize=3.5in \epsfbox{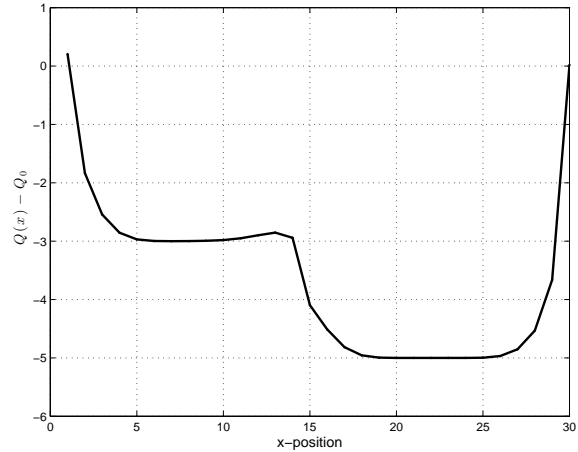}
   \caption{The charge density is plotted vs position in the x-direction with open boundary conditions. The system comprised of a Weyl semimetal with $b_{z,L}=\pi/5$ for $0<x<L_{x}/2$ and $b_{z,R}=\pi/3$ for $L_{x}/2<x<L_{x}$. The total number of sites in the $x$-direction was $L_{x}=30$ with magnetic flux per unit cell in the $x-y$ plane $\phi=-2\pi/30$. Also, $L_{z}=30$ and $L_{y}=30$. The bulk charge density is given by $N_{x}=-L_{z}L_{y}b_{z}B_{z}/4\pi^{2}=-3, -5$ as is predicted by the action.}
\end{figure}

Another simple effect to test is the density response at an interface where $\vec{b}$ changes. With $B_{z}\neq 0$, we should have
\be
\rho=\frac{eb_{z}B_{z}}{\pi h}.
\ee
So, if we vary $b_{z}$ in the $x$-direction (with open boundary conditions the $xz$ surfaces host nontrivial surface states)  the resultant charge is plotted in Fig. \ref{fig:srho}. The bulk charge follows what is predicted by the action.

\subsection{Electromagnetic Response of a 3D Dirac Semi-metal}
\label{subsec:threeDDSM}
There has been a lot of recent work predicting and measuring materials candidates for 3D Dirac semi-metals\cite{youngteo2012,liu2013,neupane2013,wang2013,wang2012}. In this section we discuss an interesting electromagnetic probe of the DSM and connect it to the response properties of the 2D time-reversal invariant quantum spin Hall insulator\cite{kane2005quantum,BHZ2006,kane2005z,konig2007}. Thus, we begin  by first examining the response of the Quantum Spin Hall(QSH) insulator itself. Analogous to all of our previous constructions, we can think of the 3D DSM  as a layered 2D topological insulator, and in this case it is formed from coupled layers of the QSH system. The layer construction has aided the discussion and analysis of the other topological semi-metals and we will see that it is very helpful in this case as well. After reviewing the response of the QSH insulator, we will discuses the analogous properties of the DSM and numerically validate our analytical calculations. 

The QSH system has an unusual electromagnetic response given by\cite{qhz2008a,qi2008} 
\be
S[A]=\frac{e}{2\pi}\int d^{3}x\,\epsilon^{\mu\nu\sigma}A_{\mu}\del_{\nu}\Omega_{\sigma}\label{eq:QSHresponse}
\ee 
where $\Omega_\mu$ is a gauge field which encodes configurations of inhomogeneous adiabatic perturbations. We will clearly define what this means in the following section. Essentially, the configurations of $\Omega_\mu$ are related to possible mass-inducing perturbations of a Dirac-type Hamiltonian. As a consequence of this response term, a magnetic film deposited at the edge of the QSH insulator can generate a localized charge density or adiabatic current if the magnetization is space or time dependent respectively\cite{qhz2008a,qi2008}. The edge of the QSH insulator is itself a robust 1D massless Dirac fermion if we preserve time-reversal symmetry. A magnetization on the edge will open a gap, and through the well-known Jackiw-Rebbi mechanism\cite{jackiw1976}, a spatial domain-wall in the magnetization will trap a low-energy mid-gap mode. This mode signals a bound charge of $Q_b=\pm e/2.$ If the magnetization on one side of the domain wall begins to rotate as a function of time, a quantized adiabatic charge current can flow along the edge through the magnetic junction. Ref. \onlinecite{qi2008} showed that both of these phenomena could be derived from Eq. \ref{eq:QSHresponse}.  This is the electromagnetic signature of the QSH insulator, and is closely tied to the response of the 3D DSM. 

Now we can construct a stack of QSH insulators. If the layers are weakly coupled then we will get the conventional WTI state\cite{FKM2007,MooreBalents07,Roy07}. If we increase the strength of the inter-layer coupling  so that we close the bulk gap we will generate the 3D DSM phase.  Just as with the WSM, the edge states of the QSH layers forming the DSM will survive in a certain region of momentum space and will connect the various 3D Dirac nodes with Fermi-surface arcs.  We can easily extrapolate the response action of the QSH insulator to the 3D DSM to find 
\be
S[A]=\frac{e}{2\pi^{2}}\int d^{4}x\,\epsilon^{\mu\nu\rho\sigma}b_{\mu}A_{\nu}\del_{\rho}\Omega_{\sigma}.
\ee
We will discuss the consequences of this action below, but first we will more carefully recount the analysis for the 2D QSH insulator. 

\subsubsection{Response from the Second Chern number}
\label{subsec:2chern}
The discussion in this Section closely follows the arguments in Ref. \onlinecite{qi2008} although we will only reproduce the necessary ingredients for our discussion of the 3D DSM and leave out some of the details which can be found in the aforementioned reference. In general the response of the QSH insulator is derived from the second Chern number $C_2,$ which is a four dimensional topological invariant. Since the QSH exists in 2D, the Bloch Hamiltonian is only parameterized by two numbers $k_x, k_y,$ which is not enough to generate a non-zero $C_2.$ Thus, to probe the electromagnetic response properties of the QSH state we need to couple the system to two additional parameters $\theta({\bf{x}},t), \phi({\bf{x}},t)$ which represent adiabatic parameters which vary slowly in space and time so that momentum space is still approximately well-defined. The gauge field $\Omega_\mu$ introduced above is a function of space and time, but only through its dependence on $\theta$ and $\phi.$

To be explicit, consider the QSH Hamiltonian given by 
\bea
\label{eq:qshham}
H_{QSH}(\mathbf{k},\hat{n})&=&\sin k_{x}\Gamma^{1}+\sin k_{y}\Gamma^{2}\\ \nonumber
&&+(\cos k_{x}+\cos k_{y}-2)\Gamma^{0}+m\sum_{a=0,3,4}\hat{n}_{a}\Gamma^{a}
\eea
in which $m>0$, $\Gamma^a$ are the $4\times 4$ Dirac matrices,  and $\hat{n}=(n_{3},n_{4},n_{0})$ is a 3D unit vector. The unperturbed QSH insulator will have $n_3=n_4=0$ but $n_0\neq 0.$ If we let $\hat{n}$ vary slowly as a function of space-time we can parameterize it using two adiabatic space-time dependent parameters via 
$\hat{n}({\bf{x}},t)=(\sin\theta({\bf{x}},t)\cos\phi({\bf{x}},t),\sin\theta({\bf{x}},t)\sin\phi({\bf{x}},t),\cos\theta({\bf{x}},t)).$ The results of Ref.~\onlinecite{qi2008} show that in the low-energy continuum limit of $H_{QSH}$ expanded around the $\Gamma$-point, the gauge curvature of $\Omega$ is directly related to the skyrmion density of the unit vector $\hat{n}$ as 
\be
\partial_{\mu}\Omega_{\nu}-\partial_{\nu}\Omega_{\mu}=\frac{1}{2}\hat{n}\cdot\partial_{\mu}\hat{n}\times\partial_{\nu}\hat{n}.
\ee\noindent Using Eq. \ref{eq:QSHresponse} we can write down the current in terms of this skyrmion density as 
\be
j^{\mu}=\frac{e}{8\pi}\epsilon^{\mu\nu\rho}\hat{n}\cdot\partial_{\nu}\hat{n}\times\partial_{\rho}\hat{n}.
\ee

Now let us consider an important example case. Assume that we have a QSH sheet with a static edge parallel to the $y$-direction and a pair of static magnetic films next to each other on the edge. If the magnetizations of the two films are opposite this will produce a domain wall on the edge with a magnetization that varies as a function of $y.$ In that case we find the parameterization $\theta=\theta(x)$ and $\phi=\phi(y).$ At the location of a $\theta$ domain wall between $\theta=0$ and $\theta= \pi$ there will be an edge. At the location of a $\phi$ domain wall between $\phi=0$ and $\phi=\pi$ there will be  a magnetic domain wall. In this geometry we find
\bea
\label{eq:qsh_charge}
j^{0}&=&\frac{e}{4\pi}\hat{n}\cdot\del_{x}\hat{n}\times\del_{y}\hat{n}\\ \nonumber
&=&\frac{e}{4\pi}\sin\theta\times\frac{d\theta}{dx}\frac{d\phi}{dy}.
\eea
Due to the dependence on the derivatives of $\theta$ and $\phi,$ the charge density is localized wherever $\theta(x)$ and $\phi(y)$ are changing. If we have a sharp magnetic domain wall on a sharp edge, then all of the charge density will be localized at the magnetic domain-wall, i.e., where the $\theta$ and $\phi$ domain walls intersect. The total charge in the neighborhood of this intersection can be calculated by integrating over $x,y.$ The integration is easily performed since the integrand is a total derivative in $x$ and $y.$ We just get the integral over the solid angle swept out by $\theta$ and $\phi,$ which for this configuration is half the sphere, i.e. $2\pi.$ This yields  a bound charge $Q_b=2\pi\frac{e}{4\pi}=\frac{e}{2}.$ 

We can similarly find an adiabatic pumping current by having a static edge ($\theta=\theta(x)$) and sweeping the relative magnetization between the two magnetic films on the edge as a function of time ($\phi=\phi(t)$)\cite{qhz2008a}. Everything carries through in exactly same way and we find
\be
j^{y}=\frac{e}{4\pi}\sin\theta\times\frac{d\theta}{dx}\frac{d\phi}{dt}.\label{eq:qsh_current}
\ee
We can again integrate over $x,t$ to get the total charge transported as the relative magnetization angle sweeps through a full cycle to find as $\phi:0\rightarrow 2\pi$  we have $\Delta Q=e.$  This current is localized  wherever $\theta$ has a sharp change in its value, i.e., on the edge. 

\begin{figure}[t!]
\label{fig:dsmSchematic}
\centering
   \epsfxsize=5cm \epsfbox{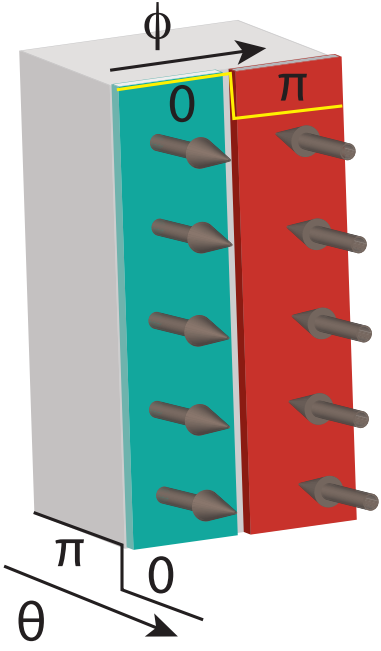}
   \caption{Setup to generate an electromagnetic response in a 3D Dirac semi-metal. To get a non-zero response there must be two adiabatic parameters $\theta$ and $\phi.$ The parameter $\theta$ represents an interpolation between a 3D Dirac semi-metal with $b_z\neq 0$ to a trivial insulator with $b_z=0.$ The parameter $\phi$ represents a magnetization domain-wall on the $xz$ surface plane. There will be a branch of low-energy fermion modes trapped on the domain wall which can bind charge or can carry current if $b_0\neq 0.$}
\end{figure}

We can understand the physics underlying the QSH response from the microscopic behavior of the edge states. In the low energy limit near the Dirac point, we can write down the Hamiltonian for one of the edges of the QSH system (say an edge at $x=0$) as
\be
H_{edge}(k)=k\sigma^{z}
\ee
where $k$ is the momentum of the coordinate along the edge, and we have set the edge velocity to unity. Coupling the magnetic layer to the edge will induce a gap from the proximity exchange (Zeeman) coupling. If the magnetization lies in the plane then the effective Hamiltonian becomes 
\be
H_{edge}+H'=k\sigma^{z}+m_x\sigma^{x}+m_y\sigma^y.
\ee Let us choose a configuration with $m_x=0$ and $m_y=m(y)$ is a shifted step-function which goes from a negative value to a positive value at $y=0.$  It is well-known\cite{jackiw1976} that this Hamiltonian has an exponentially localized zero mode at the domain wall of $m(y)$ given by 
\be
\psi=\mbox{e}^{-\int_{0}^{y} m(y')dy'}\frac{1}{\sqrt{2}}\begin{pmatrix} 1\\ 1\end{pmatrix}
\ee\noindent when the mass jumps from negative to positive as $y$ increases.
On a finite open or periodic edge, $m(y)$ will have to have two domain walls to maintain the proper boundary conditions, and the edge will have two zero modes, one at each domain wall. These localized zero modes carry a half charge each. This is the same result found from Eq. \ref{eq:qsh_charge}. To complete the story in the language above, the QSH system itself has a non-trivial value of $\theta=\pi.$ Thus, its boundary gives a natural place where $\theta$ has a jump from $\pi$ to $0$. The spatial dependence of the $\phi$ parameter is due to the magnetization induced mass. 

We can also generate an adiabatically pumped current. To see this we can add a slow, time-dependent perturbation to the edge Hamiltonian in the following way
\be
H_{edge}(k)=k\sigma^{z}+m\sin\phi(t)\,\sigma^{y}+m\cos\phi(t)\,\sigma^{x}
\ee
where $\phi(t)=2\pi t/T.$ The mass terms are periodic in time with a period of $T$. From the original work by Thouless\cite{thouless1983} we know that as $\phi\to\phi+2\pi$ an integer amount of charge will be pumped, in this case just a single electron per cycle. This is the same current which is reported in Eq. \ref{eq:qsh_current}.

\subsubsection{Response of the Dirac semi-metal}

\begin{figure}[t!]
\label{fig:qsh_plot_charge}
\centering
   \epsfxsize=9cm \epsfbox{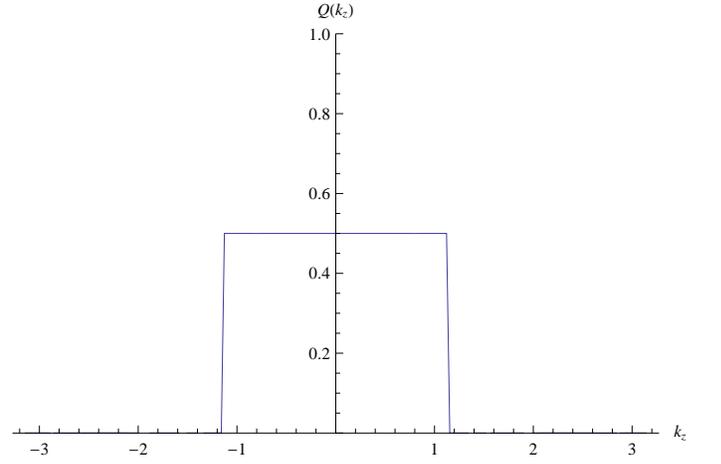}
   \caption{The localized charge on a magnetic domain wall on the surface of a 3D DSM resolved vs. $k_z$, i.e., the direction in which the Dirac nodes are separated in momentum space.  We note that there is a half charge bound at the domain wall only for each state satisfying $|k_{z}|<\cos^{-1}m$. In the plot, we have used $m=0.5$ which means $b_{z}=\tfrac{\pi}{3}$. }
\end{figure}

\begin{figure}[t!]
\label{fig:dsm_b0}
\centering
   \epsfxsize=9cm \epsfbox{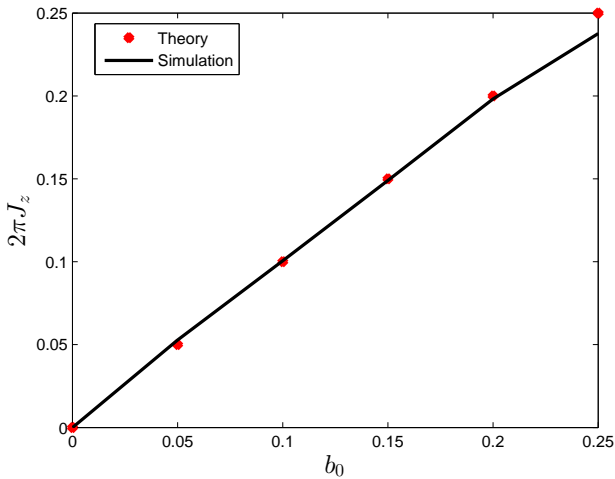}
   \caption{The total current localized at the magnetic domain wall is plotted vs $b_{0}$ for the 3D DSM. The expected value of the total current localized on the domain wall is $\frac{eb_{0}}{2\pi}$. The system size is a cube of $L=30$ lattice sites in every direction with $b_{z}=\tfrac{\pi}{2}$. We used open boundary conditions in both the $x,y$ directions and periodic boundary conditions in the $z$ direction. The red dots are the theoretical result and the black line is the numerical result. The deviation arises due to the importance of  lattice effects at larger values of $b_0.$ }
\end{figure}

Now that we have finished the discussion for a single QSH layer we are ready to move on to the 3D DSM.
We can start from the QSH Hamiltonian, but we need to modify it to include tunneling in the $z$-direction due to the coupled layers. The following model can be used
\bea
\label{eq:dsmham}
&&H_{DSM3}(\mathbf{k},\hat{n})=\sin k_{x}\Gamma^{1}+\sin k_{y}\Gamma^{2}\\ \nonumber
&&+(\cos k_{x}+\cos k_{y}+t_z\cos k_{z}-3)\Gamma^{0}+m\sum_{a=0,3,4}\hat{n}_{a}\Gamma^{a}.
\eea If the 2D layers are in the QSH phase, then when the tunneling term $t_z$ is weak, the system will be in a WTI phase. As it becomes stronger eventually the gap will close at one of the time-reversal invariant momenta along the $k_z$ axis and generate a pair of Dirac nodes to enter the 3D DSM phase. In a recent work\cite{yang2014} this is named a $Z_2$ non-trivial Dirac semi-metal. From the previous patterns of the electromagnetic response we can immediately write the response action
\be
S_{3D}=\frac{e}{2\pi^{2}}\int d^{3}x\,dt\,\epsilon^{\mu\nu\rho\sigma}b_{\mu}A_{\nu}\del_{\rho}\Omega_{\tau}\label{eq:3DSMFinalResponse}
\ee for the 3D DSM where $2b_{\mu}$ is the energy-momentum separation of the Dirac nodes. We now have a natural family of 2D  Bloch Hamiltonians parameterized by $k_z$ $H_{k_z}(k_x,k_y).$ Each of the 2D Hamiltonians, for $k_z$ not at a Dirac node, represents at 2D time-reversal invariant insulator and is classified by the same $Z_2$ invariant as the QSH insulator. As $k_z$ passes through a Dirac node the $Z_2$ invariant jumps from trivial to non-trivial or vice-versa. Thus, one of the regions of $k_z$ between the Dirac nodes will harbor non-trivial topological QSH insulators and thus generate edge states. For each $k_z$ in the topological range we will have a contribution of one layer of QSH to the total electromagnetic response. This is the meaning of Eq. \ref{eq:3DSMFinalResponse}. Ref. \onlinecite{yang2014} has shown that this type of semi-metal requires a uniaxial rotation symmetry to locally stabilize the Dirac nodes. Our model has such a symmetry ($C_4$  rotation around the $z$-axis), and thus represents a stable $Z_2$ non-trivial DSM.  We will leave a more general symmetry analysis of the electromagnetic response to future work.

Let us look at some examples of the physical phenomena associated to Eq. \ref{eq:3DSMFinalResponse}. Just like the case of a single QSH layer, to get a non-trivial response we need to apply a magnetic film to a boundary with non-trivial surface states. As shown in Fig. \ref{fig:dsmSchematic}, for Dirac nodes separated in $k_z$ we can coat the $xz$ boundary plane with a magnetic layer. A translationally invariant magnetic domain wall parallel to the $z$-axis in the magnetic layer (see Fig. \ref{fig:dsmSchematic}) will create a line of low-energy modes which do not disperse with $k_{z}$. Thus for each $k_z$ that contributes a boundary mode we will bind a half charge. We numerically calculated the bound charge at a domain wall as a function of $k_z$ and the result is shown in Fig. \ref{fig:qsh_plot_charge}. We used the mass parameter $m=0.5$ and varied $\phi$ and $\theta$ as functions of $y$ and $x$ respectively according to \ref{eq:dsmham}.

The bound charge response will also occur in a time-reversal invariant WTI system, however a new phenomenon which is not available in the WTI is the generation of a current along the domain wall in the direction along which the Dirac nodes are separated. This can occur if the Dirac nodes are not at the same energy. We can generate this energy difference $2b_{0}$ in our Hamiltonian by adding the term $\gamma\sin k_{z}\mathbb{I}$ to the Hamiltonian in Eq. \ref{eq:dsmham}. When we have a magnetic domain wall and a non-zero $b_{0},$ the localized domain wall states will disperse with energy $E_{dw}=2\gamma\sin k_z$ and this leads to a non-zero current. We calculated this current numerically  as shown in Fig. \ref{fig:dsm_b0}. With a $b_{0}\neq0$, the current  is being generated due to the dispersion of the localized edge modes which now have to traverse between the two Dirac nodes in a continuous fashion. The total current localized on the domain wall is given by 
\bea
J_{z}&=&\frac{eb_{0}}{2\pi^{2}}\int d^{2}x\,(\del_{x}\Omega_{y}-\del_{y}\Omega_{x})\\\nonumber
&=&\frac{eb_{0}}{2\pi^{2}}\int d\theta d\phi\,\frac{1}{2}\sin\theta=\frac{eb_{0}}{2\pi}
\eea\noindent which matches the numerical calculation well until $b_0$ is large enough for lattice effects to become important. This mechanism for current generation is very similar to what occurs to generate the bound edge current due to the bulk magnetization in the 2D DSM as it is related to the dispersion of a 1D band attached to the Dirac nodes. 


 \section{Discussion and Conclusions}\label{sec:conclusion}
In this paper, we have explored the electromagnetic responses of semimetals with point like Fermi surfaces in various spatial dimensions. The response depended generally on a 1 form $b_{\mu}=(b_{0},\vec{b})$ produced by an energy difference $2b_{0}$ between the nodes and a momentum separation $2b_{i}.$ We first introduced a simple 1D model of a metal which helped us understand  some response properties of the WSM by mapping the low-energy behavior of the WSM in a uniform B-field onto copies of the 1D model. This approach works because of the fact that this 1D response is embedded in the 3D WSM response. We then moved onto the case of the 2D DSM which was constructed from layered 1D TIs that are coupled together. The gapless Dirac nodes  which occur in this model each have a Chern-Simons response which, when written in terms of the electromagnetic gauge field, gives us the usual quantum Hall response along with a polarization/magnetization which was encoded in $b_{\mu}.$ In this case, an energy difference between the nodes led to an edge current (bulk orbital magnetization) and a momentum separation between the nodes led to a a boundary charge (bulk polarization). The $\mathcal{TI}$ symmetry which ensures that the Dirac nodes are locally stable also led to the quantization of polarization. The 3D DSM was then analyzed from the perspective that it is a layered QSH system. The Dirac nodes separate trivial regions of momentum space from non-trivial regions and the resultant response follows from the existence of these nontrivial QSH layers. As such,  when a magnetic film is applied to a boundary with non-trivial surface states, we get boundary modes which are localized on  domain walls of the magnetization. Additionally a nonzero $b_{0}$ gives us a localized current which runs along the domain wall. 

There are several natural areas to pursue from this point.  We have shown that we can understand some topological semi-metals, i.e., those with point Fermi surfaces, by stacking topological states in one dimension lower. We only considered the simplest cases in this article and we have barely scratched the surface of the different 1D and 2D states that could be coupled together to form 2D and 3D semi-metal states. Additionally one could take 1D topological wires and stack them into planes, and then take those planes and stack them into 3D to get a secondary WTI, or, if the inter-wire coupling is strong enough, a 3D semi-metal with line Fermi-surfaces. In this case the Lorentz violation enters as an $2$-form $b_{\mu\nu}$ and couples to the EM field via $\int d^4 x \epsilon^{\mu\nu\rho\sigma}b_{\mu\nu}F_{\rho\sigma}.$ In the simplest case this will give rise to lines of Dirac nodes which will have a polarization and magnetization response controlled by $b_{\mu\nu}.$   

In a $D$-dimensional sample, a conventional Fermi-surface is a $D-1$-dimensional surface in momentum space. The response of this metal is given by a $D$ form $b_{\mu_1\ldots \mu_D}$ which is equivalent to an current via $j^{\alpha}_{(b)}\sim\epsilon^{\alpha\mu_{1}\ldots\mu_D}b_{\mu_1\ldots \mu_D}.$ Generically when the Fermi surface is has dimension $D-q$ (codimension $q$) then the response is controlled by a $D-q+1$-form. This type of construction is also useful for discussing the properties of dislocations in WTIs and topological semi-metals\cite{ran2009,HughesYaoQi13}. We will discuss both of these further in Ref. \onlinecite{inpreparation}.

Another immediate application of our results is to the bulk response action of the 3D topological crystalline insulator protected by mirror symmetry \cite{futci,hsiehtci,tanakatci}.  It has been shown that alloys of PbSnTe exhibit a mirror-symmetry protected topological phase. If we consider the $[001]$ surface then there will be four Dirac nodes which all have the same helicity\cite{hsiehtci}, i.e., in our notation for the 2D Dirac semi-metal $\chi_a=+1$ for $a=1, 2, 3, 4.$ To define the response we also need to know the momentum positions of the Dirac nodes, and the sign of the local mass terms at the Dirac nodes. Since the four nodes are symmetrically arranged in the surface BZ let us parameterize their 2D momenta as $\vec{K}_1=(K,0), \vec{K}_2=(0,L), \vec{K}_3=(-K,0),$ and $\vec{K}_4=(0,-L).$ 

The two relevant possibilities for the response coefficients are the Chern number $C_1=\frac{1}{2}\sum_{a=1}^{4} g_a\chi_a$ and $\vec{b}=\frac{1}{2}\sum_{a=1}^{4}g_a\chi_a\vec{K}_a.$ Since the chiralities are all the same we can replace these by $C_1=\frac{1}{2}\sum_{a=1}^{4} g_a$ and $\vec{b}=\sum_{a=1}^{4}g_a\vec{K}_a,$  where we recall that $g_a$ is the sign of the local mass term at the $a$-th Dirac node. Ref. \onlinecite{hsiehtci} showed that there are four possibilities for the $g_a$ due to inversion breaking perturbations, one particular case being $g_1=g_4=-g_2=-g_3=1.$ For this set of mass signs $C_1=0$ and $\vec{b}=(-K,L).$ If we include the other choices of mass sign we get the four possibilities $\vec{b}=(\pm K, \pm L).$ This is interesting because if the top surface and bottom surface have different values of $\vec{b}$ then there will be an interfacial region where the polarization changes and there will be bound charge proportional to the difference. This bound charge arises because on the domain wall between the two regions of the surface there will be low-energy fermion modes. It would be interesting to explore this further to develop the full response theory, but we will leave this for future work.



\emph{Acknowledgements}
We acknowledge useful conversations with B. A. Bernevig, G. Y. Cho, V. Chua, V. Dwivedi,  and especially O. Parrikar.  We acknowledge support from ONR award N0014-12-1-0935.


%


\appendix
\section{Transformation from a Dirac Semi-metal on the Square Lattice to the Honeycomb Lattice}\label{app:graphene}
In this section, we show that graphene can be thought of as an array  of $1+1$-d TI wires. Let us begin with the one dimensional TI given by the following Bloch Hamiltonian:
\be
H(\mathbf{k})= t_{x}(1+m-\cos k_{x}a)\sigma^{x}+t_{x}\sin k_{x}a\,\sigma^{y}
\ee\noindent where $t_x, m$ are parameters and $a$ is the lattice constant.
The system is gapped for all values of $m$ except $m=0$ or $m=1.$ Let us now induce tunneling in the $y$ direction. In the following, the assumption of $y$ being perpendicular to $x$ is not needed. We could have this tunneling along an oblique direction and orthogonality is not required. In this case the Brillouin zone is not a simple square, but it can be a parallelogram. With hopping in the $y$-direction consider the modified Hamiltonian: 
\bea
\label{eq:layered_ti}
H(\mathbf{k})&=&\left[t_{x}+t_{x}m-t_{x}\cos k_{x}a\right.\nonumber\\
&-&\left.t_{\theta}\cos (k_{x}a\cos\theta+k_{y}a\sin\theta)\right]\sigma^{x}\nonumber\\
&+&\left[t_{x}\sin k_{x}a+\overline{t}_{\theta}\sin (k_{x}a\cos\theta+k_{y}a\sin\theta)\right]\sigma^{y}.\nonumber\\
\eea\noindent where we have parameterized the $y$-direction by an angle $\theta$ with respect to the initial $x$-axis. 

Let us now look at the graphene Hamiltonian. It is given by 
\bea
H_{G}(\mathbf{k})&=&-(t_{1}+t_{2}\cos \vec{k}\cdot\vec{a}_{1}+t_{3}\cos\vec{k}\cdot\vec{a}_{2})\sigma^{x}\\ \nonumber
&&+(t_{2}\sin\vec{k}\cdot\vec{a}_{1}+t_{3}\sin\vec{k}\cdot\vec{a}_{2})\sigma^{y}
\eea
where $\vec{a}_{1,2}=\sqrt{3}a\,(\cos(\pi/6),\pm\sin(\pi/6))$. For an easier comparison let us rotate this system in the counter-clockwise direction in real space by an angle $\pi/6$. The two lattice vectors are now given by $\vec{a}_{1}=\sqrt{3}a\,(\cos(\pi/3),\sin(\pi/3))$ and $\vec{a}_{2}=\sqrt{3}a\,(1,0)$. Labeling $\sqrt{3}a=b$, we reduce the Hamiltonian to 
\bea
\label{eq:graphene}
H_{G}(\mathbf{k})&=&-(t_{1}+t_{2}\cos(k_{x}b\cos\pi/3+k_{y}b\sin\pi/3)\\ \nonumber
&&+t_{3}\cos k_{x}b)\sigma^{x}+t_{3}\sin(k_{x}b))\sigma^{y}\\ \nonumber
&&+(t_{2}\sin(k_{x}b\cos\pi/3+k_{y}b\sin\pi/3).
\eea We note that the Hamiltonians in Eq. \ref{eq:graphene} and Eq. \ref{eq:layered_ti} are the same with the following identifications. $t_{1}\rightarrow -(t_{x}+t_{x}m)$, $t_{2}\rightarrow t_{\theta}$, $t_{3}\rightarrow t_{x}$ with the additional constraint $t_{\theta}=\overline{t}_{\theta}$. 

Let us now set all parameters in our model \ref{eq:layered_ti} to 1 except for $\overline{t}_{\theta}$. From our previous statement we know that this will be exactly the same as graphene when $\overline{t}_{\theta}=t_{\theta}=1.$ We want to show that the effect of deforming $\overline{t}_{\theta}$ away from this point is to move the Dirac nodes around in the BZ. Let us look at the gapless points of our model which are the solutions to 
\bea
&&\sin(k_{x}a)+\overline{t}_{\theta}\sin(k_{x}a\cos\theta+k_{y}a\sin\theta)=0\\
&&\cos(k_{x}a)+\cos(k_{x}a\cos\theta+k_{y}a\sin\theta)=1+m.
\eea
In the limit that $\overline{t}_{\theta}=1$, we have $(\tfrac{\pm 1}{a}\cos^{-1}(\tfrac{1+m}{2}),\tfrac{\mp(1+\cos\theta)}{a\sin\theta}\cos^{-1}(\tfrac{1+m}{2}))$ as the gapless points. On the other hand, if $\overline{t}_{\theta}=0$, we have $(0,\pm\cos^{-1}(m))$ as the gapless points. As long as $|1+m|<2$, and $0\leq\overline{t}_{\theta}\leq1$, we get two gapless points in the spectrum but their location depends generically on the model parameters. In this paper, we always use the model in \ref{eq:layered_ti} in the limit of  $t_{x}=1$, $t_{\theta}=1$, $\overline{t}_{\theta}=0$ for describing Dirac semi-metal physics with two bands.

\section{Exact Solution for Boundary States in Topological Semimetal Lattice Models}
\label{sec:edgestuff}
In this Appendix we will study the edge states of the various topological semi-metal lattice models. The solution can be found analytically for the Dirac-type models we have been using following the results of Refs. \onlinecite{creutz,konig2008}. We will begin by solving for the edge states of the two-band lattice Dirac model, i.e., the minimal model for 1+1-d and 2+1-d topological insulators. We will then go on to modify these models to form Dirac and Weyl semi-metal states and solve for their boundary modes. 
\subsection{Exact solution for edge states of the lattice Dirac model}
Consider the model given by 
\bea
\label{eq:2ddham}
\mathcal{H}&=&\epsilon(k)I_{2\times2}+d_{a}(k)\sigma^{a}\\
d_{a}(k)&=&(A\sin(k_{1}),d_{2}(k_2),M(k))\nonumber\\
M(k)&=&M-2B[2-\cos(k_{1})-\cos(k_{2})]\nonumber
\eea where $d_{2}(k_2)$ is an unspecified, but odd, function of $k_2$, and $A, B, M$ are model parameters. Let us fix the sign of $A>0$ and $B>0.$
Additionally, we assume that $\epsilon(k)=0$ for now, but we will add it back in later. Note that with $\epsilon(k)=0$ and $d_{2}(k_2)=-d_{2}(-k_2)$ the model is particle-hole symmetric with the symmetry operator $C=\sigma^x$; it is also inversion symmetric with ${\cal{I}}=\sigma^z.$ The energy eigenvalues are given by 
\begin{eqnarray}
E_{\pm}&=&\pm\sqrt{d_{a}d_{a}}\nonumber\\
&=&\pm\sqrt{A^{2}\sin^{2}(k_{1})+d_{2}^{2}(k_2)+M^{2}(k)}.
\end{eqnarray}
 This spectrum is a gapped insulator as long as $\sqrt{d_{a}d_{a}}\neq 0$. One gapless critical point of this model occurs when $k_{1}=k_{2}=M=0$ and for $M<0$ ($M>0$) the model is in a trivial (topological) insulator phase.  
 
 When the system is tuned to the non-trivial phase there are gapless edge states which can be shown explicitly in a finite strip geometry or a cylinder geometry.  Let us assume that the system has boundaries at $x_{1}=0,L$ and is infinite in the $x_{2}$ direction. Since we have an inhomogeneous system with open boundaries we need to Fourier transform the Bloch Hamiltonian back from $k_{1}$ to $x_{1}$ via the substitution
\be
c_{\vec{k}}=\frac{1}{\sqrt{L}}\sum_{j}e^{ik_{1}x_{1}(j)}c_{k_{2},j}.
\ee 
This reduces the Hamiltonian to
\bea
\mathcal{H}&=&\sum_{k_2,j}(\mathcal{M}c^{\dagger}_{k_2,j}c_{k_2,j}+\mathcal{T}c^{\dagger}_{k_2,j}c_{k_2,j+1}+\mathcal{T}^{\dagger}c^{\dagger}_{k_2,j+1}c_{k_2,j})\nonumber\\
\mathcal{M}&=&A\sin(k_2)\sigma^{2}-2B\left[2-\frac{M}{2B}-\cos(k_2)\right]\sigma^{3}\nonumber\\
\mathcal{T}&=&\frac{iA}{2}\sigma^{1}+B\sigma^{3}.\label{eq:DiracFTLattice}
\eea

Since we are interested in the exponentially localized edge states, we will focus on a solution ansatz of the form
\be
\psi_{\alpha}(j)=\lambda^{j}\phi_{\alpha}
\ee
where $\lambda$ is a complex number, $j$ is the site index in the $x_{1}$ direction, and $\phi_{\alpha}$ is a $2$ component spinor with $\alpha=1,2$. We will first look for a solution at $k_2=0$, and since the Hamiltonian is  particle-hole symmetric, the mid-gap edge state for this momentum will occur at $E=0.$  Acting with the Hamiltonian at $k_2=0$ on our ansatz yields the equation
\be
\left[\frac{iA}{2}(\lambda^{-1}-\lambda)\sigma^{1}+B(\lambda+\lambda^{-1})\sigma^{3}+\mathcal{M}(0)\sigma^{3}\right]\phi=0.\nonumber
\ee
Multiplying this equation on both sides by $\sigma^{3}$ gives us
\be
\label{eq:neg}
\frac{A}{2}(\lambda^{-1}-\lambda)(i\sigma^{3}\sigma^{1})\phi=-[B(\lambda+\lambda^{-1})+\mathcal{M}(0)]\phi.
\ee
The operator $i\sigma^{3}\sigma^{1}$ has eigenvalues $\pm1$. First consider $i\sigma^{3}\sigma^{1}\phi=-\phi$, under which Eq. \ref{eq:neg} becomes a quadratic equation in $\lambda$ which can be solved to find:
\be
\lambda_{(1,2)}=\frac{-\mathcal{M}(0)\pm\sqrt{\mathcal{M}^{2}(0)+(A^{2}-4B^{2})}}{A+2B}.
\ee
Thus, from the quadratic equation we have two $\lambda$ solutions  for the $-1$ eigenvalue (chirality) of $i\sigma^{3}\sigma^{1}.$ For every solution $\lambda$ we find that $\lambda^{-1}$ is a solution for $i\sigma^{3}\sigma^{1}\phi=+\phi,$ and thus for each eigenvalue of  $i\sigma^{3}\sigma^{1}$ there are two possible values of $\lambda.$ Let us label the eigenstates of  $i\sigma^{3}\sigma^{1}$ as $\phi_{\pm}$ corresponding to the chiralities. The most general edge state can by written as 
\be
\label{eq:edge_lambda}
\psi_{j}(k_2=0)=\left(a\lambda_{(1)}^{j}+b\lambda_{(2)}^{j}\right)\phi_{+}+\left(c\lambda_{(1)}^{-j}+d\lambda_{(2)}^{-j}\right)\phi_{-}
\ee\noindent but to satisfy open boundary conditions we must have $a=-b$ and $c=-d$ since $\phi_{\pm}$ are linearly independent. Additionally, since the mode must be normalizable, we can only keep positive or negative powers of $\lambda$ and thus only one normalizable mode exists (on each edge) as long as the $\lambda$ do not lie on the unit circle, i.e. $|\lambda_{(1,2)}|\neq1$. If $\vert\lambda_{(1,2)}|=1$ an edge state solution does not exist at all. We also note that solutions with eigenvalues $\lambda$ and $\lambda^{-1}$ are localized on opposite edges of the system based on the form of Eq. \ref{eq:edge_lambda}.

Now, let us generalize this solution for $k_{2}\neq0$. We see that the term $\cos(k_{2})$ simply acts as a shift of the parameter $M$ and can be easily accounted for. We also see that $[i\sigma^{3}\sigma^{1},\sigma^{2}]=0$  and clearly $[i\sigma^{3}\sigma^{1},I_{2\times 2}]=[\sigma^{2},I_{2\times 2}]=0$. So, the terms $d_{2}(k_2)\sigma^{2}$ and $\epsilon(k_2)I_{2\times 2}$ can simply be included as $k_2$ dependent shifts of the energy. These terms change the energy dispersion of the edge states, but the eigenstates remain the same. The energy for the edge state for any $k_{2}$ is given by 
\be
E_{\pm}(k_{2})=\epsilon(k_2)\mp d_{2}(k_2).
\ee 
Importantly, this dispersion does not hold across the entire $k_2$ Brillouin zone because there will exist some values of $k_2$ where the values of $\lambda$ coming from a solution of 
\bea
\lambda_{(1,2)}(k_2)&=&\frac{-m(k_2,M)\pm\sqrt{m(k_2,M)^{2}+(A^{2}-4B^{2})}}{A+2B}\nonumber\\
m(k_2,M)&=&-2B[2-M/2B-\cos(k_{2})]
\eea\noindent do not yield normalizable modes. 
For the edge states to be normalizable, we have to satisfy the condition that $|\lambda_{(1,2)}|\neq 1$ which can be reduced to 
\be
-2B<m(k_2,M)<2B
\ee\noindent for each $k_2.$
The special points in $k_2$-space where the inequalities become equalities are places in the energy spectrum where the edge states merge with the delocalized bulk states. Beyond these special values of $k_2$ the edge states no longer exist. This result, which consists of the dispersion, wavefunctions, and conditions for normalizablity represents the full analytic solution of the lattice edge states.

\subsection{Edge theory for two dimensional semimetal}\label{app:edgeDSM}
Based on the solution for the 2-band Dirac model we can immediately adapt it to the case of topological semi-metal states with minor modifications. First, let us consider the 2+1-d Dirac semi-metal including the possibility of the inversion breaking ($m_A$) and time-reversal breaking ($m_B$) mass terms discussed in Section \ref{sec:twoDDSM}. The Hamiltonian takes the form
\bea
\label{eq:2ddham}
\mathcal{H}&=&\epsilon(k)I_{2\times2}+d_{a}(k)\sigma^{a}\nonumber\\
d_{a}(k)&=&(A\sin k_{1},m_{A}+m_{B}\sin k_{2} ,M(k))\nonumber\\
M(k)&=&M-2B[1-\cos k_{1}-\cos k_{2}]\nonumber\\
\epsilon(k)&=&\gamma\sin(k_{2}).\nonumber
\eea
Depending on the values of $M$ and $B$ this Hamiltonian can have Dirac nodes at $(0,\pm k_{0})$ where $k_{0}=\cos^{-1}(-M/2B).$  For a cylinder geometry with open boundary in the $x_1$ direction and periodic boundary conditions in the $x_2$ direction,  this model will have edge states when the Dirac nodes exist. The edge states will occur between the Dirac nodes, but depending on the values of $M$ and $B$ they either stretch between the nodes within the Brillouin zone or across the Brillouin zone boundaries. For a choice such that they stretch within the Brillouin zone, the energies of the edge state branches on the two edges are given by 
\be
E_{\pm}=\gamma\sin(k_{2})\mp\vert m_{A}+m_{B}\sin(k_{2})\vert\quad |k_{2}|<k_{0}.\label{eq:DSMedge}
\ee
 The restriction on the range of $k_{2}$ arises from a modified condition on normalizability through the relation 
\bea
-2B&<&m(k_2,M)<2B\nonumber\\
 m(k_2,M)&=&-2B[1-M/2B-\cos(k_{2})].
\eea

We can observe several interesting details from Eq. \ref{eq:DSMedge}. First we see that if we let $m_A=\gamma=0$ but $m_B\neq0,$ then the dispersion matches that of the edge states of the 2+1-d Chern insulator\cite{haldane1988} as it must since the $m_B$ term is exactly the mass term required to convert a 2D Dirac semi-metal into a Chern insulator. If only $m_A$ is non-zero and $m_B=\gamma=0,$ then we get two flat bands, one band on each edge. Finally, if we have $\gamma\neq 0$ and $m_A\neq 0$ but $m_B=0,$ then the two flat bands from the previous case will each disperse, and at half-filling there will be bound currents on each edge that, in the limit $m_B\to 0$ give rise to the magnetization discussed in  Section \ref{sec:twoDDSM}. This matches our expectation because if $M$ and $B$ are tuned to values where $k_0\neq 0$ as we have assumed, then for non-zero $\gamma$ there will be an energy difference between the two Dirac nodes given by $\Delta E = 2\vert \gamma\sin k_0\vert.$

\subsection{Edge theory in the case of the Weyl semimetal}

The Weyl semi-metal also has a Hamiltonian which is given by the form of Eq. \ref{eq:2ddham} where
\bea
\label{eq:2band_wsm}
\mathcal{H}&=&\epsilon(k_{2},k_{3})I_{2\times2}+d_{a}(k)\sigma^{a}\nonumber\\
d_{a}(k)&=&(A\sin k_{1},A\sin k_{2},M(k))\nonumber\\
M(k)&=&M-2B[2-\cos k_{1}-\cos k_{2}-\cos k_{3} ]\nonumber
\eea where we can let $\epsilon(k)$ be a generic function of $k_2,k_3.$
 This Hamiltonian has two gapless Weyl nodes for $|M/2B|<1$ at $(k_{1},k_{2},k_{3})=(0,0,\pm k_0)$ where $k_0= \cos^{-1}(-M/2B)).$  Let us assume again that our system has boundaries at $x_{1}=0,L$ and that it is periodic in the other two directions. The main change between this case and the previous ones is that the condition for existence of these edge states at each momentum gets modified because the mass $m(k,M)$ is now parameterized by $k_2$ and $k_3.$ The new normalizability condition that must be satisfied is given by
\bea
\label{eq:constr}
-2B&<&m(k,M)<2B\\
m(k,M)&=&-2B[2-M/2B-\cos k_{2}-\cos k_{3}].\nonumber
\eea
The edge state energies in this case are given by $E_{\pm}=\epsilon(k_2,k_3)\mp \vert A\sin k_{2}\vert$.

Let us consider a simple case first where $\epsilon(k)\equiv 0.$ We want to consider the structure of the boundary modes on a surface with the normal vector in the $x$-direction and the surface Brillouin zone is the $(k_2, k_3)$ plane.   If we set the chemical potential to zero, we see that there exist Fermi arcs in this plane when $E_{\pm}=\mp \vert A\sin k_{2}\vert=0$ which allows for $k_2=0, \pi$ and does not explicitly depend on $k_3.$ The correct value of $k_2$ depends on the particular choice of $M$ and $B,$ so without loss of generality let us choose $k_2=0.$ The boundary state existence condition of Eq. \ref{eq:constr}, which does depend on $k_3$, can be simplified to give us the condition that boundary states are only present when  $|k_{3}|< k_{0}$. Thus, for this case there exist Fermi arcs that are straight lines which go from $(k_{2},k_{3})=(0,-k_{0})$ to $(k_{2},k_{3})=(0,k_{0})$ in the surface Brillouin zone.  

To get more non-trivial Fermi-arc shapes inversion symmetry needs to be broken to lift the degeneracy between the arcs on the two edges.
Let us consider the Hamiltonian given by \ref{eq:2band_wsm} with $\epsilon(k)=\gamma\sin k_{3}$. The energy is given by $E_{\pm}=\gamma\sin k_{3}\mp \vert A\sin k_{2}\vert$. With the chemical potential again set at $\mu=0$ and, for example $\gamma=A/2$, we see that the points in the Fermi arc must satisfy $\sin k_3=\pm 2\sin k_2$ and Eq. \ref{eq:constr}. The solutions to these constraints are complicated functions of $(k_2,k_3)$ and must, in general, be solved numerically.



\subsection{Tunneling Between Edge States}
In this section, we will use our model of the boundary states for the topological semimetals to study properties at interfaces between semimetals with different Lorentz violating parameters, and thus different boundary state structures. Let us consider the interface between two, semi-infinite 2D DSMs first. Assume that the interface is at $x=0$ with parameters for $x\leq0$ given by $A, B, M, \gamma$ and for $x>0$ given by $A', B', M', \gamma'$.  


The lattice Hamiltonian for $x\leq0$ is given by
\bea
\mathcal{H}&=&\left(\sum_{j,k_2=-\infty}^{j=-1}H_{j}(k_2)\right)+\mathcal{M}c^{\dagger}_{0,k_2}c_{0,k_2}\nonumber\\
&+&\mathcal{T}c^{\dagger}_{0,k_2}c_{1,k_2}+\mathcal{T}^{\dagger}c^{\dagger}_{1,k_2}c_{0,k_2}
\eea
where $H_{j}$ is the lattice Hamiltonian we have been previously using. To be specific, 
\bea
\mathcal{M}&=&\gamma\sin k_{2}\mathbb{I}+(m_{A}+m_{B}\sin k_2)\sigma^{2}\nonumber\\&-&2B\left[1-\frac{M}{2B}-\cos k_2\right]\sigma^{3}\nonumber\\
\mathcal{T}&=&\frac{iA}{2}\sigma^{1}+B\sigma^{3}.\label{eq:DiracFTLattice}
\eea The Hamiltonian for $x>0$ has a similar form, just with different parameters. 
We notice that there is a natural hopping term to connect the two systems.  The matrix element for tunneling from site $0$ to site $1$ is $\mathcal{T}^{\dagger}$ and the matrix element for tunneling from site $1$ to site $0$ is $\mathcal{T}.$ 

Assume that the edge states are of chiralities $c,c'$ which take on the values $+1,-1$. The chirality of the state is simply defined as its eigenvalue under the $i\sigma^3\sigma^1$ matrix discussed in the previous section. The state on the left edge and right edge are given by $\phi_{c},\phi_{c'}$ respectively. The Hamiltonian in the edge subspace is given by 
\be
H=
\begin{pmatrix}
\langle\phi_{c}|\mathcal{M}|\phi_{c}\rangle & \langle\phi_{c}|\mathcal{T}|\phi_{c'}\rangle \\
\langle\phi_{c'}|\mathcal{T}^{\dagger}|\phi_{c}\rangle & \langle\phi_{c'}|\mathcal{M}'|\phi_{c'}\rangle
\end{pmatrix}.
\ee
We can evaluate the matrix elements in each case by using the fact that $|\phi_{\pm}\rangle$ are eigenstates of $-\sigma^{2}$. When the chiralities are opposite, i.e $cc'<0$, we have $\langle\phi_{\pm}\vert\mathcal{M}\vert\phi_{\pm}\rangle=\gamma\sin k_{2}\mp(m_{A}+m_{B}\sin k_{2})$, $\langle\phi_{+}|\mathcal{T}|\phi_{-}\rangle=\langle\phi_{+}\vert\mathcal{T}|\phi_{-}\rangle^{\dagger}=B-A/2$. Off diagonal terms turn out to be zero if $cc'>0$ i.e. we have $\langle\phi_{+}|\mathcal{T}|\phi_{+}\rangle=\langle\phi_{-}|\mathcal{T}|\phi_{-}\rangle=0$. So, in the case of $cc'>0$, which is to say we have the same chirality for the edge states the tunneling Hamiltonian is given by 
\be
H=\frac{\gamma+\gamma'}{2}\sin k_{2}\mathbb{I}\pm\left(m_{A}+(m_{B}+\frac{\gamma-\gamma'}{2})\sin k_{2}\right)\sigma^3.\label{eq:B18}
\ee
We see that the edges don't mix and are only completely gapped when the inversion symmetry is broken (i.e. $m_A$  non-zero). When they are of opposite chiralities, the tunneling Hamiltonian is given by 
\begin{align}
H&=\frac{\gamma+\gamma'}{2}\sin k_{2}\mathbb{I}\pm \label{eq:B19}\\\nonumber
&\left(m_{A}+(m_{B}+\frac{\gamma-\gamma'}{2})\sin k_{2}\right)\sigma^{3}+(B-A/2)\sigma^{1}.
\end{align}
We see that the term $B-A/2$ when nonzero acts like a mass term and gaps the edge out in this case. In the models we consider, $A=1,B=-1/2$ and $A-B/2\neq0$. In the case when the edge modes have the same chirality the $\pm$ signs in Eq. \ref{eq:B18} refer to the chirality itself. In the case when the edge modes have the opposite chirality the $\pm$ signs in Eq. \ref{eq:B19} refer to whether the left edge has $+$ or $-$ chirality. 

An important thing to notice is that $M$ and $M'$ do not enter the edge Hamiltonians, however it still has an important effect. The above analysis tells us that the edge modes can gap each other out when they both exist at the same momentum $k_{2}$. However, it is $M$ and $M'$ that control where the Dirac nodes are  and therefore the domain of existence of the edge states in $k_{2}.$ So, those edge states on one edge with a momentum $k_{2}$ which do not have a counterpart on the other edge will remain gapless regardless. Thus, the edge states will only be removed if the domain of existence overlaps in the two systems. 

\subsection{Tunneling in Weyl semimetals}
\label{sec:tunnel}
Let us start off with the Hamiltonian which is of the same flavor as before with
\bea
\mathcal{M}&=&A\sin k_{2}\sigma^{2}-2B\left[2-\frac{M}{2B}-\cos k_2-\cos k_{3}\right]\sigma^{3}\nonumber\\
\mathcal{T}&=&\frac{iA}{2}\sigma^{1}+B\sigma^{3}.\label{eq:DiracFTLatticeA}
\eea
Let us assume that again that we have an edge at $x=0$ and the same setup as the 2D Dirac semi-metal. For $x\leq 0$ we have parameters $A,B,M$ and for $y>0$ we have parameters $A',B',M'$. Our analysis from the previous subsection helps us immensely here. The edge states $\vert\phi_{c}\rangle$ are again eigenvectors of $-\sigma^{2}$. The edge Hamiltonian when we have same chiralities on the two edges is again given by 
\be
H=\pm A\sin k_{2}\mathbb{I}.
\ee
On the other hand, when the edge states have opposite chiralities, the edge Hamiltonian is
\be
H=\pm A\sin k_{2}\sigma^{3}+(B-A/2)\sigma^{1}.
\ee
So, yet again, when the edges have opposite chiralities, the term $B-A/2$ acts like a mass term and gaps the modes out. This is of course only valid if the edge states exist at the same $k_{3}$. Edge states with a momentum $k_{3}$ which do not have a counterpart on the other edge will remain gapless. The $\pm$ signs are related to the same definitions in the previous subsection. 
There could be more complications when a term $\epsilon(k_{2},k_{3})\mathbb{I}$ is added to the Hamiltonian. This modifies the surface Fermi arcs from being straight lines to some other complicated structure. When this happens, only those states on the surface which are degenerate at the same momenta $k_{2},k_{3}$ gap each other out.

\section{Details on Choice of Brillouin Zone for 4-node Dirac Semimetal}\label{app:BZ}
In the case where the BZ is spanned by $\mathbf{G}_{F}=2\pi(1,1)$ and $\mathbf{G}_{N}=2\pi(m+1,m)$, for non-zero $m$, imagine the BZ as having the four corners given by the points $\mathbf{O}, \mathbf{G}_{F}, \mathbf{G}_{N},\mathbf{G}_{F}+\mathbf{G}_{N}$ where $\mathbf{O}$ is the origin. We denote this new BZ as $\Lambda$. We note that the nodes at $\pm(\tfrac{\pi}{2},\tfrac{\pi}{2})$ lie on the edge connecting $\mathbf{O}$ and $\mathbf{G}_{F}$ at $(\tfrac{\pi}{2},\tfrac{\pi}{2})$ and $(\tfrac{3\pi}{2},\tfrac{3\pi}{2})$ respectively. So, their separation when projected onto $\mathbf{O}\rightarrow\mathbf{G}_{F}$ edge of the BZ is given by $0\,\mbox{mod}\,2\pi$ if the helicities of these two nodes are the same (as in $H_{1}^{(4)}$) and $\pi\,\mbox{mod}\,2\pi$ when they are opposite (as in  $H_{2}^{(4)}$). When projected onto the other edge, they coincide and give us no polarization. 

The nodes $\pm(\tfrac{\pi}{2},-\tfrac{\pi}{2})$ however are harder to analyze. We are allowed to shift these nodes by $2\pi(h,k)$ where $h,k\in\mathbb{Z}$ to bring them into the BZ $\Lambda$ we consider here. The way we find $h,k$ is to compute the slope `$t$' of the vector $(\pm\tfrac{\pi}{2}+2\pi h,\mp\tfrac{\pi}{2}+2\pi k)$ and make sure that 
\be
\frac{m}{m+1}\leq t\leq 1
\ee
We note that after carrying out this procedure, we have $(-\tfrac{\pi}{2},\tfrac{\pi}{2})\rightarrow(-\tfrac{\pi}{2}+2\pi(m+1),\tfrac{\pi}{2}+2\pi m)$ and $(\tfrac{\pi}{2},-\tfrac{\pi}{2})\rightarrow(\tfrac{\pi}{2}+2\pi m,-\tfrac{\pi}{2}+2\pi m)$ where the corresponding slopes are given by $\tfrac{4m+1}{4m+3}$ and $\tfrac{4m-1}{4m+1}$ which are both greater than $\tfrac{m}{m+1}$ for $m\geq 1$. So, we now know where these nodes sit inside our new BZ. These new nodes have a momentum difference given by the vector $\pi(1,1)$, which means that their separation is parallel to the BZ edge connecting $\mathbf{O}\rightarrow\mathbf{G}_{F}$. So, they must project to $(\tfrac{3\pi}{2},\tfrac{3\pi}{2})$ and $(\tfrac{\pi}{2},\tfrac{\pi}{2})$ inevitably. Which node projects to which of these two points is different for $m$ even and $m$ odd, but in either case, we must have that the separation of these nodes along the $\mathbf{O}\rightarrow\mathbf{G}_{F}$ edge to be $0\,\mbox{mod}\,2\pi$ when they have the same helicities ($H_{1}^{(4)}$)  or $\pi\,\mbox{mod}\,2\pi$ when they have opposite helicities ($H_{2}^{(4)}$) . Also, because they lie along a line parallel to the $\mathbf{O}\rightarrow\mathbf{G}_{F}$ edge, they project to the same point on the $\mathbf{O}\rightarrow\mathbf{G}_{N}$ edge of the BZ and don't give rise to any polarization on that edge.

So, putting together everything, we see that whichever configuration of helicities we consider and whichever edge we consider, the total polarization due to all four nodes will be always add up to $0\,\mbox{mod}\,\pi$.

\section{K-matrix formalism}\label{app:Kmatrix}
The action in Eq. \ref{eq:CI} can be rewritten as 
\begin{equation}
S_{eff}=\frac{e^2}{4h}\int d^3 x K_{ab}\epsilon^{\mu\nu\rho}A_{(a)\mu}\partial_\nu A_{(b)\rho}
\end{equation}\noindent where $K_{ab}=\chi_a g_a \delta_{ab}.$ From these independent currents and gauge fields we can extract the electromagnetic response which couples democratically to each Dirac cone via a 2N-dimensional  ``charge"-vector $t_{EM}=(e,e,\ldots,e,e)^{T}$ where $e$ is the electron charge. The Hall conductivity is then $\sigma_{xy}=\tfrac{1}{2h}t_{EM}^{T}Kt_{EM}.$ We can also define a valley charge vector $t_{V}=(\chi_1,\chi_2,\ldots, \chi_{2N})^T.$ We can define a valley Hall conductivity via $\sigma^{V}_{xy}=\tfrac{1}{h}t_{EM}^{T} K t_{V}$ which determines the valley current in response to an electromagnetic field. Finally, we can define a valley-valley Hall conductivity via $\sigma^{VV}_{xy}=\tfrac{1}{2h}t_{V}^{T} K t_V$ which determines the amount of valley current that flows in response to a valley electromagnetic field (generated, for example, by strain). 

In general we may have other interesting types of charge vectors $t_{S}$ if we have more symmetries, e.g., spin-rotation symmetry, or point-group symmetries, that correspond to the quantum numbers carried by the corresponding Dirac cones. We can define charge and valley Hall conductivities of those additional quantum numbers by $\sigma^{S}_{xy}=\tfrac{1}{h}t_{EM}^T Kt_S$ and $\sigma^{VS}_{xy}=\tfrac{1}{h}t_{V}^T Kt_S.$ As an example, suppose that we have translation symmetry in spacetime, which gives rise to conserved momentum and energy. For translation along the x-direction each Dirac cone has a momentum component $k_{(i)}^x$ leading to a charge vector $t_{x}=\hbar(k_{(1)}^x, k_{(2)}^x,\ldots, k_{(2N)}^x)^T$. We could use this to define the charge polarization along the $y$ direction as $P_{1}^{y}= \tfrac{1}{2h}t_{EM}^{T}Kt_{x}$. This can be written in a more covariant way as $P_{1}^{a}=\tfrac{1}{2h}\epsilon^{ab}t_{EM}^{T}Kt_{b}$ and $M=\tfrac{1}{2h}t_{EM}^{T}Kt_{\epsilon}$ where $t_{\epsilon}=\hbar(\epsilon_{(1)}, \epsilon_{(2)},\ldots, \epsilon_{(2N)})^T$.

Let us consider a few explicit examples. The simplest case is $N=1$ where the the Dirac cones are specified, without loss of generality by $(+,{\bf{P}}_{(1)},\epsilon_1, g_1)$and  $(-,{\bf{P}}_{(2)},\epsilon_2, g_2).$ Up to global signs, the two possible K-matrices are $K_1=\mathbb{I}$ and $K_2=\sigma^z.$ The K-matrix $K_1$ ($K_2$) corresponds to the case of a time-reversal symmetry (inversion symmetry) breaking anomalous response. The electromagnetic and valley charge vectors for both K-matrices are $t_{EM}=(e,e)^{T}$ and $t_{V}=(1,-1)^{T}.$ For $K_1$ we easily find $\sigma_{xy}=e^2/h$, $\sigma^{V}_{xy}=0$ and $\sigma^{VV}_{xy}=1/h.$ For $K_2$ we have $\sigma_{xy}=\sigma^{VV}_{xy}=0$ and $\sigma^{V}_{xy}=\frac{e}{h}.$ 

Now let us consider translation invariance so that we can construct a charge vector associated to the energy and momentum of each Dirac point $t_{x}=(k_{(1)x}, k_{(2)x}),$ $t_{y}=(k_{(1)y}, k_{(2)y}),$ $t_{\epsilon}=(\epsilon_{(1)}, \epsilon_{(2)}).$ We can see that the Polarization would be $P_{1}^{a}=\tfrac{1}{4\pi}\epsilon^{ab}(k_{(1)b}+k_{(2)b})$ when $K=\mathbb{I}$ and $P_{1}^{a}=\tfrac{e}{4\pi}\epsilon^{ab}(k_{(1)b}-k_{(2)b})$ when $K=\sigma^{z}$. The Magnetization would be given by $M=\tfrac{e}{4\pi}(\epsilon_{(1)}-\epsilon_{(2)})$ when $K=\sigma^{z}$ and $M=\tfrac{e}{4\pi}(\epsilon_{(1)}+\epsilon_{(2)})$ when $K=\mathbb{I}$.

We can also consider a more complicated example with $N=2$ which will have four Dirac cones. Without loss of generality take $\chi_1=\chi_2=1$ and $\chi_3=\chi_4=-1.$ The electromagnetic and valley charge vectors are $t_{EM}=(e, e, e, e)^{T}$ and $t_{V}=(1, 1, -1, -1).$ We can also define two other useful, linearly-independent charge vectors $t_{U}=(1, -1, -1, 1)$ and $t_{W}=(1, -1, 1, -1).$ There are $2^4=16$ possible K-matrices but we only need to consider eight since the other eight differ by an overall sign. These eight are
\begin{eqnarray}
K_1&={\textrm{diag}}[ 1, 1, 1, 1]\;\;\;\; &K_2={\textrm{diag}}[ 1, 1, -1, -1]\nonumber\\
K_3&={\textrm{diag}}[ 1, -1, -1, 1]\;\;\;\; &K_4={\textrm{diag}}[ -1, 1, -1, 1]\nonumber\\
K_5&={\textrm{diag}}[ 1, 1, -1, 1]\;\;\;\; &K_6={\textrm{diag}}[ 1, -1, 1, 1]\nonumber\\
K_7&={\textrm{diag}}[ -1, 1, 1, 1]\;\;\;\; &K_8={\textrm{diag}}[ 1, 1, 1, -1].
\end{eqnarray}\noindent We can tabulate their (dimensionless) electromagnetic responses via $\tfrac{1}{2}t_{EM}^{T}Kt_{\alpha}$ where $\alpha=$ EM, V, U, and W. We find :
\begin{equation}
\left[\begin{array}{ccccc}
&EM & V & U & W \\
K_1& 2 & 0 & 0 & 0\\
K_2& 0 & 2 & 0 & 0\\
K_3& 0 & 0 & 2 & 0\\
K_4& 0 & 0 & 0 & 2\\
K_5& 1 & 1 & 1 & 1\\
K_6& 1 & -1 & 1 & -1\\
K_7& 1 & -1 & -1 & 1\\
K_8& 1& 1 & -1 & -1\\
\end{array}\right].
\end{equation}

\end{document}